%% file: main.tex
\documentclass[accepted]{uai2023} 

\usepackage[american]{babel}

\usepackage{natbib} 
    \renewcommand{\bibsection}{\subsubsection*{References}}
\usepackage{mathtools} 
\usepackage{booktabs} 
\usepackage{tikz} 

\usepackage{xr}

\usepackage{array}
\usepackage{caption}
\usepackage{graphicx}
\usepackage{siunitx}
\usepackage[normalem]{ulem}
\usepackage{colortbl}
\usepackage{multirow}
\usepackage{hhline}
\usepackage{calc}
\usepackage{tabularx}
\usepackage{threeparttable}
\usepackage{wrapfig}
\usepackage{adjustbox}
\usepackage{hyperref}

\usepackage{import}
\usepackage{amsmath, amssymb, amsfonts}
\usepackage{xcolor}
\usepackage{color}
\usetikzlibrary{tikzmark}
\usepackage{multirow}
\usepackage{bm}
\usepackage{graphicx}
\usepackage{enumerate}
\usepackage[makeroom]{cancel}
\usepackage{hyperref}
\usepackage{xcolor}

\usepackage{amsthm}

\newtheorem{example}{Example}

\newtheorem{lemma}{Lemma}
\newtheorem{theorem}{Theorem}

\newtheorem{condition}{Condition}

\newcommand{\E}{\mathbb{E}}
\newcommand{\N}{\mathbb{N}}
\DeclareMathOperator{\pa}{pa}

\title{Sufficient Identification Conditions and Semiparametric Estimation under Missing Not at Random Mechanisms}

\author[1]{Anna~Guo\vspace{-0.3cm}}
\author[2]{Jiwei~Zhao}
\author[1]{Razieh~Nabi}
 \affil[$\text{\textcolor{white}{1}}$]{%
	{\hspace{.cm} anna.guo@emory.edu \hspace{1.7cm} jiwei.zhao@wisc.edu \hspace{1.7cm} razieh.nabi@emory.edu \vspace{0.25cm}}
}
\affil[1]{%
    Dept. of Biostatistics and Bioinformatics\\
    Emory University\\
    Atlanta, Georgia, USA
}
\affil[2]{%
    Dept. of Biostatistics \& Medical Informatics\\
    University of Wisconsin\\
    Madison, Wisconsin, USA
}

\begin{document}
\maketitle

\begin{abstract}
  Conducting valid statistical analyses is challenging in the presence of missing-not-at-random (MNAR) data, where the missingness mechanism is dependent on the missing values themselves even conditioned on the observed data. Here, we consider a MNAR model that generalizes several prior popular MNAR models in two ways: first, it is less restrictive in terms of statistical independence assumptions imposed on the underlying joint data distribution, and second, it allows for all variables in the observed sample to have missing values. This MNAR model corresponds to a so-called \textit{criss-cross} structure considered in the literature on graphical models of missing data that prevents nonparametric identification of the entire missing data model. Nonetheless, part of the complete-data distribution remains nonparametrically identifiable. By exploiting this fact and considering a rich class of exponential family distributions, we establish sufficient conditions for identification of the complete-data distribution as well as the entire missingness mechanism. We then propose methods for testing the independence restrictions encoded in such models using odds ratio as our parameter of interest. We adopt two semiparametric approaches for estimating the odds ratio parameter and establish the corresponding asymptotic theories: one involves maximizing a conditional likelihood with order statistics and the other uses estimating equations. The utility of our methods is illustrated via simulation studies. 
\end{abstract}

\section{Introduction}
\label{sec:intro}

Conducting valid statistical analyses is challenging in the presence of missing data as the observed data may not be representative of the population of interest. According to the terminology of \cite{rubin1976inference}, a missingness mechanism is called missing-at-random (MAR) if it only depends on the observed data values, and it is called missing-not-at-random (MNAR) if it is dependent on the missing values themselves even conditioned on the observed data. Under a MAR model, identification of a target parameter as a function of the observed data is a relatively straightforward task, and estimation strategies are well-studied, ranging from likelihood-based methods such as expectation maximization \citep{dempster77maximum, little2002statistical}, to multiple imputation \citep{rubin87multiple}, inverse probability weighting \citep{robins1994estimation, li2013weighting}, and semiparametric methods closely related to the estimation of causal parameters \citep{robins1995analysis, tsiatis06missing}. On the other hand, MNAR mechanisms are substantially more complicated and under-studied, yet they are construed as the most prevalent form of missingness mechanisms in practice.  

In the presence of MNAR mechanisms, it is generally not possible to express the underlying \textit{complete-data} distribution as a function of the \textit{observed data} distribution without imposing additional assumptions. A lack of identification result implies that there exist at least two models that differ in their respective complete-data distribution but share the same observed data distribution. A well-known example of a non-identified MNAR mechanism is the non-ignorable non-response model in survey sampling, where the response variable directly causes its own missingness, often referred to as a \textit{self-censoring} missingness mechanism. Other MNAR models include scenarios where missingness of a variable depends on other variables that themselves could be missing. 

Common approaches for making progress in non-identified MNAR models include imposing, often untestable, (semi)parametric assumptions that yield identification \citep{wu1988estimation, little2002statistical, zhao2015semiparametric}. For instance, in order to deal with the self-censoring mechanism involving a univariate response variable, several authors have considered the presence of a fully observed variable along with certain assumptions to identify and estimate distributional quantities involving the response variable -- e.g., \cite{wang2014instrumental} considers a \textit{shadow variable}\footnote{The authors refer to $X$ as the instrumental variable. However, following the work of \citep{miao2015identification}, we believe it is more appropriate to label $X$ as the shadow variable.} that is not determinant of the underlying missingness, and \cite{sun2018semiparametric} considers an \textit{instrumental variable} that is dependent with the missingness indicator of the response variable but independent of the response variable itself (marginally or conditioned on other fully observed variables). Other approaches include conducting sensitivity analysis \citep{rotnitzky1998semiparametric, scharfstein2003generalized, scharfstein2021semiparametric} or obtaining nonparametric bounds for parameters of interest \citep{horowitz2000nonparametric}. A recent line of work considers missing data models with a collection of independence restrictions among variables and corresponding missingness indicators that can be represented by directed acyclic graphs (DAGs); see \cite{nabi2022causal} for a detailed discussion. 

In this work, we consider a MNAR model that corresponds to a graphical characterization, the \textit{criss-cross} structure discussed in \cite{nabi2022testability}, where missingness of the response variable depends on the missingness of covariates and vice versa. This kind of missingness is common in cross-sectional and survey studies. Unlike most prior work, all variables in our model can be subject to missingness, i.e., our results do not rely on the presence of fully observed variables. Furthermore, the MNAR model under study generalizes several prior popular missing data models, including the permutation model \citep{robins97non-a}, the block-conditional MAR model \citep{zhou10block}, and the block-parallel model \citep{mohan13missing}, making it less restrictive in terms of statistical independence assumptions imposed on the underlying joint data distribution. 

The criss-cross MNAR structure prevents nonparametric identification of the entire missing data model. We show, however, part of the complete-data distribution remains nonparametrically identifiable. We consider a quantitative measure, based on the rank of a \textit{Jacobian matrix}, to examine the amount of information in the identifiable part that would be sufficient for recovering the entire complete-data law, a.k.a. the \textit{target law}, as a function of only partially observed data. We explore these sufficient conditions extensively in the rich class of exponential family distributions. We further extend these results to higher dimensional parameter spaces and explore identifiability conditions for the entire missingness selection model, studied under \textit{full law} identification. Aside from identification arguments, we explore procedures for testing independence relations among variables that are themselves missing in terms of an \textit{odds ratio} parameterization of the complete-data law, as well as other model assumptions. We propose semiparametric estimating equations and conditional likelihoods based on order statistics to compute parameters that can be used for model selection purposes. Asymptotic properties of these two approaches are studied. We  show empirically that the estimating equation approach is more efficient compared to the conditional likelihood approach while the latter is more robust to misspecifications of the missingness selection model. 

The paper is organized as follows. We describe our notation and a brief overview of missing data DAGs in Section~\ref{sec:prelim}, and formally define the MNAR model under study in Section~\ref{sec:model}. We first consider univariate settings and discuss our (non)parametric identification and semiparametric estimation results in Sections~\ref{sec:ID} and \ref{sec:estimation}, respectively, followed by generalizations to multidimensional covariate spaces in Section~\ref{sec:high-dim}. The simulation results are provided in Section~\ref{sec:sims}, followed by conclusions in Section~\ref{sec:conc}.  All proofs are deferred to supplementary materials. 

\section{Preliminaries}  
\label{sec:prelim}

Let $Z$ be a vector of random variables with finite support and probability density $p(Z).$ Given a finite sample, variables in $Z$, indexed here by $k$, may have missing instances. 
Let $R$ be the corresponding vector of binary missingness indicators where $R_{k} = 1$ if $Z_{k}$ is observed and $R_{k}=0$ if $Z_{k}$ is missing. We only observe a coarsened version of $Z$ in our sample, which we denote by $Z^*.$ Each $Z^*_k \in Z$ is deterministically defined as follows: $Z^*_k = Z_k$ if $R_k = 1$ and $Z^* = \ ``?"$ if $R_k=0.$ $Z$ has a counterfactual connotation as it corresponds to variables ``had they been fully observed"  or  ``had $R$  been set to one" (no missingness) -- see \cite{bhattacharya19mid}. We use lowercase $z$ to denote the observed realization of $Z$.

\begin{figure}[t] 
	\begin{center}
		\scalebox{0.7}{
			\begin{tikzpicture}[>=stealth, node distance=1.5cm]
				\tikzstyle{format} = [thick, minimum size=1.0mm, inner sep=2pt]
				\tikzstyle{square} = [draw, thick, minimum size=4.5mm, inner sep=2pt]
					
				\begin{scope}[xshift=0cm]
					\path[->, thick]
					node[] (y) {$Y$}
					node[below of=y] (r) {$R_y$}
					node[right of=r] (ys) {$Y^*$}
					
					(y) edge[blue] (r) 
					
					(y) edge[gray] (ys)
					(r) edge[gray] (ys)
					
					node[format, below of=r, xshift=0.75cm, yshift=0.75cm] (a) {(a)} ;
				\end{scope}
				
				\begin{scope}[xshift=2.75cm]
					\path[->, thick]
					node[] (x) {$X$}
					node[right of=x, xshift=0.25cm] (y) {$Y$}
					node[below of=x, xshift=0.cm] (w) {$W$}
					node[below of=y] (r) {$R_y$}
					node[right of=r] (ys) {$Y^*$}
					
					(x) edge[blue] (y) 
					(y) edge[blue] (r) 
					
					(y) edge[gray] (ys)
					(r) edge[gray] (ys)
					
					(w) edge[blue, dashed] (x)
					(w) edge[blue, dashed] (y)
					(w) edge[blue, dashed] (r)
					
					node[format, below of=r, xshift=0.0cm, yshift=0.75cm] (b) {(b)} ;
				\end{scope}
				
				\begin{scope}[xshift=7.25cm]
					\path[->, thick]
					node[] (x) {$X$}
					node[right of=x, xshift=0.25cm] (y) {$Y$}
					node[below of=x, xshift=0.cm] (w) {$W$}
					node[below of=y] (r) {$R_y$}
					node[right of=r] (ys) {$Y^*$}
					
					(x) edge[blue] (r) 
					(y) edge[blue] (r) 
					
					(y) edge[gray] (ys)
					(r) edge[gray] (ys)
					
					(w) edge[blue, dashed] (x)
					(w) edge[blue, dashed] (y)
					(w) edge[blue, dashed] (r)
					
					node[format, below of=r, xshift=0.0cm, yshift=0.75cm] (c) {(c)} ;
				\end{scope}
				
			\end{tikzpicture}
		}
		\caption{(a) Self-censoring MNAR mechanism; (b) Shadow variable setup considered in \cite{wang2014instrumental}; (c)  Instrumental variable setup considered in \cite{sun2018semiparametric}. A dashed edge implies potential dependence between the endpoint variables.}
		\label{fig:nonignor}
	\end{center}
\end{figure}
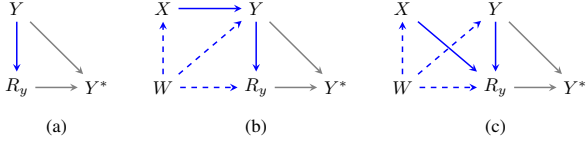

Following the literature on graphical models of missing data, it is descriptive to use directed acyclic graphs (DAGs) to encode assumptions in a given missing data model. A DAG ${\cal G}(V)$ is a set of vertices $V$ connected by directed edges such that there are no directed cycles. The statistical model of a DAG ${\cal G}(V)$ is a set of distributions that factorize as $p(V) = {\prod}_{V_i \in V}p(V_i \mid \pa_{\cal G}(V_i) )$, where $\pa_{\cal G}(V_i)$ denotes parents (direct causes) of $V_i$ in ${\cal G}(V)$; when the vertex set is clear from the context, ${\cal G}(V)$ is abbreviated as ${\cal G}$. Using the conventions in \cite{mohan13missing, bhattacharya19mid}, a missing data DAG (or mDAG for short) is defined over the set of vertices that correspond to variables in $V = \{Z, R, Z^*\}.$ In addition to acyclicity, a mDAG restricts the presence of certain edges: each $Z^*_k \in Z^*$ has only two parents ($Z_k$ and $R_k$), $Z^*_k$ does not have any outgoing edges and variables in $R$ cannot point to variables in $Z$. As an example, Fig.~\ref{fig:nonignor} illustrates the self-censoring mechanism in (a), the shadow variable setup in (b), and the instrumental variable approach in (c). Here, $Y$ is the non-response variable, and $X, W$ are fully observed variables. Deterministic edges are drawn in gray in all mDAGs. 

A missing data model associated with a mDAG ${\cal G}$ is the set of distributions $p(Z, R, Z^*)$ that factorize as 
\begin{align}
	\prod_{V_i \in Z} \ p(V_i \mid \pa_{\cal G} (V_i))  \times \prod_{R_k \in R} \ p(R_k \mid \pa_{\cal G}(R_k)).  
    \label{eq:mDAG_fact}
\end{align}%
We exclude the factors $p(Z^*_k \mid Z_k, R_k)$ which are deterministically defined.
Similar to a DAG, a mDAG encodes a set of ordinary conditional independence restrictions which can be easily read via Markov properties and d-separation rules: given disjoint subsets of vertices $A, B, C,$ the DAG global Markov property states that if $A \perp_\text{d-sep} B \mid C$ in ${\cal G}(V)$, then $A \perp B \mid C$ in $p(V)$ \citep{pearl09causality}. We refer to $p(Z)$ as  the \emph{target law}, $p(R  \mid Z)$  as  the \emph{missingness mechanism}, and $p(R, Z^*)$ as the \emph{observed data law}. The product of target law and missingness mechanism, i.e., $p(Z, R)$, is  referred to as the \emph{full law}. Note that in addition to partially missing variables, we may also have variables that are fully observed. However, in this work, we allow for the possibility of having all variables be partially missing in our model. 

Aside from the mDAG factorization, an \textit{odds ratio} parameterization  of the full law (or parts of it) can be useful in handling missing data models as it is illustrated by our methods in later sections; for more use of such parameterization see \cite{nabi20completeness, malinsky2021semiparametric}. Given disjoint sets of variables $A,B,C$ and reference values $A=a_0, B=b_0,$ the odds ratio parameterization of $p(A=a, B=b \mid C)$, given by \cite{chen07semiparametric}, is as follows: 
\begin{align}
    \frac{1}{Z(C)} \times p(a \mid b_0, C) \times p(b \mid a_0, C) \times \text{OR}(a, b \mid C),
	\label{eq:odds_ratio}
\end{align}	
where $\text{OR}(A = a, B = b \mid C)$ is defined as 
\begin{align*}
	&\frac{p(A = a \mid B = b, C)}{p(A = a_0 \mid B = b, C)} \times \frac{p(A = a_0 \mid B = b_0, C)}{p(A = a \mid B = b_0, C)},
\end{align*}%
and $Z(C) = \sum_{A,B} \ p(A | B = b_0, C) \times p(B | A = a_0, C) \times \text{OR}(A, B \mid C)$ is the normalizing term.

\section{The MNAR missing data model}  
\label{sec:model}

\begin{figure}[t] 
	\begin{center}
		\scalebox{0.7}{
			\begin{tikzpicture}[>=stealth, node distance=1.2cm]
				\tikzstyle{format} = [thick, circle, minimum size=1.0mm, inner sep=2pt]
				\tikzstyle{square} = [draw, thick, minimum size=4.5mm, inner sep=2pt]

				\begin{scope}[xshift=0cm]
					\path[->, thick]
					node[] (x) {$X$}
					node[right of=x, xshift=0.5cm] (y) {$Y$}
					node[below of=y] (r) {$R_y$}
					node[below of=x] (rx) {$R_x$}
					node[below of=rx, yshift=0.cm] (xs) {$X^*$}
					node[below of=r, yshift=0.cm] (ys) {$Y^*$}
					
					(x) edge[blue] (r) 
					(rx) edge[blue] (r) 
					(y) edge[blue] (rx) 
					
					(x) edge[blue] (y) 
					
					(y) edge[gray, bend left] (ys)
					(r) edge[gray] (ys)
					
					(x) edge[gray, bend right] (xs)
					(rx) edge[gray] (xs)
					
					node[format, below of=xs, xshift=1.cm, yshift=0.55cm] (a) {(a)} ;
				\end{scope}
				
				\begin{scope}[xshift=3cm, yshift=0cm]
					\path[->, thick]
					node[] (x) {$X$}
					node[right of=x, xshift=0.5cm] (y) {$Y$}
					node[below of=y] (r) {$R_y$}
					node[below of=x] (rx) {$R_x$}
					node[below of=rx, yshift=0.cm] (xs) {$X^*$}
					node[below of=r, yshift=0.cm] (ys) {$Y^*$}
					
					(xs) edge[blue] (r) 
					(rx) edge[blue] (r) 
					(y) edge[blue] (rx) 
					
					(x) edge[blue] (y) 
					
					(y) edge[gray, bend left] (ys)
					(r) edge[gray] (ys)
					
					(x) edge[gray, bend right] (xs)
					(rx) edge[gray] (xs)
					
					node[format, below of=xs, xshift=1.cm, yshift=0.55cm] (b) {(b)} ;
				\end{scope}
				
				\begin{scope}[xshift=6cm, yshift=0cm]
					\path[->, thick]
					node[] (x) {$X$}
					node[right of=x, xshift=0.5cm] (y) {$Y$}
					node[below of=y] (r) {$R_y$}
					node[below of=x] (rx) {$R_x$}
					node[below of=rx, yshift=0.cm] (xs) {$X^*$}
					node[below of=r, yshift=0.cm] (ys) {$Y^*$}
					
					(x) edge[blue, -] (y) 
					(x) edge[blue] (r) 
					(y) edge[blue] (rx) 
					
					(x) edge[blue] (y) 
					
					(y) edge[gray, bend left] (ys)
					(r) edge[gray] (ys)
					
					(x) edge[gray, bend right] (xs)
					(rx) edge[gray] (xs)
					
					node[format, below of=xs, xshift=1.cm, yshift=0.55cm] (c) {(c)} ;
				\end{scope}
				
				\begin{scope}[xshift=9cm, yshift=0cm]
					\path[->, thick]
					node[] (x) {$X$}
					node[right of=x, xshift=0.5cm] (y) {$Y$}
					node[below of=y] (r) {$R_y$}
					node[below of=x] (rx) {$R_x$}
					node[below of=rx, yshift=0.cm] (xs) {$X^*$}
					node[below of=r, yshift=0.cm] (ys) {$Y^*$}
					
					(x) edge[blue] (r) 
					(rx) edge[blue] (r) 
					
					(x) edge[blue] (y) 
					
					(y) edge[gray, bend left] (ys)
					(r) edge[gray] (ys)
					
					(x) edge[gray, bend right] (xs)
					(rx) edge[gray] (xs)
					
					node[format, below of=xs, xshift=1.cm, yshift=0.55cm] (d) {(d)} ;
				\end{scope}
				
			\end{tikzpicture}
		}
		\caption{(a) Criss-cross MNAR model; (b) Permutation model \citep{robins97non-a}; (c) Block-parallel model \citep{mohan13missing}; (d) Block-conditional MAR model \citep{zhou10block}.} 
		\label{fig:cross_model}
	\end{center}
\end{figure}
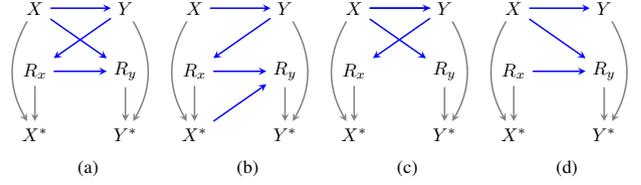

We partition $Z$ into two disjoint sets $X$ and $Y$, where the missingness of $X$ and $Y$ depend on each other as follows: 
\begin{align}
    (i) \ R_x \perp X \mid Y \qquad 
    (ii) \ R_y \perp Y \mid X, R_x
    \label{eq:criss_cross_assump}
\end{align}
The above set of assumptions can be represented via the mDAG shown in Fig.~\ref{fig:cross_model}(a), which corresponds to the so-called \textit{criss-cross} structure discussed in \cite{nabi2022testability}. This missing data model is a supermodel of several popular models in the literature. It relaxes the independence restrictions among variables imposed by models such as the \textit{permutation model} \citep{robins97non-a} shown in Fig.~\ref{fig:cross_model}(b), \textit{block-parallel model}  \citep{mohan13missing} shown in Fig.~\ref{fig:cross_model}(c), and \textit{block-conditional MAR model} \citep{zhou10block} shown in Fig.~\ref{fig:cross_model}(d). For instance, the permutation model implies the following set of independence restrictions: \textit{(i)} $R_x \perp X \mid Y$ and \textit{(ii)} $R_y \perp Y, X \mid X^*, R_x$. 
The independence restriction in \textit{(ii)} implies $R_y \perp Y \mid X, R_x = 1$  and $R_y \perp Y, X \mid R_x = 0$. These assumptions are a superset of the assumptions made in the criss-cross model, as defined in (\ref{eq:criss_cross_assump}). For more detailed comparisons across the aforementioned models, see \cite{nabi2022causal}.  

The importance of the criss-cross graphical characterization is that in the presence of such structure, the target law is not nonparametrically identifiable as a function of the observed data distribution \citep{nabi2022testability}, similar to the presence of self-censoring structure shown in Fig.~\ref{fig:nonignor}(a). See \cite{bhattacharya19mid} for sufficient conditions under which the target law is nonparametrically identifiable and \cite{nabi20completeness} for necessary and sufficient conditions under which the full law is nonparametrically identifiable, in a given mDAG. 

\section{Identification results}  
\label{sec:ID}

\subsection{Nonparametric identification}
\label{subsec:np-ID}

\cite{bhattacharya19mid} proved that the conditional density of $p(R_y  \mid R_x = 0, X)$ is not nonparametrically identifiable in the criss-cross model. This directly implies that the full law is not nonparametrically identified as a function of the observed data law. \cite{nabi2022testability} further proved that the target law is not identified either by providing a counterexample using binary variables for $X$ and $Y$. We verify the lack of nonparametric identification of the target law in Appendix~A, using continuous variables following normal distributions. 

The conditional distribution $p(X \mid Y)$ is, however, nonparametrically identified. This is because using the independence assumptions in display~(\ref{eq:criss_cross_assump}) and Bayes rule, we can write: 
\begin{align*}
    p(X \mid Y) = p(X \mid Y, R_x = 1) = \frac{p(X, Y, R_x=1)}{\int p(x, Y, R_x = 1) dx}, 
\end{align*}
where the marginal distribution $ p(X, Y, R_x = 1)$ equals: 
\begin{align*}
    \frac{p(X, Y, R_x = 1, R_y = 1)}{p(R_y =1 \mid R_x = 1, X, Y)} = \frac{p(X, Y, R_x = 1, R_y = 1)}{p(R_y =1 \mid R_x = 1, X)}, 
\end{align*}
and thus it is identified. 
The probabilistic operation of taking the full law and dividing it by the conditional density of $p(R_y \mid \pa_{\cal G}(R_y))$ (evaluated at $R=1$) corresponds to an intervention on $R_y$ that sets it to one. This provides an intuitive inverse probability weighting estimation strategy for parameters involving the conditional density of $X$ given $Y$. See Section~\ref{subsec:est_gmm} for a discussion on estimation and \cite{nabi2022causal} for more details on the interventional view to identification in graphical models of missing data.  

We take advantage of the nonparametric identification of $p(X \mid Y)$ in two ways: one is by combining this knowledge with consideration of a class of exponential family distributions to provide sufficient conditions for the identification of target and full laws (Section~\ref{subsec:par-ID}), and the other is by exploiting the knowledge in $p(X \mid Y)$ to estimate the odds ratio between $X$ and $Y$ as a method of an independence test, using either a conditional likelihood approach (Section~\ref{subsec:est_order}) or a generalized estimating equation (GEE) approach (Section~\ref{subsec:est_gmm}).

\subsection{Parametric identification}
\label{subsec:par-ID}

We first consider identification of the target law $p(X, Y)$ when $X$ is assumed to be univariate. We generalize our identification results to multivariate $X$ in Section~\ref{sec:high-dim}. 

\subsubsection{Target law identification}
Assume $p(X)$ and $p(Y \mid X)$ belong to the exponential family distribution. That is, 
\begin{align}
    &p(x) \sim \exp\left\{\frac{x\eta_x-b_x(\eta_x)}{\Phi_x}+c_x(x;\; \Phi_x)\right\}
     \label{eq:par_model} \\
    &p(y \mid x) \sim \exp\left\{\frac{y\eta-b(\eta)}{\Phi} \!+\! c(y;\; \Phi)\right\}, g(\mu(\eta)) \!=\! \alpha \!+\! \beta x, \nonumber 
\end{align}
where $b, c, b_x, c_x$ are known functions, $\Phi, \Phi_x > 0$ are dispersion parameters that may be known or unknown, and $g$ is a known one-to-one, third-order continuously differentiable link function. Let $\mu(\eta) \coloneqq \E[Y | X]$ and $\mu_x(\eta_x) \coloneqq \E[X]$. From the exponential family theory, we know that $b^{\prime}(\eta)=\mu(\eta)$ and $b^\prime_x(\eta_x) = \mu_x$. If $\mu = g^{-1}$, then $g$ is called the canonical link function and is denoted by $g_c$. We outline sufficient conditions for identifying the parameter vector $\theta = (\alpha, \beta, \Phi, \eta_x, \Phi_x)$ in the following theorem. 

\begin{theorem}\label{thm:id-par}
Assume the model in display (\ref{eq:par_model}) and $X$ takes $k+1$ distinct values $x_0, x_1,\cdots,x_k$. Let $\varphi=[g\circ \mu]^{-1}$, $\zeta=b\circ\varphi$. Define the following equations: 
\begin{align*}
    \phi_i(\theta) &= \{\varphi(\alpha+x_i\beta)-\varphi(\alpha+x_0\beta)\}/{\Phi} \\ 
    \zeta_i(\theta) &= \frac{-\zeta(\alpha+x_1\beta)+\zeta(\alpha+x_0\beta)}{\Phi}+\frac{\eta_x(x_1-x_0)}{\Phi_x} \\ 
    & +c(x_1;\;\Phi_x)-c(x_0;\;\Phi_x). 
\end{align*}
Define the Jacobian matrix $J={\partial(\Phi,\,Z)}/{\partial \theta }$, where $\Phi=\{\phi_1,\dots, \phi_k\}$ and $Z=\{\zeta_1,\dots,\zeta_k\}$. Under regularity conditions (detailed in Appendix~B.1), the target law $p(X,Y)$ is identifiable if 
\begin{align*}
    (i) \ k\geq dim(\theta), \quad
    (ii) \ \text{Jacobian matrix } J \text{ has full rank.}
\end{align*}
\end{theorem}%
See Appendix~B.1 for a proof. To provide an insight into Theorem~\ref{thm:id-par}, we emphasize the following observation: for any two distinct points of $X$, say $x_1$, and $x_0$, we have
\begin{align}\label{eq:key}
 \frac{p(x_1\mid y)}{p(x_0\mid y)}=\frac{p(y\mid x_1)}{p(y\mid x_0)}\times\frac{p(x_1)}{p(x_0)}.  
\end{align}
The left-hand side of equation (\ref{eq:key}) is identified, therefore as we vary the choice of distinct points of $X$, we are getting a series of equations that connect the identified conditional distribution $p(X\mid Y)$ to the target law. The rank of the Jacobian matrix $J$ provides a quantitative measure for the amount of information about the target law that is reflected in the conditional distribution $p(X\mid Y)$. When $J$ is full rank, we are able to obtain a unique solution of the target law, as a function of observed data law, by solving a system of equations. In the case of $J$ being rank deficient, we observe that removing some columns of $J$ can lead $J$ to be full rank. Removing columns from $J$ has the interpretation of assuming the corresponding parameters to be known, which yields sufficient conditions for identification claims. A similar argument is made by \cite{zhao2015semiparametric} in the non-ignorable non-response model (a.k.a. self-censoring) where $X$ is assumed to be fully observed and the parametric marginal density of $X$ is known. 

We highlight that our identification framework is highly generalizable. As the dimensionality of the distribution increases, the core of the theorem remains unchanged. We delve into the generalization of Theorem \ref{thm:id-par} thoroughly in Section \ref{sec:high-dim}. In addition, while the proposed method is not limited to the exponential family distributions, our emphasis on this particular family allows for clear and concise identification characterizations. We will further demonstrate in Section \ref{subsec:full-law} that the full law identification is easier to establish within the exponential family. 

In Appendix~C, we show the utilization of Theorem \ref{thm:id-par} in establishing sufficient conditions for target law identification in widely used exponential family distributions, including normal, Bernoulli, exponential, and Poisson distributions with either canonical or inverse links. The second condition in Theorem~\ref{thm:id-par}, namely that the Jacobian matrix must be of full rank, has different implications on what specific knowledge is required for $\theta$ in advance. For instance, under normal distributions with an inverse link discussed in Appendix~C.2 or exponential distributions discussed in Appendix~C.7, the target law is identified without any further restrictions on the parameter vector $\theta$. While in certain other distributions, the full-rank requirement of the Jacobian matrix implies that part of $\theta$ must be known apriori. For instance, in bivariate normal distributions with a canonical link discussed in Appendix~C.1, it is essential for identification arguments that at least the marginal mean of either $X$ or $Y$ is known. We emphasize that Theorem \ref{thm:id-par} only provides sufficient, not necessary, identification conditions. This means that stronger-than-needed characterizations might be established. 

\subsubsection{Full law identification}\label{subsec:full-law}

Under the conditions of Theorem~\ref{thm:id-par}, we can use the joint factorization of the full law in the criss-cross model to show that the conditional density of $R_x$ given $Y$, a.k.a. the propensity score of $R_x$, is identified: $p(X, Y, R_x=1, R_y=r_y) = p(X, Y) \times p(R_x=1 \mid Y) \times p(R_y=r_y \mid X, R_x=1)$, for $r_y = 0, 1$. To fully identify the full law, we need to show whether the full law evaluated at $R_x = 0$, i.e., $p(X, Y, R_x=0, R_y=r_y)$, is identified or not, or equivalently whether or not the propensity score of $R_y$ evaluated at $R_x = 0$, i.e., $p(R_y = 1 \mid R_x=0, X)$, is identified. The question of full law identification translates into the nonexistence of any two distinct propensity scores for $R_y$, e.g.,  $p_1(R_y\mid X, R_x)\neq p_2(R_y\mid X, R_x)$, such that $\int  \left[ \ p_1(R_y=1 \mid R_x=0, x) \ - \ p_2(R_y=1 \mid R_x=0, x) \ \right]$ $p(x \mid Y) \ dx=0$. Let $h(X) = p_1(R_y=1 \mid R_x=0, X)-p_2(R_y=1 \mid R_x=0, X)$. This condition then implies that if $\E[h(X) \mid Y] = 0$, then it must be the case that $h(X) = 0$ for the full law to be identified. This relates to the \textit{completeness} condition described below. 
\begin{condition}\label{cond:completeness}
For any function $h(X)$ with finite mean, $\E\{h(X)\mid Y\}=0$ implies $h(X)=0$ almost surely.
\end{condition}
\noindent With the completeness condition introduced, we can establish identification of the full law as follows. 
\begin{lemma}\label{lemma:full_law}
    Given the conditions in Theorem~\ref{thm:id-par} and  Condition~\ref{cond:completeness}, the full law $p(X,Y,R_x,R_y)$ is identified. 
\end{lemma}
See Appendix~B.3 for a proof. Identification under the completeness condition is widely seen among previous works \citep{newey2003instrumental,miao2015identification,zhao2022versatile}. As a special case, full law identification can be established from the completeness property of the exponential family distributions. More specifically, Condition \ref{cond:completeness} is guaranteed to hold if $p(X \mid Y)$ takes the following form:
\begin{align*}
    p(X \mid Y)=s\left(X\right) t(Y)\exp\left[\mu(Y)^{T}\tau\left(X\right)\right], 
\end{align*}
where $s\left(X\right)>0$, $\tau\left(X\right)$ is one-to-one in $X$, and the support of $\mu(Y)$ is an open set.

We show that the specific examples discussed in Appendices~C.1, C.3, C.4, C.5, and C.6 all have $p(X\mid Y)$ lie in the exponential family, therefore the full law is guaranteed to be identified (under conditions outlined in Theorem~\ref{thm:id-par}). In examples discussed in Appendices~C.2 and C.7, $p(X\mid Y)$ falls out of the exponential family, therefore the full law may or may not be identified. 

\section{Estimation and inference}  
\label{sec:estimation}

Our primary target of inference is the odds ratio between $X$ and $Y$, denoted by $\text{OR}(X, Y)$ and defined in (\ref{eq:odds_ratio}). Since the conditional density $p(X \mid Y)$ is nonparametrically identified, this odds ratio is also nonparametrically identified. In order to estimate this parameter, we establish two semiparametric methods outlined below.
Hereafter, we use $n$ to denote the size of the completely observed samples and $N$ the size of all samples.

\subsection{Conditional likelihood with order statistics}\label{subsec:est_order}

We assume access to $n$ i.i.d. copies of observed random variables $(X, Y),$ making up the data set $(x_1, y_1), \ldots, (x_n, y_n)$. As our first approach to estimating $\text{OR}(X,Y)$, we adopt the conditional likelihood approach based on order statistics $\widetilde{x}=(x_{(1)}, \ldots, x_{(n)})$. Let ${\cal P}$ collect all $n!$ permutations of $\{1, \ldots, n\}$. For a given permutation $P$ in ${\cal P}$, let $P(i)$ denote the $i$-th element of $P$. Consider the conditional likelihood $\prod_{i=1}^n p(x_i \mid y_i, r_{x_i}=1, r_{y_i}=1, \widetilde{x})$, which equals 
\begin{align}
    &\frac{\prod_{i=1}^{n} \  p\left(x_i \mid  y_i,r_{x_i}=1, r_{y_i}=1\right)}{\sum\limits_{P \in {\cal P}} \ \prod_{i=1}^{n} \ p\left(x_{P(i)}\mid y_i, r_{x_i}=1, r_{y_i}=1\right)} \notag \\ 
    &\hspace{0.75cm} =\frac{\prod_{i=1}^{n} p\left(x_{i} \mid y_{i}\right)}{\sum\limits_{P \in {\cal P}} \ \prod_{i=1}^{n} \ p\left(x_{P(i)} \mid y_{i}\right)}.\label{eq:pseudo-lik}
\end{align}
The last equality holds since by Bayes rule, we have: 
\begin{align*}
    &p(x_i \mid y_i, r_{x_i}=1, r_{y_i}=1) \\
    &\hspace{0.75cm} = \frac{p(x_i \mid y_i) \ p(r_{x_i}=1, r_{y_i}=1 \mid x_i, y_i)}{p(r_{x_i}=1, r_{y_i}=1 \mid y_i)}, 
\end{align*}
and given the mDAG factorization we can write $p(r_{x_i}=1, r_{y_i}=1 \mid x_i, y_i)$ as $p(r_{x_i}=1 \mid y_i) \ p(r_{y_i}=1 \mid r_{x_i}=1, x_i).$ We can rewrite $p\left(x_{P(i)} \mid y_{i},r_{x_i}=1,r_{y_i}=1\right)$ in a similar way. The terms related to the missingness mechanism remain invariant under permutations and cancel out from the numerator and denominator.

By exploiting the information available in this conditional likelihood, it is possible to estimate some parameters, such as the odds ratio, in the model of $p(X\mid Y)$.
The nice feature of applying this conditional likelihood is that for each subject $i$, the corresponding terms $p(R_x=1, R_y=1\mid Y,X)$ and $p(R_x=1, R_y=1\mid Y)$ are all canceled out during the above derivations; therefore, this conditional likelihood approach is robust to the model misspecification of the propensity scores, i.e., neither $p(R_y=1\mid R_x=1,X)$ nor $p(R_x=1\mid Y)$ need to be correctly specified in order to have a consistent estimation of the odds ratio.

Since the above conditional likelihood has the computation complexity of order $n!$, in reality, we approximate the conditional likelihood with the following pairwise pseudo-likelihood
\begin{align*}
&\prod_{i<k} \frac{p\left(x_{i} \mid y_{i}\right) p\left(x_{k} \mid y_{k}\right)}{p\left(x_{i} \mid y_{i}\right) p\left(x_{k} \mid y_{k}\right)+p\left(x_{i} \mid y_{k}\right) p\left(x_{k} \mid y_{i}\right)} \\
&\hspace{0.75cm} =\prod\limits_{i<k} \frac{1}{1+Q\left(x_{i}, y_{i} ; x_{k}, y_{k}\right)},
\end{align*}%
where $Q\left(x_{i}, y_{i} ; x_{k}, y_{k}\right)$ is the inverse of odds ratio (OR) and equals
$$\{p\left(x_{i} \mid y_{k}\right) p\left(x_{k} \mid y_{i}\right)\}/\{p\left(x_{i} \mid y_{i}\right) p\left(x_{k} \mid y_{k}\right)\}.$$
Therefore, by analyzing the completely observed subjects from the biased sample $p(X\mid Y,R_x=1,R_y=1)$, 
we are able to estimate the odds ratio $\text{OR}$ between $X$ and $Y$.
This conditional likelihood approach was first proposed in \citep{kalbfleisch1978likelihood} for hypothesis testing and then was used in a variety of statistical problems including both parameter estimation \citep{liang2000regression} and variable selection \citep{zhao2018penalized}; see \citep{chen2021semiparametric} for a more comprehensive exposition.

To illustrate the above pairwise pseudo-likelihood, we first consider a special case that $X\mid Y\sim \N(\alpha+\beta Y,\sigma^2)$, then 
$$\text{OR}=\exp\left(\frac{\beta}{\sigma^2}(x_i-x_k)(y_i-y_k)\right)=\exp \left[\frac{\beta}{\sigma^2}\left(w_j v_j\right)\right],$$
where $w_j=-\operatorname{sign}\left(y_i-y_k\right) \text { and } v_j=(x_i-x_k)\left|y_i-y_k\right|$, $j=1,\ldots,n(n-1)/2$ corresponds to each pair of $(i,k), i,k=1,\ldots,n$. Hence, the logarithm of the above pairwise pseudo-likelihood can be written as
\[
-\sum_{j} \log \left\{1+\exp \left[\frac{\beta}{\sigma^2}\left(w_j v_j\right)\right]\right\}.
\]
Thus, one can obtain the estimate of the parameter $\frac{\beta}{\sigma^2}$, denoted as $\theta$ hereafter, by performing the logistic regression with response $u_k$ and covariate $v_k$ without the intercept term, where
\[
u_k= \begin{cases}1 & \text { if }y_i-y_k>0 \\ 0 & \text { if } y_i-y_k<0.\end{cases}
\]
It is worth noting that the unknown parameter in OR pertains solely to the ratio of $\beta/\sigma^2$. As a result, this estimation approach accommodates potential misspecification of $\alpha$, which is the intercept in the regression $\E(X\mid Y)$, leading to a more comprehensive semiparametric assumption for the relationship between X and Y.

Let $\widetilde{\theta}$ denote the parameter estimate. Our result below demonstrates the asymptotic normality of $\widetilde{\theta}$.
\begin{theorem}
\label{thm:est-order}
  Denote $Q(x_i,y_i;x_k,y_k;\theta) \!=\! Q_{ik}(\theta)$ and $\zeta_{ik}(\theta) \!=\! \partial\log\{1+Q_{ik}(\theta)\}/\partial\theta$. Assume that $\E\|\zeta_{12}(\theta)\|^2<\infty$ for any $\theta$ in the parameter space. Then, 
  \begin{align*}
  \sqrt{N}(\widetilde\theta-\theta_0) \xrightarrow{d} \N(0, A^{-1}BA^{-1}),
  \end{align*}
  where $A \!=\! \E\left\{R_{x_1}R_{y_1}R_{x_2}R_{y_2}\partial\zeta_{12}(\theta_0)/\partial\theta\right\}$ and $B \!=\! 4\E\left\{R_{x_1}R_{y_1}R_{x_2}R_{y_2}R_{x_3}R_{y_3}\zeta_{12}(\theta_0)\zeta_{13}(\theta_0)\right\}$.
\end{theorem}
See Appendix~D.1 for a proof. The aforementioned pairwise pseudo-likelihood is favorable under a large sample size given its computational efficiency. However, the pairwise pseudo-likelihood estimator is generally inefficient. To improve efficiency, groupwise pseudo-likelihood can be adopted. Instead of picking two observations at a time, groupwise pseudo-likelihood uses more than two observations as a group. For example, with a group size of three, we will have
\begin{align*}
    L \propto \!\! \prod_{i<j<k} \frac{p(x_i\mid y_i) \ p(x_j\mid y_j) \ p(x_k\mid y_k)}{\sum_P p(x_{P(i)}\mid y_i) \ p(x_{P(j)}\mid y_j) \ p(x_{P(k)}\mid y_k)}
\end{align*}
where $P$ is the permutation of $(i,j,k)$.
Increased group size gives better efficiency with the cost of computational time. The final choice of group size should base on the consideration of computational time and statistical efficiency. Computational techniques with adaptive Monte Carlo approximation and Metropolis algorithm for directly maximizing the conditional likelihood are also well established and can be found in Chapter 4 of \cite{chen2021semiparametric}.


\subsection{Generalized estimating equations}\label{subsec:est_gmm} 

In the estimation approach presented in Section~\ref{subsec:est_order}, we need to specify the conditional density function $p(X\mid Y)$ either fully parametrically or semiparametrically. Alternatively, the model $p(X\mid Y)$ can be semiparametrically specified. For instance, assuming $\E(X\mid Y)=h(Y;\theta)$ with $h(\cdot)$ a known function and $\theta$ the unknown parameter of interest, we have the following estimating equation
\begin{align*}
	&\E \Big[  \frac{R_x \times R_y}{\pi(X)} \times f(Y) \times  (X - E(X \mid Y))  \Big] = 0,
\end{align*}
for any arbitrary function $f(Y)$.
Hereafter, we denote $\pi(X)=p(R_y = 1 \mid R_x = 1, X)$.
Note that the model $\pi(X)$ does not involve any missing data, so any off-the-shelf statistical method can be applied to model $\pi(X)$.
To better illustrate our proposed method, we do not particularly discuss the method for estimating $\pi(X)$ here.

Thus, the estimator of the parameter $\theta$, denoted as $\widehat\theta$, can be obtained by solving the following empirical version of the estimating equation
\[
\frac1N\sum_{i=1}^N \frac{R_{x_i} \times R_{y_i}}{\pi(x_i)} \times f(y_i) \times  (x_i - h(y_i;\theta)) = 0.
\]
In the following, we develop the asymptotic normality of the estimator $\widehat\theta$.
In particular, we also identify the optimal choice of $f(y)$, denoted by $f_{opt}(y)$, such that it achieves the best possible estimation efficiency among all choices of arbitrary function $f(y)$.
For simplicity, we denote $\Psi(X, Y, R_{x}, R_y ; \theta)=\frac{R_x \times R_y}{\pi(X)}\times f(Y) \times (X - h(Y;\theta))$.

\begin{theorem}
    \label{thm:est-gmm}
  Assume that $\E\|\Psi(X, Y, R_{x}, R_y ; \theta)\|^2<\infty$ for any $\theta$ in the parameter space. Then,
  \begin{itemize}
    \item[(a)] For any function $f(Y)$, we have
    \begin{align*}
        \sqrt{N}(\widehat\theta-\theta_0) \xrightarrow{d} \N(0, C^{-1}D(C^{-1})^T),
    \end{align*}
    where 
    \begin{align*}
        D &= \E\left\{\frac{R_x R_y}{\pi(X)^2}(X-h(Y;\theta))^2f(Y)f(Y)^T\right\}, \\
        C &= \E\left\{\! \frac{R_x R_y}{\pi(X)}a(Y)f(Y)^T\! \right\}, \text{ and } \\
        a(Y) &= \frac{\partial h(Y;\theta)}{\partial \theta} \! \bigg\rvert_{\theta=\theta_0}. 
    \end{align*}
    \item[(b)] The optimal choice of $f(Y)$ is
    \begin{align*}
        f_{opt}(Y)=\left[\E\left\{\frac{(X-h(Y;\theta))^{2}}{\pi(X)}\mid Y\right\}\right]^{-1} a(Y).
    \end{align*}
  \end{itemize}
\end{theorem}

See Appendix~D.2 for a proof.


\subsection{Alternative estimation targets} 

In addition to the associational relation between $X$ and $Y$, one might be interested in testing additional model assumptions, e.g., whether the missingness of $X$ is indeed influenced by $Y$ or not. This can be easily set up by rewriting the propensity score of $R_x$ using a parameterization that encodes the odds ratio between $R_x$ and $Y$ as $p(R_x = 1 \mid y) = \{1 + \exp(\lambda + \eta(y))\}^{-1}$ where $\eta(y) \coloneqq \log(\text{OR}(R_x = 0, y))$ and $\lambda = \log[{p(R_x = 0 \mid y_0)}/{p(R_x = 1 \mid y_0)} ]$. Under the conditions of Theorem~\ref{thm:id-par}, $\eta(y)$ would be identified. Exploring detailed estimation strategies are left to future work. 

It is worth pointing out that under the conditions of Theorem~\ref{thm:id-par} and Condition~\ref{cond:completeness}, one can simply estimate the entire parameter vector of the full law, assuming the parametric forms of the propensity scores in the missingness mechanism are known. 
More flexible estimation approaches are possible if one is willing to make additional modeling assumptions. For instance, in addition to independence restrictions in display~(\ref{eq:criss_cross_assump}), we may assume $p(R_y = 1 \mid R_x, X)$ is not a function of $X$ when $R_x = 0$. This reduces down the criss-cross model to the permutation MNAR model proposed by  \cite{robins97non-a}, where the full law is nonparametrically identified and the model is nonparametrically saturated, i.e., it imposes no restriction on the observed data law. In this case, we can proceed with nonparametric influence function based estimation, as discussed in Appendix~E.

\section{Multidimensional \ $\mathbf{X}$}
\label{sec:high-dim}

We now discuss how our identification arguments can be easily generalized to higher dimensional vector spaces. For a reasonable representation of sampling distributions, we extend Theorem~\ref{thm:id-par} to instances where $X$ follows either a multivariate normal or a multinomial distribution. The corresponding identification theories under these two scenarios are provided in Appendix~B.2; generalization to other sampling distributions can be carried out in a similar fashion. 

As two special cases, we consider $X$ to follow a multivariate normal or a multinomial distribution while $Y\mid X$ follows a normal distribution under the canonical link. We assume that the first condition in Theorem~\ref{thm:id-par} is satisfied by having sufficient observations. 
\begin{example} ($X$ is multivariate normal and $Y\mid X$ is normal under canonical link) \ Suppose 
\begin{align*}
    X \sim \N_d(\mu, \Sigma), 
    \qquad 
    Y \mid X \sim \N(\alpha + X^{T} \beta,\Phi). 
\end{align*}%
Assume the nuisance parameter $\Sigma$ is known. The unknown vector of parameters is $\theta=(\alpha, \beta, \Phi, \mu)$. A sufficient condition for identification of the target law $p(X,Y)$ is for the intercept $\alpha$ to be known. According to Lemma~\ref{lemma:full_law}, the full law is also identified.
\end{example}

\begin{example} ($X$ is multinomial and $Y\mid X$ is normal under canonical link) \ Suppose 
\begin{align*}
&X \sim \operatorname{Multinomial}_d(n, p), \quad Y \mid X \sim \N(\alpha + X^T\beta,\Phi),
\end{align*}%
where $p=(p_1,\ldots, p_d)$ is the vector of event probabilities, and $n$ is the number of trials. We can write $p(x)=\exp [x^T\eta+c(x)]$ where $\eta = \left(\log p_1, \ldots, \log p_d\right), c(x)=\log \frac{n !}{x_{1} ! \ \ldots \ x_{d} !}.$ Assume $n$ is known. The unknown vector of parameters is $\theta=(\alpha, \beta, \Phi, \eta)$. A sufficient condition for identification of the target law $p(X,Y)$ is for the intercept $\alpha$ to be known, or knowing at least one element of $\eta$. According to Lemma~\ref{lemma:full_law}, the full law is also identified.
\end{example}

\section{Experiments}
\label{sec:exp}

\subsection{Simulations}
\label{sec:sims}

We now compare the finite sample behavior of the three proposed estimation strategies, namely (i) non-optimal GEE, (ii) optimal GEE, and (iii) conditional likelihood with order statistics.\footnote{R code can be found at \url{https://github.com/annaguo-bios/criss-cross-model-code}. } We conduct simulation studies of $(X,Y)$ following bivariate normal distribution 
\begin{align*}
    \left(\begin{array}{l}Y \\ X\end{array}\right) \sim \quad \N\left[\left(\begin{array}{l}\mu_{1} \\ \mu_{2}\end{array}\right),\left(\begin{array}{cc}\sigma_{1}^{2} & \rho \sigma_{1} \sigma_{2} \\ \rho \sigma_{1} \sigma_{2} & \sigma_{2}^{2}\end{array}\right)\right],
\end{align*}
with $\mu_1=2,\,\mu_2=0.4,\,\sigma_1=1,\,\sigma_2=3,\,\rho=0.3$. The missingness mechanism is set as follows: 
\begin{align*}
    &p(R_x=1\mid Y)= \operatorname{expit}(-0.5+Y), \\
    &p(R_y=1 \mid X,R_x)= \operatorname{expit}(2-R_x+0.7X). 
\end{align*}

\begin{figure*}[!t]
    \centering
    \includegraphics[height=5.2cm, width=16cm]{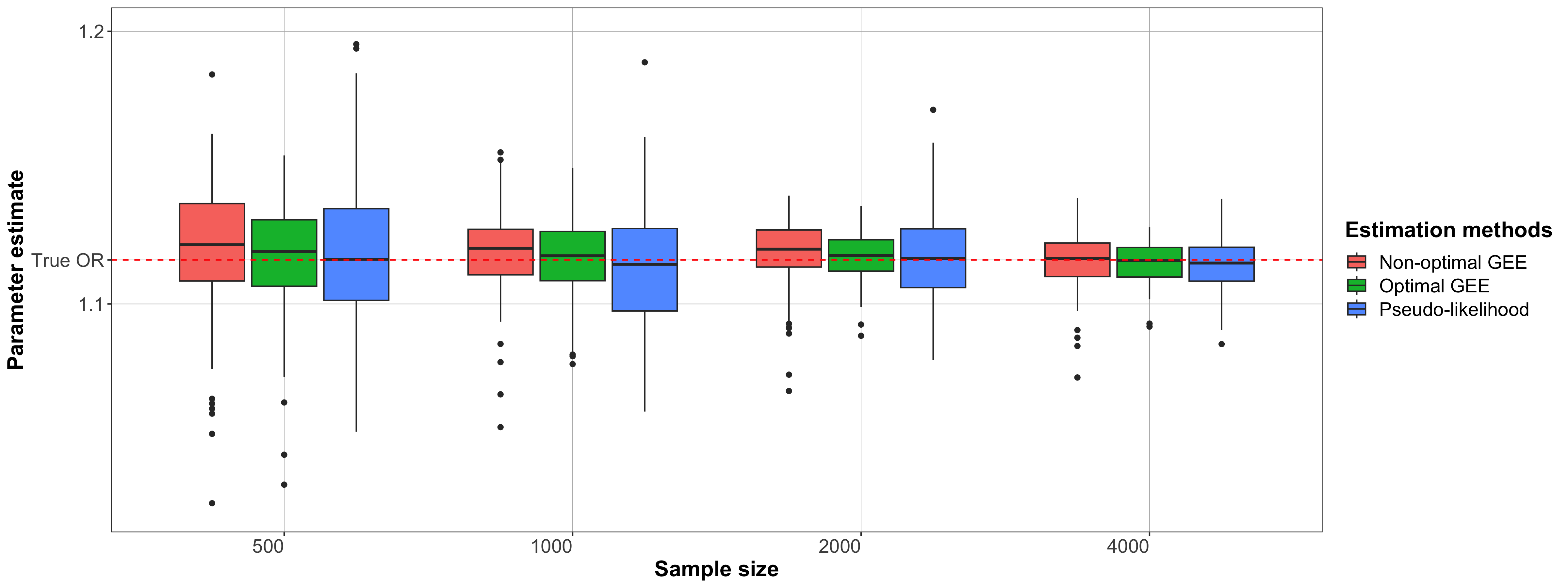}
    \caption{Odds ratio estimation with varying sample size.}
    \label{fig:sample_size}
\end{figure*}

\input{table1.tex}

Under this setup, approximately $5\%$ of observations have both $X$ and $Y$ missing, $16\%$ of observations have $X$ missing and $Y$ observed, $25\%$ of observations have $X$ observed and $Y$ missing and $54\%$ of observations have both $X$ and $Y$ observed. Under the above setup, we have 
\begin{align*}
    &X\mid Y\sim \N(\alpha+\beta Y,\sigma^2)=\N(-1.4+0.9Y,8.19)\\
    &\text{OR}=\exp\big\{\frac{\beta}{\sigma^2}(x_i-x_k)(y_i-y_k)\big\}. 
\end{align*}
Assuming the nuisance parameters $\sigma_1,\sigma_2$ are known, we aim at estimating $\alpha$ and $\beta$ with non-optimal and  optimal GEE approaches. We further estimate the odds ratio when $(x_i-x_k)(y_i-y_k)=1$ using all three aforementioned methods. For non-optimal GEE, we choose $f(Y)=(1,Y)$. Note that for the optimal GEE, $f_{opt}(Y)$ might be a function of $\alpha,\beta$. In such scenarios, to construct $\widehat{f}_{opt}(Y)$, we utilize the estimated values $\widehat{\alpha}$ and $\widehat{\beta}$, obtained as medians over 100 simulation runs from the non-optimal GEE. 
All code necessary to reproduce our simulations is included with this submission. 

We evaluate the performance of our three proposed estimators based on three main criteria: (i) finite sample behavior as sample size increases, (ii) bias behavior as a result of model misspecification for $p(R_y=1\mid X,R_x=1)$, and (iii) efficiency behavior as a result of varying the correlation between $X$ and $Y.$ For each case, we conduct 100 simulation runs. The empirical comparisons for the second and third criteria are deferred to Appendix~F due to page limits. 

Figure \ref{fig:sample_size} illustrates how the odds ratio estimation varies across a range of sample sizes from $500$ to $4000$. In order to ensure a fair comparison across the three methods, we assume that the intercept $\alpha$ of $\E(X\mid Y)$ is known for both non-optimal and optimal GEEs. The results demonstrate that all three methods yield unbiased estimates with reduced estimation uncertainty as the sample size increases. The conditional likelihood estimators are less efficient followed by non-optimal GEE, especially when the sample size is small. Overall, all three methods provide comparable OR estimates with small bias, mean-squared error (MSE), and standard deviation (SD) when the sample size is large.

Apart from OR estimation, the GEE approach is also capable of estimating the intercept $\alpha$. Table \ref{table:1} compares the performance of the two GEEs for estimating $\alpha$ and $\beta$, in terms of bias, MSE, and SD. As expected, the results show that the optimal GEE method outperforms the non-optimal GEE method in terms of smaller SD, regardless of the sample size. Additionally, for small sample sizes, the optimal GEE exhibits smaller bias and MSE than the non-optimal GEE. For additional simulations, see Appendix~F. 

\subsection{Real data application}
\label{sec:real_data}

We implemented the proposed methods in a real-world scenario involving an obesity study, where the outcome variable is binary indicating obesity status. Specifically, we analyzed the Muscatine Coronary Risk Factor Study (MCRF) \citep{woolson1984analysis}, which collected data on obesity from 4856 school children in 1977, 1979, and 1981. Our objective was to estimate the obesity rates stratified by sex. For our analysis, we focused on the data from 1977 and 1981, where only $40\%$ of the records were complete for both years.

In our study, we defined $X$ as the indicator of obesity in 1977 and $Y$ as the indicator of obesity in 1981, with values 1 representing non-obesity and 2 representing obesity. We denoted the obesity rates as $\theta_{ij} \coloneqq p(X=i, Y=j)$, where $i$ and $j$ take values from the set ${1, 2}$. We accounted for the possibility of both $X$ and $Y$ having missing-not-at-random (MNAR) patterns. This considers the potential impact of extrapolative projections, such as how the likelihood of recording obesity indications at the current follow-up may be influenced by anticipated obesity (or its absence) in the future, or how inquiries about obesity history or forecasts at one time point can lead to additional inquiries at another time point. By accommodating MNAR mechanisms for both $X$ and $Y$, our model becomes more practical and applicable to real-world scenarios.

Based on our identification results, determining the complete joint distribution requires knowledge of one parameter from the set: ${\theta_{11}, \theta_{12}, \theta_{21}, \theta_{22}}$. In our analysis, we assumed that $\theta_{11}$ is known and obtained this value from the complete-case records. To estimate the obesity rates and the log odds ratio between $X$ and $Y$ (equivalent to $\log(OR) = \log(\theta_{11}\theta_{22}/\theta_{12}\theta_{21})$), we employed generalized estimating equations (GEE). Both optimal GEE and non-optimal GEE approaches were utilized.

For the non-optimal GEE, we set $f(Y)$ as $(1, Y)$ based on Theorem~\ref{thm:est-gmm}. Additionally, we employed pseudo-likelihood estimation for estimating the $\log(OR)$ parameter. To assess the precision of the estimates, we employed bootstrap resampling with 1000 replicates. The estimation results are presented in Table~\ref{table:binary}.

\input{table_realdata_binary}

The estimates obtained from the optimal GEE approach closely align with those from the non-optimal GEE, and therefore, they are not presented in the preceding analysis. The key findings reveal a substantial temporal correlation in obesity rates. Specifically, the non-obesity status exhibits a persistence rate of 0.723 for girls and 0.71 for boys between the two years. Both girls and boys have an equal probability of 0.118 of being obese in both years. Additionally, an intriguing observation for policy intervention purposes is that non-obese girls demonstrate a higher susceptibility to obesity compared to non-obese boys. This observation calls for further careful examination to facilitate effective strategies for obesity prevention. 

Furthermore, we also analyzed a real-world dataset related to income, where the outcome variable is continuous. The detailed findings are in Appendix~F. 

\section{Conclusions}  
\label{sec:conc}

In this paper, we considered a MNAR model which, like the self-censoring missingness mechanism, is an impediment to nonparametric identification of the complete-data distribution. We provided sufficient identification assumptions for both target and full laws by examining the rich class of exponential family distributions. We provided different semiparametric estimation strategies for computing parameters of the underlying joint distribution that can be used for pairwise independence tests and model selection purposes. An interesting avenue for future work is the exploration of a doubly-robust estimation theory that would enable the use of more flexible machine learning and statistical models in computing various model parameters.  



\begin{acknowledgements} 
    This work is partly supported by the National Center for Advancing Translational Sciences of the National Institutes of Health under Award Number UL1TR002378. 
\end{acknowledgements}

\bibliography{references}

\clearpage 
\appendix

\onecolumn 

{\Large \bf APPENDIX}

\vspace{0.25cm}

The appendix is organized as follows. 
In Appendix~\ref{app:nonid-ex}, we provide a counterexample for lack of target law identification in the criss-cross MNAR model using continuous variables under normal distributions. 
Appendix~\ref{app:id-proofs} contains our identification proofs in the exponential family distribution: target law with univariate $X$ (\ref{app:sub-target-id-uni}), target law with multivariate $X$ (\ref{app:sub-target-id-multi}) and full law (\ref{app:sub-full-id}). 
In Appendix~\ref{app:parID-ex}, we include several examples on parametric identification of popular distributions in the exponential family distributions. 
Appendix~\ref{app:est-proofs} contains our proofs regarding asymptotic behaviors of our suggested estimators for conditional likelihood with order statistics (\ref{app:sub-est-order}) and generalized method of moments (\ref{app:sub-est-gmm}). In Appendix~\ref{app:est-additional}, we provide additional discussions on (non)parametric estimation approaches. Appendix~\ref{app:sims} contains additional experiments.

\section{Counterexample for lack of target law identification}\label{app:nonid-ex}

Consider two distinct distributions $p_1$ and $p_2$ defined over variables in $\{X, Y, R_x, R_y\}$  as follows: 

\textbf{Model 1:} $Y\sim \N(1,1),\, X\mid Y\sim \N(y,1),\, p_1(R_x=1\mid y)=\frac{\sqrt{5/6}}{\sqrt{5/6}+\exp \left[-\frac{1}{12}(y-1)^{2}\right]}$, and $$p_1(R_y=1\mid x, R_x) =
\begin{cases}
	\phi(x),\text{ when }R_x=1\\
	\phi(\frac{x-5}{\sqrt{5}}),\text{ when }R_x=0
\end{cases}   $$

\textbf{Model 2:} $Y\sim \N(1,\frac{6}{5}),\, X\mid Y\sim \N(y,1),\, p_2(R_x=1\mid y)=\frac{\exp \left[-\frac{1}{12}(y-1)^{2}\right]}{\sqrt{5/6}+\exp \left[-\frac{1}{12}(y-1)^{2}\right]}$, and $$p_2(R_y=1\mid x, R_x) =
\begin{cases}
	\phi(x),\text{ when }R_x=1\\
	\exp (-\frac{8}{9})*\sqrt{\frac{2}{5}}\phi(x-\frac{7}{3}),\text{ when }R_x=0.
\end{cases}$$
\label{app:counter}%
Here $\phi(.)$ denotes the standard normal CDF, and $p_i(x,y,R_x,R_y)=p_i(y) \ p(x\mid y) \ p_i(R_x\mid y) \ p_i(R_y\mid x,R_x),\,i=1,2$. Note that $p_1 \not= p_2$. In what follows, we analyze the four missingness patterns one by one and show that the above two models map to the exact same observed data distribution and thus the target law is not identifiable as a unique function of the observed data law.   
\begin{itemize}
	\item \underline{Missingness pattern $(R_x=1, R_y=1)$.} We need to prove  
	$$p_1(x, y, R_x = 1, R_y = 1) = p_2(x, y, R_x = 1, R_y = 1).$$ 
	This holds since 
	\begin{align*}
		&p_{1}(y) \ p( x\mid y) \ p_{1}(R_{x}=1 \mid y) \ p_{1}\left(R_{y}=1 \mid x, R_{x}=1\right)\\
		&=\frac{1}{\sqrt{2 \pi}} \exp \left\{-\frac{1}{2}(y-1)^{2}\right\}\times p(x \mid y) \times \frac{\sqrt{\frac{5}{6}}}{\sqrt{\frac{5}{6}}+\exp \left[-\frac{1}{12}(y-1)^{2}\right]} \times  \frac{1}{\sqrt{2 \pi}} \exp \left\{-\frac{1}{2} x^{2}\right\} \\
		&=\frac{1}{\sqrt{2 \pi}\sqrt{\frac{6}{5}}} \exp \left\{-\frac{1}{2\times\frac{6}{5}}(y-1)^{2}\right\}\times p(x \mid y) \times \frac{\exp \left[-\frac{1}{12}(y-1)^{2}\right]}{\sqrt{\frac{5}{6}}+\exp \left[-\frac{1}{12}(y-1)^{2}\right]} \times  \frac{1}{\sqrt{2 \pi}} \exp \left\{-\frac{1}{2} x^{2}\right\} \\
		&=p_{2}(y) \ p( x\mid y) \ p_{2}(R_{x}=1 \mid y) \ p_{2}\left(R_{y}=1 \mid x, R_{x}=1\right). 
	\end{align*}
	
	\item \underline{Missingness pattern $(R_x=1, R_y=0)$.} We need to prove
	$$\int p_1(x, y, R_x = 1, R_y = 0) dy = \int p_2(x, y, R_x = 1, R_y = 0) dy.$$ 
	That is, 
	\begin{align*}
		&\int p_{1}(y) p{(x \mid y)} p_{1}\left(R_{x}=1 \mid y\right) p_{1}\left(R_{y}=0 \mid x, R_{x}=1\right) d y\\
		&\hspace{1cm}=\int p_{2}(y) p{(x \mid y)} p_{2}\left(R_{x}=1 \mid y\right) p_{2}\left(R_{y}=0 \mid x, R_{x}=1\right) d y.
	\end{align*}
	Or in other words: 
	\begin{align*}
		&\int p_{1}(y) p(x\mid y) p_{1}\left(R_{x}=1 \mid y\right) d y-\int p_{1}(y) p(x\mid y) p_{1}\left(R_{x}=1 \mid y\right) p_1(R_y=1\mid x, R_x=1)d y\\
		&\hspace{0.5cm}=\int p_{2}(y) p(x\mid y) p_{2}\left(R_{x}=1 \mid y\right) d y-\int p_{2}(y) p(x\mid y) p_{2}\left(R_{x}=1 \mid y\right) p_2(R_y=1\mid x, R_x=1)d y. 
	\end{align*}
	Since $\int p_{1}(y) p(x\mid y) p_{1}\left(R_{x}=1 \mid y\right) p_1(R_y=1\mid x, R_x=1)d y =\int p_{2}(y) p(x\mid y) p_{2}\left(R_{x}=1 \mid y\right) p_2(R_y=1\mid x, R_x=1)d y$ holds by the missingness pattern $(R_x = 1, R_y = 1)$, we only need to show $$\int p_{1}(y) p(x\mid y) p_{1}\left(R_{x}=1 \mid y\right) d y=\int p_{2}(y) p(x\mid y) p_{2}\left(R_{x}=1 \mid y\right) d y.$$
	We have: 
	\begin{align*}
		&p_{1}(y) \ p(x\mid y) \ p_{1}\left(R_{x}=1 \mid y\right)\\
		&\hspace{1cm}=\frac{1}{\sqrt{2 \pi}} \exp \left\{-\frac{1}{2}(y-1)^{2}\right\} \times p(x \mid y)\times \frac{\sqrt{\frac{5}{6}}}{\sqrt{\frac{5}{6}}+\exp \left[-\frac{1}{12}(y-1)^{2}\right]} \\
		&\hspace{1cm}=\frac{1}{\sqrt{2 \pi}\sqrt{\frac{6}{5}}} \exp \left\{-\frac{1}{2\times\frac{6}{5}}(y-1)^{2}\right\}  \times  p(x \mid y)\times \frac{\exp \left[-\frac{1}{12}(y-1)^{2}\right]}{\sqrt{\frac{5}{6}}+\exp \left[-\frac{1}{12}(y-1)^{2}\right]} \\
		&\hspace{1cm}=p_{2}(y) \ p(x\mid y) \ p_{2}\left(R_{x}=1 \mid y\right).
	\end{align*}
	\item \underline{Missingness pattern ($R_x=0,\, R_y=1$).} We need to prove 
	$$\int p_1(x, y, R_x = 0, R_y = 1) dx = \int p_2(x, y, R_x = 0, R_y = 1) dx.$$ 
	For any $\mu\text{ and }\sigma>0$, it is true that 
	{\scriptsize 
		\begin{align*}
			&\int \phi(x-y)\times\phi(\frac{x-\mu}{\sigma})d x\\
			&=\int \frac{1}{\sqrt{2 \pi}} \exp \left\{-\frac{1}{2}(x-y)^{2}\right\} \times \frac{1}{\sqrt{2 \pi} \sigma} \exp \left\{-\frac{1}{2 \sigma^{2}}(x-\mu)^{2}\right\} d x\\
			&=\frac{1}{\sqrt{2} \pi} \times \frac{1}{\sqrt{2 \pi} \sigma} \int \exp \left\{-\frac{1}{2} x^{2}+x y-\frac{1}{2} y^{2}-\frac{1}{2 \sigma^{2}} x^{2}+\frac{1}{\sigma^{2}} x \mu-\frac{1}{2 \sigma^{2}} \mu^{2}\right\} d x\\
			&=\frac{1}{\sqrt{2 \pi}} \frac{1}{\sqrt{2 \pi} \sigma} \times \int \exp \left\{-\frac{1}{2 \times \frac{\sigma^{2}}{\sigma^{2}+1}}\left[x^{2}-2 x\left(y+\frac{\mu}{\sigma^{2}}\right) \frac{\sigma^{2}}{\sigma^{2}+1}+\left(y+\frac{\mu}{\sigma^{2}}\right)^{2}\left(\frac{\sigma^{2}}{\sigma^{2}+1}\right)^{2}\right]\right\} \\ 
			&\hspace{3cm} \times  \exp \left[-\frac{1}{2} y^{2}-\frac{1}{2 \sigma^{2}} \mu^{2}+\frac{1}{2 \frac{\sigma^{2}}{\sigma^{2}+1}} \times\left(y+\frac{\mu}{\sigma^{2}}\right)^{2}\left(\frac{\sigma^{2}}{\sigma^{2}+1}\right)^{2}\right] d x\\
			&=\frac{1}{\sqrt{2 \pi}} \times \sqrt{\frac{1}{1+\sigma^{2}}} \times \exp \left[-\frac{1}{2} \frac{1}{1+\sigma^{2}} y^{2}+\frac{1}{1+\sigma^{2}} \mu y-\frac{1}{2} \frac{\mu^{2}}{1+\sigma^{2}}\right]. 
	\end{align*}
}
	
	Thus, we have: 
	\begin{align*} 
		& p_{1}(y) p_{1}\left(R_{x}=0 \mid y\right) \int p(x \mid y) p_{1}\left(R_{y}=1 \mid x, R_{x}=0\right) d x \\ 
		&=\frac{1}{\sqrt{2 \pi}}\exp \left\{-\frac{1}{2}(y-1)^{2}\right\} \times \frac{\exp \left[-\frac{1}{12}(y-1)^{2}\right]}{\sqrt{\frac{5}{6}}+\exp \left[-\frac{1}{12}(y-1)^{2}\right]}\times \frac{1}{\sqrt{2 \pi}} \sqrt{\frac{1}{6}} \exp \left[-\frac{1}{12} y^{2}+\frac{5}{6} y-\frac{1}{2} \times \frac{25}{6}\right]\\
		&=\frac{1}{2 \pi} \sqrt{\frac{1}{6}} \frac{1}{\sqrt{\frac{5}{6}}+\exp \left[-\frac{1}{12}(y-1)^{2}\right]}\times \exp \left\{-\frac{7}{12}(y-1)^{2}-\frac{1}{12} y^{2}+\frac{5}{6} y-\frac{1}{2} \times \frac{25}{6}\right\}\\
		&=\frac{1}{2 \pi} \sqrt{\frac{1}{6}} \frac{1}{\sqrt{\frac{5}{6}}+\exp \left[-\frac{1}{12}(y-1)^{2}\right]}\times \exp \left\{-\frac{2}{3} y^{2}+2 y-\frac{8}{3}\right\}\\
		&=p_{2}(y) p_{2}\left(R_{x}=0 \mid y\right) \int p(x \mid y) p_{2}\left(R_{y}=1 \mid x, R_{x}=0\right) d x\\
		&=\frac{1}{\sqrt{2 \pi}}\exp \left\{-\frac{1}{2 \times \frac{6}{5}}(y-1)^{2}\right\} \times \frac{\sqrt{\frac{5}{6}}}{\sqrt{\frac{5}{6}}+\exp \left[-\frac{1}{12}(y-1)^{2}\right]}\times \exp(-\frac{8}{9})\sqrt{\frac{2}{5}}\frac{1}{\sqrt{2 \pi}} \sqrt{\frac{1}{2}} \exp\left[-\frac{1}{4} y^{2}+\frac{7}{6} y-\frac{49}{36}\right]\\
		&=\frac{1}{2 \pi} \sqrt{\frac{1}{6}}\exp(-\frac{8}{9})\frac{1}{\sqrt{\frac{5}{6}}+\exp \left[-\frac{1}{12}(y-1)^{2}\right]}\exp \left\{-\frac{5}{12}(y-1)^{2}-\frac{1}{4} y^{2}+\frac{7}{6} y-\frac{49}{36}\right\}\\
		&=\frac{1}{2 \pi} \sqrt{\frac{1}{6}}\frac{1}{\sqrt{\frac{5}{6}}+\exp \left[-\frac{1}{12}(y-1)^{2}\right]}\exp \left\{-\frac{2}{3} y^{2}+2 y-\frac{8}{3}\right\}.
	\end{align*}
	
	\item \underline{Missingness pattern ($R_x=0,\, R_y=0$).} We need to prove
	$$\int p_1(x, y, R_x=0,R_y=0) dxdy = \int p_2(x, y, R_x=0,R_y=0) dxdy,$$ 
	which is guaranteed to hold since  the previous three missingness patterns yield the same observed data law and the fact that probabilities should integrate to one. 
	
\end{itemize}

This concludes the claim that the target law is not identified in the criss-cross MNAR model. 

\newpage
\section{Identification Proofs}
\label{app:id-proofs} 

\subsection{Theorem~\ref{thm:id-par} \quad {\small (Target law parametric identification: univariate $X$)}}\label{app:sub-target-id-uni}

We have 
\vspace{-0.5cm}
\begin{align*}
	X &\sim \exp\left\{\frac{x\eta_x-b_x(\eta_x)}{\Phi_x}+c_x(x;\; \Phi_x)\right\}
	\\
	Y\mid X &\sim \exp\left\{\frac{y\eta-b(\eta)}{\Phi}+c(y;\; \Phi)\right\}, \quad g(\mu(\eta)) = \alpha + \beta x. 
\end{align*}

The parameters of interest are $\theta=(\alpha,\beta,\Phi,\eta_x,\Phi_x)$. Since $p(x\mid y)$ is nonparametrically (np)-identified, we can select two distinct points of $X$, say $x_1$ and $x_0$ and write
\begin{align*}
	\frac{p(x_1\mid y)}{p(x_0\mid y)}&={\frac{p(y\mid x_1)p(x_1)}{p(y)}} \div {\frac{p(y\mid x_0)p(x_0)}{p(y)}}
	=\frac{p(y\mid x_1)}{p(y\mid x_0)}\times\frac{p(x_1)}{p(x_0)}\\
	&=\exp\left\{ \frac{y(\eta_1-\eta_0)-[b(\eta_1)-b(\eta_0)]}{\Phi} \right\}\times\exp\left\{\frac{\eta_x(x_1-x_0)}{\Phi_x}+c(x_1;\;\Phi_x)-c(x_0;\;\Phi_x)\right\}. 
\end{align*}
We take a $log$ on both sides. The left-hand side is only a function of $y$. Suppose the coefficient of $y$ on the left-hand side is $\phi_1$ and the intercept is $\zeta_1$. For the ease of notation, define $\varphi=[g\circ \mu]^{-1}$ and $\zeta=b([g\circ \mu]^{-1})$. We can then write the following: 
$$\begin{aligned}
	\phi_1(\theta) &= \frac{\eta_1-\eta_0}{\Phi}=\frac{[g\circ \mu]^{-1}(\alpha+x_1\beta)-[g\circ \mu]^{-1}(\alpha+x_0\beta)}{\Phi}=\frac{\varphi(\alpha+x_1\beta)-\varphi(\alpha+x_0\beta)}{\Phi}\\
	\zeta_1(\theta) &= \left\{ \frac{-[b(\eta_1)-b(\eta_0)]}{\Phi}+\frac{\eta_x(x_1-x_0)}{\Phi_x}+c(x_1;\;\Phi_x)-c(x_0;\;\Phi_x)\right\}\\
	&=\left\{ \frac{-\left[b\left([g\circ \mu]^{-1}(\alpha+x_1\beta)\right)-b\left([g\circ \mu]^{-1}(\alpha+x_0\beta)\right)\right]}{\Phi}+\frac{\eta_x(x_1-x_0)}{\Phi_x}+c(x_1;\;\Phi_x)-c(x_0;\;\Phi_x)\right\}\\
	&=\left\{ \frac{-\zeta(\alpha+x_1\beta)+\zeta(\alpha+x_0\beta)}{\Phi}+\frac{\eta_x(x_1-x_0)}{\Phi_x}+c(x_1;\;\Phi_x)-c(x_0;\;\Phi_x)\right\}. 
\end{aligned}$$
Suppose we have $k+1$ distinct values of $x$. We can then create $2k$ equations like above, say $\phi_i$ and $\zeta_i$ with $i=1,\dots, k$. The core of our identification proof relies on the \textit{implicit function theorem}. In order to use this theorem, the above equations need to satisfy the followings: 
\begin{itemize}
	\item There exists at least one solution $\theta_0$ that satisfies the above equations, 
	\item $\phi_i(\theta)$ and $\zeta_i(\theta)$ are continuous in $\Theta$, i.e., the parameter space with $\theta_0$ as an inner point, 
	\item $\phi_i(\theta)$ and $\zeta_i(\theta)$ are first order partially differentiable in $\Theta$, 
	\item Let $\Phi=\{\phi_1,\dots, \phi_k\}$ and $Z=\{\zeta_1,\dots,\zeta_k\}$. Define the Jacobian matrix $J$ as $J=\frac{\partial(\Phi,\,Z)}{\partial(\theta)}$, which is described below: 
\end{itemize}
{	\scriptsize
	\begin{align*}
		J=\begin{bmatrix}
			\varphi^{\prime}\left(\alpha+x_{1} \beta\right)-\varphi^{\prime}\left(\alpha+x_{0} \beta\right) & \varphi^{\prime}\left(\alpha+x_{1} \beta\right) x_{1}-\varphi^{\prime}\left(\alpha+x_{0} \beta\right) x_{0} & \varphi\left(\alpha+x_{1} \beta\right)-\varphi\left(\alpha+x_{0} \beta\right) & 0 & 0\\
			\vdots & \vdots & \vdots & \vdots & \vdots \\
			\varphi^{\prime}\left(\alpha+x_{k} \beta\right)-\varphi^{\prime}\left(\alpha+x_{0} \beta\right) & \varphi^{\prime}\left(\alpha+x_{k} \beta\right) x_{k}-\varphi^{\prime}\left(\alpha+x_{0} \beta\right) x_{0} & \varphi\left(\alpha+x_{k} \beta\right)-\varphi\left(\alpha+x_{0} \beta\right) & 0 & 0\\
			\zeta^{\prime}\left(\alpha+x_{1} \beta\right)-\zeta^{\prime}\left(\alpha+x_{0} \beta\right) & \zeta^{\prime}\left(\alpha+x_{1} \beta\right) x_{1}-\zeta^{\prime}\left(\alpha+x_{0} \beta\right) x_{0} & \zeta\left(\alpha+x_{1} \beta\right)-\zeta\left(\alpha+x_{0} \beta\right) & x_{1}-x_{0}& -\frac{\eta_{x}\left(x_{1}-x_{0}\right)}{\Phi^2_x}+\frac{\partial c\left(x_{1} , \Phi_{x}\right)}{\partial \Phi_x}-\frac{\partial c\left(x_{0} , \Phi_{x}\right)}{\partial \Phi_x}\\
			\vdots & \vdots & \vdots & \vdots & \vdots \\
			\zeta^{\prime}\left(\alpha+x_{k} \beta\right)-\zeta^{\prime}\left(\alpha+x_{0} \beta\right) & \zeta^{\prime}\left(\alpha+x_{k} \beta\right) x_{k}-\zeta^{\prime}\left(\alpha+x_{0} \beta\right) x_{0} & \zeta\left(\alpha+x_{k} \beta\right)-\zeta\left(\alpha+x_{0} \beta\right) & x_{k}-x_{0} & -\frac{\eta_{x}\left(x_{k}-x_{0}\right)}{\Phi^2_x}+\frac{\partial c\left(x_{k} , \Phi_{x}\right)}{\partial \Phi_x}-\frac{\partial c\left(x_{0} , \Phi_{x}\right)}{\partial \Phi_x}
		\end{bmatrix}
	\end{align*}
}

\begin{itemize}
	\item[]$J$ must be of full rank under $\left(\theta_0,\phi_i(\theta_0),\zeta_i(\theta_0)\right)$, 
	
	\item The number of equations must be greater or equal to the number of unknown parameters, i.e.,  $2k\geq dim(\theta)$.
\end{itemize}


Under the above conditions, there exists neighborhood $U$ around the true parameters $\theta_0$ as $U=B\left(\theta_0, \epsilon\right) \subset \Theta$, and the neighborhood $V$ around $(\phi_i(\theta_0),\zeta_i(\theta_0))$ as $V=B\left((\phi_1(\theta_0),\dots,\phi_k(\theta_0),\zeta_1(\theta_0),\dots,\zeta_k(\theta_0)),\eta\right)\subset R^{2k}$ with $\epsilon, \eta>0$, and uniquely defined functions $g=\left(g_1, \ldots, g_{2 k}\right)$ on $V$ that each $g_i$ is first-order continuously differentiable. We have $$\theta=g\left(\phi_1(\theta),\dots,\phi_k(\theta),\zeta_1(\theta),\dots,\zeta_k(\theta)\right),$$ where $\left(\phi_1(\theta),\dots,\phi_k(\theta),\zeta_1(\theta),\dots,\zeta_k(\theta)\right)\in V$, with $\theta\in U$. Given that the $\left(\phi_1,\dots,\phi_k,\zeta_1,\dots,\zeta_k\right)$ we observed is generated under the true value $\theta_0$, which is
$\operatorname{observed}\left(\phi_1,\dots,\phi_k,\zeta_1,\dots,\zeta_k\right)=\left(\phi_1(\theta_0),\dots,\phi_k(\theta_0),\zeta_1(\theta_0),\dots,\zeta_k(\theta_0)\right)$, by applying $g$, we can uniquely find $\theta_0=g\left(\phi_1(\theta_0),\dots,\phi_k(\theta_0),\zeta_1(\theta_0),\dots,\zeta_k(\theta_0)\right).$

\vspace{0.5cm}
\subsection{Target law parametric identification: multivariate \ $\bf X$}\label{app:sub-target-id-multi}

\subsubsection{Multivariate normal \ $\bf X$}\label{proofs:target_id-multinormal} 

Suppose $$\begin{aligned}
	&X \sim \N_d(\mu, \Sigma)\\
	& Y\mid X \sim \exp\left\{\frac{y\eta-b(\eta)}{\Phi}+c(y;\; \Phi)\right\}, \quad g(\mu(\eta)) = \alpha + x^{T}\beta. 
\end{aligned}$$
Assume the nuisance parameter $\Sigma$ is known and $\theta=(\alpha, \beta, \Phi, \mu)$. We can write down the following equation: 
{ \scriptsize
	\begin{align*}
		\frac{p\left(x_1 \mid y\right)}{p\left(x_0 \mid y\right)}&=\frac{p\left(y \mid x_1\right)}{p\left(y \mid x_0\right)} \times \frac{p\left(x_1\right)}{p\left(x_0\right)}\\
		&=\exp \left\{\frac{y\left(\eta_1-\eta_0\right)-\left[b\left(\eta_1\right)-b \left(\eta_0\right)\right]}{\Phi}\right\} \exp \left\{-\frac{1}{2}\left(x_1-\mu\right)^{T} \Sigma^{-1}\left(x_1-\mu\right)+\frac{1}{2}\left(x_0-\mu\right)^{T} \Sigma^{-1}\left(x_0-\mu\right)\right\}. 
	\end{align*}
}

Taking a log on both sides yields the following equation: 
{\scriptsize
	\begin{align*}
		\log \left[p\left(x_1 \mid y\right)\right]-\log\left[p\left(x_0 \mid y\right)\right]&=y \times \frac{\eta_1-\eta_0}{\Phi}-\frac{b\left(\eta_1\right)-b\left(\eta_0\right)}{\Phi}-\frac{1}{2}\left(x_1-\mu\right)^{T} \Sigma^{-1}\left(x_1-\mu\right)+\frac{1}{2}\left(x_0-\mu\right)^{T} \Sigma^{-1}\left(x_0-\mu\right). 
	\end{align*}
}

The left-hand side is only a function of $y$. Suppose the coefficient of $y$ is $\phi_1$ and the intercept is $\zeta_1$.  For the ease of notation, define $\varphi=[g\circ \mu]^{-1}$ and $\zeta=b([g\circ \mu]^{-1})$. Then, we obtain the following equation: 
$$\begin{aligned}
	\phi_1(\theta)&=\frac{\eta_1-\eta_0}{\Phi}=\frac{\left[g\circ \mu\right]^{-1}\left(\alpha+x_1^{T} \beta\right)-\left[g\circ \mu\right]^{-1}\left(\alpha+x_0^{T} \beta\right)}{\Phi}=\frac{\varphi\left(\alpha+x_1^{T} \beta\right)-\varphi\left(\alpha+x_0^{T} \beta\right)}{\Phi}\\
	\zeta_1(\theta) &= -\frac{b\left(\eta_1\right)-b\left(\eta_0\right)}{\Phi}-\frac{1}{2}\left(x_1-\mu\right)^{T} \Sigma^{-1}\left(x_1-\mu\right)+\frac{1}{2}\left(x_0-\mu\right)^{T} \Sigma^{-1}\left(x_0-\mu\right)\\
	& =  -\frac{\zeta(\alpha+x_1^{T} \beta)-\zeta(\alpha+x_0^{T} \beta)}{\Phi}-\frac{1}{2}\left(x_1-\mu\right)^{T} \Sigma^{-1}\left(x_1-\mu\right)+\frac{1}{2}\left(x_0-\mu\right)^{T} \Sigma^{-1}\left(x_0-\mu\right). 
\end{aligned}$$

Suppose we have $k+1$ distinct values of $x$. Thus, we can construct $2k$ equations, $\phi_i$ and $\zeta_i$ with $i=1,\dots, k$. In order to use this theorem, the above equations
need to satisfy the followings:
\begin{itemize}
	\item There exists at least one solution $\theta_0$ that satisfies the above equations, 
	\item $\phi_i(\theta)$ and $\zeta_i(\theta)$ are continuous on $\Theta$, i.e., the parameter space with $\theta_0$ as an inner point, 
	\item $\phi_i(\theta)$ and $\zeta_i(\theta)$ are first order partially differentiable on $\Theta$, 
	\item Let $\Phi=\{\phi_1,\dots, \phi_k\}$ and $Z=\{\zeta_1,\dots,\zeta_k\}$. Define then Jacobian matrix $J$ as $J=\frac{\partial(\Phi,\,Z)}{\partial(\theta)}$, described below: 
	
{\scriptsize 
	\begin{align*}
		J=\begin{bmatrix}
			\varphi^{\prime}\left(\alpha+x^{T}_1\beta\right)-\varphi^{\prime}\left(\alpha+x^{T}_0\beta\right) & \varphi^{\prime}\left(\alpha+x^{T}_1\beta\right) x^{T}_{1}-\varphi^{\prime}\left(\alpha+x^{T}_0\beta\right) x^{T}_{0} & \varphi\left(\alpha+x^{T}_1\beta\right)-\varphi\left(\alpha+x^{T}_0\beta\right) & 0\\
			\vdots & \vdots & \vdots & \vdots  \\
			\varphi^{\prime}\left(\alpha+x^{T}_{k} \beta\right)-\varphi^{\prime}\left(\alpha+x^{T}_0\beta\right) & \varphi^{\prime}\left(\alpha+x^{T}_{k} \beta\right) x^{T}_{k}-\varphi^{\prime}\left(\alpha+x^{T}_0\beta\right) x^{T}_{0} & \varphi\left(\alpha+x^{T}_{k} \beta\right)-\varphi\left(\alpha+x^{T}_0\beta\right) & 0 \\
			\zeta^{\prime}\left(\alpha+x^{T}_1\beta\right)-\zeta^{\prime}\left(\alpha+x^{T}_0\beta\right) &\zeta^{\prime}\left(\alpha+x^{T}_1\beta\right) x^{T}_{1}-\zeta^{\prime}\left(\alpha+x^{T}_0\beta\right) x^{T}_{0} & \zeta\left(\alpha+x^{T}_1\beta\right)-\zeta\left(\alpha+x^{T}_0\beta\right) & \left(x_1-x_0\right)^{T} \Sigma^{-1}\\
			\vdots & \vdots & \vdots & \vdots \\
			\zeta^{\prime}\left(\alpha+x^{T}_{k} \beta\right)-\zeta^{\prime}\left(\alpha+x^{T}_0\beta\right) & \zeta^{\prime}\left(\alpha+x^{T}_{k} \beta\right) x^{T}_{k}-\zeta^{\prime}\left(\alpha+x^{T}_0\beta\right) x^{T}_{0}& \zeta\left(\alpha+x^{T}_k\beta\right)-\zeta\left(\alpha+x^{T}_0\beta\right) & \left(x_k-x_0\right)^{T} \Sigma^{-1}
		\end{bmatrix}
		\end{align*}
	}
	
	$J$ must be of full rank under $\left(\theta_0,\phi_i(\theta_0),\zeta_i(\theta_0)\right)$, 
	
	\item The number of equations must be greater or equal to the number of unknown parameters, i.e.,  $2k\geq dim(\theta)$.
\end{itemize}

Under the special case where $Y\mid X\sim \N\left(\alpha+x^{T}\beta,\Phi\right)$, we have: 
$$\begin{aligned}
	&\phi_i(\theta)=\frac{\left(x_i-x_0\right)^{T} \beta}{\Phi}\\
	&\zeta_i(\theta)=-\frac{\left(\alpha+x_i^{T} \beta\right)^2-\left(\alpha+x_0^{T} \beta\right)^2}{2 \Phi}-\frac{1}{2}\left(x_i-\mu\right)^{T} \Sigma^{-1}\left(x_i-\mu\right)+\frac{1}{2}\left(x_0-\mu\right)^{T} \Sigma^{-1}\left(x_0-\mu\right),\\
	&\text{where } i\in (1,\ldots,k), \text{ and }
\end{aligned}$$

{\scriptsize 
$$
	J=\begin{bmatrix}
		0 & \frac{(x_{1}-x_{0})^{T}}{\Phi} & -\frac{(x_{1}-x_{0})^{T}\beta}{\Phi^2} & 0\\
		\vdots & \vdots & \vdots & \vdots  \\
		0 & \frac{(x_{k}-x_{0})^{T}}{\Phi} & -\frac{(x_{k}-x_{0})^{T}\beta}{\Phi^2} & 0\\
		-\frac{(x_{1}-x_{0})^{T}\beta}{\Phi} & -\frac{\alpha(x_{1}-x_{0})^{T}+\beta^{T}(x_1x^{T}_1-x_0x^{T}_0)}{\Phi} & \frac{(\alpha+x^{T}_1\beta)^2-(\alpha+x^{T}_0\beta)^2}{2\Phi^2} & \left(x_{1}-x_{0}\right)^{T} \Sigma^{-1}\\
		\vdots & \vdots & \vdots & \vdots \\
		-\frac{(x_{k}-x_{0})^{T}\beta}{\Phi} & -\frac{\alpha(x_{k}-x_{0})^{T}+\beta^{T}(x_kx^{T}_k-x_0x^{T}_0)}{\Phi} & \frac{(\alpha+x^{T}_k\beta)^2-(\alpha+x^{T}_0\beta)^2}{2\Phi^2} & \left(x_{k}-x_{0}\right)^{T} \Sigma^{-1}\\
	\end{bmatrix}
$$
}
After performing some rank-preserving modifications to this matrix, we have
$$
\scriptsize
	J=\begin{bmatrix}
		0 & (x_{1}-x_{0})^{T} & -(x_{1}-x_{0})^{T}\beta & 0\\
		\vdots & \vdots & \vdots & \vdots  \\
		0 & (x_{1}-x_{0})^{T} & -(x_{1}-x_{0})^{T}\beta & 0\\
		(x_{1}-x_{0})^{T}\beta & -\left[\alpha(x_{1}-x_{0})^{T}+\beta^{T}(x_1x^{T}_1-x_0x^{T}_0)\right] & \frac{(\alpha+x^{T}_1\beta)^2-(\alpha+x^{T}_0\beta)^2}{2} & \left(x_{1}-x_{0}\right)^{T} \Sigma^{-1}\\
		\vdots & \vdots & \vdots & \vdots \\
		(x_{k}-x_{0})^{T}\beta & -\left[\alpha(x_{k}-x_{0})^{T}+\beta^{T}(x_kx^{T}_1-x_0x^{T}_0)\right] & \frac{(\alpha+x^{T}_k\beta)^2-(\alpha+x^{T}_0\beta)^2}{2} & \left(x_{k}-x_{0}\right)^{T} \Sigma^{-1}\\
	\end{bmatrix}
$$

The dimension of $J$ is $dim(J)=2k\times (2+2d)$. Assume $2k\geq (2+2d)$. A sufficient condition to make $J$ full rank is knowing at least $\alpha$. 

Note that in this example \textbf{$p(X\mid Y)$ is in the exponential family}, since:
{\scriptsize
$$
	\begin{aligned}
		&p(x \mid y) =\frac{p(y \mid x) p(x)}{p(y)}\\
		&=\exp \left\{-\frac{\left[y-\left(\alpha+x^{T} \beta\right)\right]^2}{2 \Phi}+\log \frac{1}{\sqrt{2 \pi \Phi}}-\frac{1}{2}(x-\mu)^{T} \Sigma^{-1}(x-\mu)+\log \frac{1}{\sqrt{(2 \pi)^d|\Sigma|}}-\log (y)\right\}\\
		&=\exp\bigg\{-\frac{(y-\alpha)^2}{2 \Phi}+\frac{(y \beta-\alpha \beta)^{T}}{\Phi} x-\frac{\operatorname{tr}\left(\beta \beta^{T} x x^{T}\right)}{2 \Phi}+\log \frac{1}{\sqrt{2 \pi \Phi}}
		+\mu^{T} \Sigma^{-1} x-\frac{1}{2} x^{T} \Sigma^{-1} x-\frac{1}{2} \mu^{T} \Sigma^{-1} \mu+\log \frac{1}{\sqrt{(2 \pi)^{d}|\Sigma|}}-\log (y)\bigg\}\\
		&=\exp\bigg\{ \left[\frac{(y \beta-\alpha \beta)^{T}}{\Phi}+\mu^{T} \Sigma^{-1}, -\frac{\operatorname{vec}\left(\beta \beta^{T}\right)^{T}}{2 \Phi}\right]\left(\begin{array}{c}
			x \\
			\operatorname{vec}\left(x x^{T}\right)
		\end{array}\right)-\frac{(y-\alpha)^2}{2 \Phi}+\log \frac{1}{\sqrt{2 \pi \Phi}}
		-\frac{1}{2} x^{T} \Sigma^{-1} x-\frac{1}{2} \mu^{T} \Sigma^{-1} \mu+\log \frac{1}{\sqrt{(2 \pi)^{d}|\Sigma|}}-\log (y)\bigg\}.
\end{aligned}
$$
}

Here \textit{tr}(.) denotes the trace of the input matrix and \textit{vec}(.) refers to the vectorization operation applied to the input matrix, e.g., $A_{n\times m}$, as stacking the rows of the matrix one by one to form a long column vector with size $nm\time 1$, i.e., 
$$\operatorname{vec}[A]=\operatorname{vec}\left[\left(\begin{array}{ccc}
	a_{11} & \cdots & a_{1 m} \\
	\vdots & \ddots & \vdots \\
	a_{n 1} & \cdots & a_{n m}
\end{array}\right)\right]=\left(\begin{array}{c}
	a_{11} \\
	\vdots \\
	a_{1 m} \\
	\vdots \\
	a_{n m}
\end{array}\right).$$

\vspace{0.5cm}
\subsubsection{Multinomial \ $\bf X$}\label{proofs:target_id-multinomial}

Suppose $$\begin{aligned}
	&X\sim \operatorname{Multinomial}_d(n,p),\\
	&Y\mid X \sim \exp\left\{\frac{y\eta-b(\eta)}{\Phi}+c(y;\; \Phi)\right\}, \quad g(\mu(\eta)) = \alpha + x^{T}\beta,
\end{aligned}$$

where $p=(p_1,\ldots, p_d)$ is the vector of event probabilities, and $n$ is the number of trials. We can write $p(x)=\exp [x^{T} \eta+c(x)]\text { with } \eta=\left(\log p_1, \ldots, \log _{p_d}\right), c(x)=\log \frac{n !}{x_{1} ! \cdots x_{d} !}$. Assume the nuisance parameter $n$ is known and $\theta=(\alpha, \beta, \Phi, \eta)$. We can write down the following:  
\begin{align*}
	\frac{p\left(x_1 \mid y\right)}{p\left(x_0 \mid y\right)}&=\frac{p\left(y \mid x_1\right)}{p\left(y \mid x_0\right)} \times\frac{p\left(x_1\right)}{p\left(x_0\right)} \\
	& =\exp \left\{\frac{y\left(\eta_1-\eta_0\right)-\left[b\left(\eta_1\right)-b\left(\eta_0\right)\right]}{\Phi}\right\} \exp \left\{\left(x_1-x_0\right)^{T} \eta+c\left(x_1\right)-c\left(x_0\right)\right\}. 
\end{align*}
Taking a $\log$ on both sides yields the following: 
$$\log \left[p\left(x_1 \mid y\right)\right]-\log \left[p\left(x_0 \mid y\right)\right]=y \frac{\eta_1-\eta_0}{\Phi}-\frac{b\left(\eta_1\right)-b\left(\eta_0\right)}{\Phi}+\left(x_1-x_0\right)^{T}\eta+c\left(x_1\right)-c\left(x_0\right)$$
The left-hand side is only a function of $y$. Suppose the coefficient of $y$ is $\phi_1$ and the intercept is $\zeta_1$.  For the ease of notation, define $\varphi=[g\circ \mu]^{-1}$ and $\zeta=b([g\circ \mu]^{-1})$. Thus, we obtain the following: 
$$\begin{aligned}
	\phi_1(\theta)&=\frac{\eta_1-\eta_0}{\Phi}=\frac{\left[g\circ \mu\right]^{-1}\left(\alpha+x_1^{T} \beta\right)-\left[g\circ \mu\right]^{-1}\left(\alpha+x_0^{T} \beta\right)}{\Phi}=\frac{\varphi\left(\alpha+x_1^{T} \beta\right)-\varphi\left(\alpha+x_0^{T} \beta\right)}{\Phi}\\
	\zeta_1(\theta) &= -\frac{b\left(\eta_1\right)-b\left(\eta_0\right)}{\Phi}+\left(x_1-x_0\right)^{T}\eta+c\left(x_1\right)-c\left(x_0\right)\\
	& =  -\frac{\zeta(\alpha+x_1^{T} \beta)-\zeta(\alpha+x_0^{T} \beta)}{\Phi}+\left(x_1-x_0\right)^{T} \eta+c\left(x_1\right)-c\left(x_0\right). 
\end{aligned}$$

Suppose we have $k+1$ distinct values of $x$. Thus, we can construct $2k$ equations, $\phi_i$ and $\zeta_i$ with $i=1,\dots, k$. To apply the implicit function theorem, the equations need to satisfy the following conditions: 
\begin{itemize}
	\item There exists at least one solution $\theta_0$ that satisfies the above equations, 
	\item $\phi_i(\theta)$ and $\zeta_i(\theta)$ are continuous on $\Theta$, i.e., the parameter space with $\theta_0$ as an inner point, 
	\item $\phi_i(\theta)$ and $\zeta_i(\theta)$ are first order partially differentiable on $\Theta$, 
	\item Let $\Phi=\{\phi_1,\dots, \phi_k\}$ and $Z=\{\zeta_1,\dots,\zeta_k\}$. Define then Jacobian matrix $J$ as $J=\frac{\partial(\Phi,\,Z)}{\partial(\theta)}$, described below:
	 
	$$
\scriptsize
		J=\begin{bmatrix}
			\varphi^{\prime}\left(\alpha+x^{T}_1\beta\right)-\varphi^{\prime}\left(\alpha+x^{T}_0\beta\right) & \varphi^{\prime}\left(\alpha+x^{T}_1\beta\right) x^{T}_{1}-\varphi^{\prime}\left(\alpha+x^{T}_0\beta\right) x^{T}_{0} & \varphi\left(\alpha+x^{T}_1\beta\right)-\varphi\left(\alpha+x^{T}_0\beta\right) & 0\\
			\vdots & \vdots & \vdots & \vdots  \\
			\varphi^{\prime}\left(\alpha+x^{T}_{k} \beta\right)-\varphi^{\prime}\left(\alpha+x^{T}_0\beta\right) & \varphi^{\prime}\left(\alpha+x^{T}_{k} \beta\right) x^{T}_{k}-\varphi^{\prime}\left(\alpha+x^{T}_0\beta\right) x^{T}_{0} & \varphi\left(\alpha+x^{T}_{k} \beta\right)-\varphi\left(\alpha+x^{T}_0\beta\right) & 0 \\
			\zeta^{\prime}\left(\alpha+x^{T}_1\beta\right)-\zeta^{\prime}\left(\alpha+x^{T}_0\beta\right) & \zeta^{\prime}\left(\alpha+x^{T}_1\beta\right) x^{T}_{1}-\zeta^{\prime}\left(\alpha+x^{T}_0\beta\right) x^{T}_{0} & \zeta\left(\alpha+x^{T}_1\beta\right)-\zeta\left(\alpha+x^{T}_0\beta\right) & \left(x_1-x_0\right)^{T} M\\
			\vdots & \vdots & \vdots & \vdots \\
			\zeta^{\prime}\left(\alpha+x^{T}_{k} \beta\right)-\zeta^{\prime}\left(\alpha+x^{T}_0\beta\right) & \zeta^{\prime}\left(\alpha+x^{T}_{k} \beta\right) x^{T}_{k}-\zeta^{\prime}\left(\alpha+x^{T}_0\beta\right) x^{T}_{0}& \zeta\left(\alpha+x^{T}_k\beta\right)-\zeta\left(\alpha+x^{T}_0\beta\right) & \left(x_k-x_0\right)^{T} M
		\end{bmatrix}
	$$
	
	where $M_{d\times d-1}=\begin{bmatrix}
		\mathbb{I}_{d-1\times d-1} \\
		(-1,-1,\cdots,-1)_{1\times d-1} \\
	\end{bmatrix},\, \mathbb{I}\text{ is the identity matrix}.$ 
	
	The Jacobian matrix $J$ must be of full rank under $\left(\theta_0,\phi_i(\theta_0),\zeta_i(\theta_0)\right)$.

	\item The number of equations must be greater or equal to the number of unknown parameters, i.e.,  $2k\geq dim(\theta)$.
\end{itemize}

Under the special case where $Y\mid X\sim \N\left(\alpha+x^{T}\beta,\Phi\right)$, we have: 
$$\begin{aligned}
	&\phi_i(\theta)=\frac{\left(x_i-x_0\right)^{T} \beta}{\Phi}\\
	&\zeta_i(\theta)=-\frac{\left(\alpha+x_i^{T} \beta\right)^2-\left(\alpha+x_0^{T} \beta\right)^2}{2 \Phi}+\left(x_i-x_0\right)^{T} \eta+c\left(x_i\right)-c\left(x_0\right), \quad i\in (1,2,\cdots,k), 
\end{aligned}$$

$$
\scriptsize
	J=\begin{bmatrix}
		0 & \frac{(x_{1}-x_{0})^{T}}{\Phi} & -\frac{(x_{1}-x_{0})^{T}\beta}{\Phi^2} & 0\\
		\vdots & \vdots & \vdots & \vdots  \\
		0 & \frac{(x_{k}-x_{0})^{T}}{\Phi} & -\frac{(x_{k}-x_{0})^{T}\beta}{\Phi^2} & 0\\
		-\frac{(x_{1}-x_{0})^{T}\beta}{\Phi} & -\frac{\alpha(x_{1}-x_{0})^{T}+\beta^{T}(x_1x^{T}_1-x_0x^{T}_0)}{\Phi} & \frac{(\alpha+x^{T}_1\beta)^2-(\alpha+x^{T}_0\beta)^2}{2\Phi^2} & \left(x_{1}-x_{0}\right)^{T}M\\
		\vdots & \vdots & \vdots & \vdots \\
		-\frac{(x_{k}-x_{0})^{T}\beta}{\Phi} & -\frac{\alpha(x_{k}-x_{0})^{T}+\beta^{T}(x_kx^{T}_k-x_0x^{T}_0)}{\Phi} & \frac{(\alpha+x^{T}_k\beta)^2-(\alpha+x^{T}_0\beta)^2}{2\Phi^2} & \left(x_{k}-x_{0}\right)^{T}M\\
	\end{bmatrix}
$$

After performing some rank-preserving modifications to this matrix, we get: 
$$
\scriptsize
	J=\begin{bmatrix}
		0 & (x_{1}-x_{0})^{T} & -(x_{1}-x_{0})^{T}\beta & 0\\
		\vdots & \vdots & \vdots & \vdots  \\
		0 & (x_{k}-x_{0})^{T} & -(x_{k}-x_{0})^{T}\beta & 0\\
		(x_{1}-x_{0})^{T}\beta & -\left[\alpha(x_{1}-x_{0})^{T}+\beta^{T}(x_1x^{T}_1-x_0x^{T}_0)\right] & \frac{(\alpha+x^{T}_1\beta)^2-(\alpha+x^{T}_0\beta)^2}{2} & \left(x_{1}-x_{0}\right)^{T}M\\
		\vdots & \vdots & \vdots & \vdots \\
		(x_{k}-x_{0})^{T}\beta & -\left[\alpha(x_{k}-x_{0})^{T}+\beta^{T}(x_kx^{T}_k-x_0x^{T}_0)\right] & \frac{(\alpha+x^{T}_k\beta)^2-(\alpha+x^{T}_0\beta)^2}{2} & \left(x_{k}-x_{0}\right)^{T}M\\
	\end{bmatrix}
$$

The dimension of $J$ is $dim(J)=2k\times (1+2d)$. Assume $2k\geq (1+2d)$. A sufficient condition to make $J$ full rank is knowing $\alpha$ or at least one element of $\eta$. 

Note that in this example, \textbf{$p(X \mid Y)$ is in the exponential family}, since:
\begin{align*}
	p(x \mid y)&=\frac{p(y \mid x) p(x)}{p(y)}\\
	&=\exp \left\{-\frac{\left[y-\left(\alpha+x^{T} \beta\right)\right]^2}{2 \Phi}+\log \frac{1}{\sqrt{2 \pi \Phi}}+x^{T} \eta+c(x)-\log p(y)\right\}\\
	&=\exp\left\{\left[\frac{(y \beta-\alpha \beta)^{T}}{\Phi}+\eta^{T}, -\frac{\operatorname{vec}\left(\beta \beta^{T}\right)^{T}}{2 \Phi}\right]\left(\begin{array}{c}
		x \\
		\operatorname{vec}\left(x x^{T}\right)
	\end{array}\right)-\frac{(y-\alpha)^2}{2 \Phi}+c(x)-\log p(y)\right\}. 
\end{align*}   

\vspace{0.5cm}
\subsection{Lemma~\ref{lemma:full_law} \quad {\small (Full law identification)}}\label{app:sub-full-id}

Using the DAG factorization we have
$$p\left(X, Y, R_x=1, R_y=1\right)= p(X, Y) \ p(R_x=1 \mid Y) \ p(R_y=1 \mid X, R_x=1).$$

Given the above relation and the fact that the target law $p(X,Y)$ is identified, it is straightforward to conclude that 
$p(R_x\mid Y)$ is also identified. We now prove under the completeness condition, $p(R_y\mid X,R_x)$ is also identified. Therefore the full law is identified.
The full observed data law can be written down as follows: 
\begin{align*}
	\mathcal{L}_\text{full}(Z_\text{obs}, R; \theta, \psi) 
	&= \prod_{R_x=1, R_y=1} p(X, Y, R_x=1, R_y=1) \times \prod_{R_x=1, R_y=0}  \int p(X, Y, R_x=1, R_y=0) dy \\
	&\ \times \prod_{R_x=0, R_y=1} \int p(X, Y, R_x=0, R_y=1) dx \times \prod_{R_x=0, R_y=0}  \int p(X, Y, R_x=0, R_y=0) dxdy.
\end{align*}
Given the fact that $p(X,Y)$, $p(R_x=1\mid Y)$, and $p(R_x=0,R_y=0)$ are all identified, the following would stay the same across different models: 

{\small 
	\begin{align*}
		&\prod_{R_x=1, R_y=1} p(X, Y, R_x=1, R_y=1) \times \prod_{R_x=1, R_y=0}  \int p(X, Y, R_x=1, R_y=0) dy \times \prod_{R_x=0, R_y=0}  \int p(X, Y, R_x=0, R_y=0) dxdy. 
	\end{align*}
}

Suppose there exist $p_1(Ry\mid X, R_x)$ and $p_2(Ry\mid X, R_x)$ such that 
$$\int p(X, Y)p(R_x=0\mid Y)p_1(R_y=1\mid R_x=0,X) dx=\int p(X, Y)p(R_x=0\mid Y)p_2(R_y=1\mid R_x=0,X) dx$$
Let $g(X)=p_1(R_y=1\mid R_x=0,X)-p_2(R_y=1\mid R_x=0,X)$, we have
$$p(R_x=0\mid Y = y)\ p(Y = y)\int p(x \mid Y = y) \ g(x) \ dx=0,\, \forall y$$
This must mean that $E[g(X)\mid y]=0,\, \forall y$. 
In our case, $g(X)$ is bounded, thus is with finite mean. Based on the completeness condition, $g(X)=0$ almost surely, which implies $p_1(R_y \mid X, R_x) = p_2(R_y\mid X, R_x)$ almost surely. This concludes that the full law is indeed identified.

\newpage
\section{Examples from the exponential family distributions}\label{app:parID-ex}

In order to better illustrate the implications of Theorem~\ref{thm:id-par}, we provide explicit sufficient identification conditions in a variety of examples in the class of exponential family distributions. In all subsequent examples, we assume that if $X$ is continuous, a sufficient number of unique $X$ values have been observed such that the first condition in Theorem~\ref{thm:id-par}, namely that $k \geq dim(\theta)$, is satisfied. If $X$ is discrete, it is assumed that every category of $X$ is observed in the sample. 


\subsection{$X$ and $Y$ are bivariate normal}\label{app:parID-ex-bivar}

Suppose  
$$\left(\begin{array}{l} Y \\ X\end{array}\right) \quad \sim \quad \N\left[\left(\begin{array}{l}\mu_{1} \\ \mu_{2}\end{array}\right), 
\left(\begin{array}{cc}\sigma_{1}^{2} & \rho \sigma_{1} \sigma_{2} \\ \rho \sigma_{1} \sigma_{2} & \sigma_{2}^{2} \end{array}\right)\right].$$

According to Theorem~\ref{thm:id-par},  $p(X,Y)$ is identifiable if at least $\mu_1$ or $\mu_2$ is known, in addition to knowing at least one more parameter in $\{\sigma_1, \sigma_2, \rho\}$. As special cases, when either the marginal distribution of $X$ or $Y$ is known, we can identify $p(X,Y)$.

The above claim can be proven as follows. First, we note that $p(X \mid Y)$ also follows a normal distribution: 
\begin{align*}
	X \mid Y \sim \N\left[\mu_{2}+\rho \frac{\sigma_{2}}{\sigma_{1}}\left(y-\mu_{1}\right),\left(1-\rho^{2}\right) \sigma_{2}^{2}\right]. 
\end{align*}
Since $p(X \mid Y)$ is nonparametrically identified, it means the mean and variance are both identifiable, i.e., $\mu_{2}+\rho \frac{\sigma_{2}}{\sigma_{1}}\left(y-\mu_{1}\right)$ and $\left(1-\rho^{2}\right) \sigma_{2}^{2}$. Thus the following three parameters are identified: 
\begin{align*}
	\mu_{2}-\rho \frac{\sigma_{2}}{\sigma_{1}}\mu_{1},\quad\rho \frac{\sigma_{2}}{\sigma_{1}},\quad \left(1-\rho^{2}\right) \sigma_{2}^{2}
\end{align*}
Let $\theta=(\mu_1,\mu_2,\sigma_1,\sigma_2,\rho)$. By taking derivative with respect to $\theta$, we obtain the following Jacobian matrix: 
\begin{align*}
	J=\left[\begin{array}{ccccc}-\rho \frac{\sigma_{2}}{\sigma_{1}} & 1 & \rho \frac{\sigma_{2}}{\sigma_{1}^{2}} \mu_{1} & -\rho \frac{1}{\sigma_{1}} \mu_{1} & -\frac{\sigma_{2}}{\sigma_{1}} \mu_{1} \\ 0 & 0 & -\rho \frac{\sigma_{2}}{\sigma_{1}^{2}} \mu_{1} & \rho \frac{1}{\sigma_{1}} & \frac{\sigma_{2}}{\sigma_{1}} \\ 0 & 0 & 0 & 2\left(1-\rho^{2}\right) \sigma_{2} & -2 \rho \sigma_{2}^{2}\end{array}\right]
\end{align*}
The number of unknown parameters is greater than the number of equations. To establish target law identification, we need to assume two of the five parameters are known. However, not every pair of parameters will be useful in establishing identification. We go over different options one by one: ($|J|$ denotes the determinant of matrix $J$.)
\begin{itemize}
	\item Assume $\mu_1, \mu_2$ are known, then $|J|\neq 0\Longrightarrow \text{target law is identified}$
	$$J=\left[\begin{array}{ccc} \rho \frac{\sigma_{2}}{\sigma_{1}^{2}} \mu_{1} & -\rho \frac{1}{\sigma_{1}} \mu_{1} & -\frac{\sigma_{2}}{\sigma_{1}} \mu_{1} \\  -\rho \frac{\sigma_{2}}{\sigma_{1}^{2}} \mu_{1} & \rho \frac{1}{\sigma_{1}} & \frac{\sigma_{2}}{\sigma_{1}} \\  0 & 2\left(1-\rho^{2}\right) \sigma_{2} & -2 \rho \sigma_{2}^{2}\end{array}\right]$$
	
	\item Assume $\mu_1, \sigma_1$ are known, then $|J|\neq 0\Longrightarrow \text{target law is identified}$
	$$J=\left[\begin{array}{ccc} 1 &  -\rho \frac{1}{\sigma_{1}} \mu_{1} & -\frac{\sigma_{2}}{\sigma_{1}} \mu_{1} \\  0 &  \rho \frac{1}{\sigma_{1}} & \frac{\sigma_{2}}{\sigma_{1}} \\  0 &  2\left(1-\rho^{2}\right) \sigma_{2} & -2 \rho \sigma_{2}^{2}\end{array}\right]$$
	
	\item Assume $\mu_1, \sigma_2$ are known, then $|J|\neq 0\Longrightarrow \text{target law is identified}$
	$$J=\left[\begin{array}{ccc} 1 & \rho \frac{\sigma_{2}}{\sigma_{1}^{2}} \mu_{1} &  -\frac{\sigma_{2}}{\sigma_{1}} \mu_{1} \\  0 & -\rho \frac{\sigma_{2}}{\sigma_{1}^{2}} \mu_{1} &  \frac{\sigma_{2}}{\sigma_{1}} \\ 0 & 0 &  -2 \rho \sigma_{2}^{2}\end{array}\right]$$
	
	\item Assume $\mu_1, \rho$ are known, then $|J|\neq 0\Longrightarrow \text{target law is identified}$
	$$J=\left[\begin{array}{ccc} 1 & \rho \frac{\sigma_{2}}{\sigma_{1}^{2}} \mu_{1} & -\rho \frac{1}{\sigma_{1}} \mu_{1}  \\  0 & -\rho \frac{\sigma_{2}}{\sigma_{1}^{2}} \mu_{1} & \rho \frac{1}{\sigma_{1}} \\  0 & 0 & 2\left(1-\rho^{2}\right) \sigma_{2} \end{array}\right]$$
	
	\item Assume $\mu_2, \sigma_1$ are known, then $|J|\neq 0\Longrightarrow \text{target law is identified}$
	$$J=\left[\begin{array}{ccc} -\rho \frac{\sigma_{2}}{\sigma_{1}} &  -\rho \frac{1}{\sigma_{1}} \mu_{1} & -\frac{\sigma_{2}}{\sigma_{1}} \mu_{1} \\ 0 &  \rho \frac{1}{\sigma_{1}} & \frac{\sigma_{2}}{\sigma_{1}} \\ 0 & 2\left(1-\rho^{2}\right) \sigma_{2} & -2 \rho \sigma_{2}^{2}\end{array}\right]$$
	
	\item Assume $\mu_2, \sigma_2$ are known, then $|J|\neq 0\Longrightarrow \text{target law is identified}$ 
	$$J=\left[\begin{array}{ccc}-\rho \frac{\sigma_{2}}{\sigma_{1}} &  \rho \frac{\sigma_{2}}{\sigma_{1}^{2}} \mu_{1}  & -\frac{\sigma_{2}}{\sigma_{1}} \mu_{1} \\ 0 &  -\rho \frac{\sigma_{2}}{\sigma_{1}^{2}} \mu_{1}  & \frac{\sigma_{2}}{\sigma_{1}} \\ 0 &  0 &  -2 \rho \sigma_{2}^{2}\end{array}\right]$$
	
	This recovers the case studied in \cite{zhao2015semiparametric}. 
	
	\item Assume $\mu_2, \rho$ are known, then $|J|\neq 0\Longrightarrow \text{target law is identified}$
	$$J=\left[\begin{array}{ccc}-\rho \frac{\sigma_{2}}{\sigma_{1}} &  \rho \frac{\sigma_{2}}{\sigma_{1}^{2}} \mu_{1} & -\rho \frac{1}{\sigma_{1}} \mu_{1}  \\ 0 &  -\rho \frac{\sigma_{2}}{\sigma_{1}^{2}} \mu_{1} & \rho \frac{1}{\sigma_{1}}  \\ 0 &  0 & 2\left(1-\rho^{2}\right) \sigma_{2} \end{array}\right]$$
	
	\item Assume $\sigma_1, \sigma_2$ are known, then $|J|= 0\Longrightarrow \text{target law is \textbf{not} identified}$
	$$J=\left[\begin{array}{ccc}-\rho \frac{\sigma_{2}}{\sigma_{1}} & 1 &  -\frac{\sigma_{2}}{\sigma_{1}} \mu_{1} \\ 0 & 0 &  \frac{\sigma_{2}}{\sigma_{1}} \\ 0 & 0 &  -2 \rho \sigma_{2}^{2}\end{array}\right]$$
	
	\item Assume $\sigma_1, \rho$ are known, then $|J|= 0\Longrightarrow \text{target law \textbf{is not} identified}$
	$$J=\left[\begin{array}{ccc}-\rho \frac{\sigma_{2}}{\sigma_{1}} & 1 & -\rho \frac{1}{\sigma_{1}} \mu_{1} \\ 0 & 0 &  \rho \frac{1}{\sigma_{1}}  \\ 0 & 0  & 2\left(1-\rho^{2}\right) \sigma_{2} \end{array}\right]$$
	
	\item Assume $\sigma_2, \rho$ are known, then $|J|= 0\Longrightarrow \text{target law \textbf{is not} identified}$
	$$J=\left[\begin{array}{ccc}-\rho \frac{\sigma_{2}}{\sigma_{1}} & 1 &\rho \frac{\sigma_{2}}{\sigma_{1}^{2}} \mu_{1} \\ 0 & 0 & -\rho \frac{\sigma_{2}}{\sigma_{1}^{2}} \mu_{1} \\ 0 & 0 & 0 \end{array}\right]$$
\end{itemize}

This concludes that under the bivariate normal distribution, the target law is identified if either $\mu_1$ or $\mu_2$ is known, in addition to knowing at least one more parameter in $\{\sigma_1, \sigma_2, \rho\}$.

It is straightforward to show that \textbf{$p(X\mid Y)$ lies in the exponential family}.


\subsection{$X$ and $Y\mid X$ are normal under inverse link}\label{app:parID-ex-normal-inverse}

Suppose 
\begin{align*}
	X \sim \N\left(\mu, \phi_x\right), \qquad 
	Y \mid X \sim \N\left((\alpha+\beta x)^{-1}, \phi\right). 
\end{align*} 

According to Theorem~\ref{thm:id-par}, $p(X,Y)$ is identifiable without any additional assumptions on the unknown parameter vector $\theta=(\alpha,\beta,\phi,\mu,\phi_x)$. This can be proven as follows: based on Theorem~\ref{thm:id-par}, we have the following equations,
\begin{align*}
	\phi_i(\theta) &=\frac{\left(\alpha+\beta x_{i}\right)^{-1}-\left(\alpha+\beta x_{0}\right)^{-1}}{\phi} \\ 
	\zeta_i(\theta) &=\left\{-\frac{b\left[\left(\alpha+\beta x_{i}\right)^{-1}\right]-b\left[\left(\alpha+\beta x_{0}\right)^{-1}\right]}{\phi}+\frac{\mu\left(x_{i}-x_{0}\right)}{\phi_{x}}+c\left(x_{i}, \phi_{x}\right)-c\left(x_{0}, \phi_{x}\right)\right\} \\ 
	&=-\frac{\left(\alpha+\beta x_{i}\right)^{-2}-\left(\alpha+\beta x_{0}\right)^{-2}}{2 \phi}+\frac{\mu\left(x_{i}-x_{0}\right)}{\phi_{x}}-\frac{x_{i}^{2}-x_{0}^{2}}{2 \phi_{x}}, \quad 
	\text{where } i\in (1,\ldots, k). 
\end{align*} 

The Jacobian matrix is as follows: 

$$\begin{bmatrix}
	-\frac{\left(\alpha+\beta x_{1}\right)^{-2}-\left(\alpha+\beta x_{0}\right)^{-2}}{\phi} & -\frac{\left(\alpha+\beta x_{1}\right)^{-2} x_{1}-\left(\alpha+\beta x_{0}\right)^{-2} x_{0}}{\phi} & -\frac{\left(\alpha+\beta x_{1}\right)^{-1}-\left(\alpha+\beta x_{0}\right)^{-1}}{\phi^{2}} & 0 & 0\\
	\vdots & & & & \vdots\\
	-\frac{\left(\alpha+\beta x_{k}\right)^{-2}-\left(\alpha+\beta x_{0}\right)^{-2}}{\phi} & -\frac{\left(\alpha+\beta x_{k}\right)^{-2} x_{k}-\left(\alpha+\beta x_{0}\right)^{-2} x_{0}}{\phi} & -\frac{\left(\alpha+\beta x_{k}\right)^{-1}-\left(\alpha+\beta x_{0}\right)^{-1}}{\phi^{2}} & 0 & 0\\
	2 \frac{\left(\alpha+\beta x_{1}\right)^{-3}-\left(\alpha+\beta x_{0}\right)^{-3}}{2 \phi} & 2 \frac{\left(\alpha+\beta x_{1}\right)^{-3} x_{1}-\left(\alpha+\beta x_{0}\right)^{-3} x_{0}}{2 \phi} & \frac{\left(\alpha+\beta x_{1}\right)^{-2}-\left(\alpha+\beta x_{0}\right)^{2}}{2 \phi^2} & \frac{x_{1}-x_{0}}{\phi_{x}} & \frac{\left(x_{1}-x_{0}\right)\left(x_{1}+x_{0}-2 \mu\right)}{2 \phi_{x}^{2}}\\
	\vdots & & & & \vdots\\
	2 \frac{\left(\alpha+\beta x_{k}\right)^{-3}-\left(\alpha+\beta x_{0}\right)^{-3}}{2 \phi} & 2 \frac{\left(\alpha+\beta x_{k}\right)^{-3} x_{k}-\left(\alpha+\beta x_{0}\right)^{-3} x_{0}}{2 \phi} & \frac{\left(\alpha+\beta x_{k}\right)^{-2}-\left(\alpha+\beta x_{0}\right)^{2}}{2 \phi^2} & \frac{x_{k}-x_{0}}{\phi_{x}} & \frac{\left(x_{k}-x_{0}\right)\left(x_{k}+x_{0}-2 u\right)}{2 \phi_{x}^{2}}
\end{bmatrix}$$

After performing some rank-preserving modifications to this matrix, we get: 

$$
\scriptsize
	\begin{bmatrix}
		\left(\alpha+\beta x_{1}\right)^{-2}-\left(\alpha+\beta x_{0}\right)^{-2} & \left(\alpha+\beta x_{1}\right)^{-2} x_{1}-\left(\alpha+\beta x_{0}\right)^{-2} x_{0} &\left(\alpha+\beta x_{1}\right)^{-1}-\left(\alpha+\beta x_{0}\right)^{-1} & 0 & 0\\
		\vdots & & & & \vdots\\
		\left(\alpha+\beta x_{k}\right)^{-2}-\left(\alpha+\beta x_{0}\right)^{-2} & \left(\alpha+\beta x_{k}\right)^{-2} x_{k}-\left(\alpha+\beta x_{0}\right)^{-2} x_{0} &\left(\alpha+\beta x_{k}\right)^{-1}-\left(\alpha+\beta x_{0}\right)^{-1} & 0 & 0\\
		\left(\alpha+\beta x_{1}\right)^{-3}-\left(\alpha+\beta x_{0}\right)^{-3} & \left(\alpha+\beta x_{1}\right)^{-3} x_1 -\left(\alpha+\beta x_{0}\right)^{-3} x_0 & \frac{1}{2}\left[\left(\alpha+\beta x_{1}\right)^{-2}-\left(\alpha+\beta x_{0}\right)^{-2}\right] & x_1-x_0 & \left(x_{1}-x_{0}\right)\left(x_{1}+x_{0}-2 \mu\right)\\
		\vdots & & & & \vdots\\
		\left(\alpha+\beta x_{k}\right)^{-3}-\left(\alpha+\beta x_{0}\right)^{-3} & \left(\alpha+\beta x_{k}\right)^{-3} x_k -\left(\alpha+\beta x_{0}\right)^{-3} x_0 & \frac{1}{2}\left[\left(\alpha+\beta x_{k}\right)^{-2}-\left(\alpha+\beta x_{0}\right)^{-2}\right] & x_k-x_0 & \left(x_{k}-x_{0}\right)\left(x_{k}+x_{0}-2 \mu\right)
\end{bmatrix}
$$
which is of full rank. 

It is worth pointing out that unlike the example in (\ref{app:parID-ex-bivar}), \textbf{$p(X\mid Y)$ in this example is not in the exponential family}, since: 
\begin{align*}
	p(x \mid y)&=\frac{p(y \mid x) p(x)}{p(y)}=\frac{\N\left((a+b x)^{-1}, \sigma_{y}^{2}\right) \N\left(\mu, \sigma_{x}^{2}\right)}{p(y)}\\
	&=\exp \left\{-\frac{\left(y-\frac{1}{a+b x}\right)^{2}}{2\sigma_{y}^{2}}+\log \frac{1}{\sqrt{2 \pi} \sigma_{y}}-\frac{(x-\mu)^{2}}{2 \sigma_{x}^{2}}+\log \frac{1}{\sqrt{2 \pi} \sigma_{x}}-\log p(y)\right\}\\
	&=\exp \left\{-\frac{\frac{1}{(a+b x)^{2}}-\frac{2 y}{a+b x}+y^{2}}{2\sigma_{y}^{2}}+\log \frac{1}{\sqrt{2 \pi} \sigma_{y}}-\frac{(x-\mu)^{2}}{2 \sigma_{x}^{2}}+\log \frac{1}{\sqrt{2 \pi} \sigma_{x}}-\log p(y)\right\}. 
\end{align*}


\subsection{$X$ and $Y$ are binary}\label{app:parID-ex-binary}

Suppose $p(X = 0, Y = 1) = p_1$, $p(X = 1, Y = 0) = p_2$, $p(X = 0, Y = 0) = p_3$, and $p(X = 1, Y = 1) = p_4$, where $\sum_{i = 1}^4 p_i=1, p_i \neq 0$. 
The unknown parameters of interest are $\theta=(p_1,p_2,p_3,p_4)$. 

In this binary case, there are at most two distinct values of $X$ as $0$ or $1$. According to Theorem~\ref{thm:id-par}, $p(X,Y)$ is identifiable if any one of $p_i$ is known or marginal distribution of either $X$ or $Y$ is known. 

In order to prove the above claim, we look at two distinct parameterizations of $p(X, Y)$. 

\subsubsection{Parameterization 1}
Suppose $p_1 = p(X = 0, Y = 1)$, $p_2 = p(X = 1, Y = 0)$, $p_3 = p(X = 0, Y = 0)$, $p_4(X = 1, Y = 1)$, $p_i \neq 0,\, i=1,\ldots, 4$. 

Since $p(X \mid Y)$ is nonparametrically identified, we obtain the following three equations with four unknowns: 
\begin{align*}
	p(X=1 \mid Y=1) =\frac{p_{4}}{p_{1}+p_{4}},  \qquad  
	p(X=1 \mid Y=0) =\frac{p_{2}}{p_{2}+p_{3}}, \qquad
	\sum_{i=1}^{4} p_{i} =1
\end{align*}

In order to possibly achieve identification, we need to assume one parameter is known. We consider the four different scenarios one by one. 
\begin{itemize}
	\item Assume $p_1$ is known, then $|J|\neq 0 \Longrightarrow \text{target law is identified}$
	$$J=\left[\begin{array}{ccc}0 & 0 & \frac{p_{1}}{\left(p_{1}+p_{4}\right)^{2}} \\ \frac{p_{3}}{\left(p_{2}+p_{3}\right)^{2}} & \frac{-p_{2}}{\left(p_{2}+p_{3}\right)^{2}} & 0 \\ 1 & 1 & 1\end{array}\right]$$
	
	\item Assume $p_2$ is known, then $|J|\neq 0 \Longrightarrow \text{target law is identified}$
	$$J=\left[\begin{array}{ccc}\frac{-p_{4}}{\left(p_{1}+p_{4}\right)^{2}} & 0 & \frac{p_{1}}{\left(p_{1}+p_{4}\right)^{2}} \\ 0 & \frac{p_{3}}{\left(p_{2}+p_{3}\right)^{2}} & 0 \\ 1 & 1 & 1\end{array}\right]$$
	
	\item Assume $p_3$ is known, then $|J|\neq 0 \Longrightarrow \text{target law is identified}$
	$$J=\left[\begin{array}{ccc}\frac{-p_{4}}{\left(p_{1}+p_{4}\right)^{2}} & 0 & \frac{p_{1}}{\left(p_{1}+p_{4}\right)^{2}} \\ 0 & \frac{p_{3}}{\left(p_{2}+p_{3}\right)^{2}} & 0 \\ 1 & 1 & 1\end{array}\right]$$
	
	\item Assume $p_4$ is known, then $|J|\neq 0 \Longrightarrow \text{target law is identified}$
	$$J=\left[\begin{array}{ccc}\frac{-p_{4}}{\left(p_{1}+p_{4}\right)^{2}} & 0 & 0 \\ 0 & \frac{p_{3}}{\left(p_{2}+p_{3}\right)^{2}} & \frac{-p_{2}}{\left(p_{2}+p_{3}\right)^{2}} \\ 1 & 1 & 1\end{array}\right]$$
\end{itemize}
In the binary case, it is also useful to assume
\begin{itemize}
	\item Assume $p(Y=1)=p_1+p_4$ is known, then $|J|\neq 0 \Longrightarrow \text{target law is identified}$
	$$J=\left[\begin{array}{cccc}-\frac{p_{4}}{\left(p_{1}+p_{4}\right)^{2}} & 0 & 0 & \frac{p_{1}}{\left(p_{1}+p_{4}\right)^{2}} \\ 0 & \frac{p_{3}}{\left(p_{2}+p_{3}\right)^{2}} & -\frac{p_{2}}{\left(p_{2}+p_{3}\right)^{2}} & 0 \\ 1 & 1 & 1 & 1 \\ 1 & 0 & 0 & 1\end{array}\right]$$
	
	\item Assume $p(X=1)=p_2+p_4$ is known, then $|J|\neq 0 \Longrightarrow \text{target law is identified}$
	$$J=\left[\begin{array}{cccc}-\frac{p_{4}}{\left(p_{1}+p_{4}\right)^{2}} & 0 & 0 & \frac{p_{1}}{\left(p_{1}+p_{4}\right)^{2}} \\ 0 & \frac{p_{3}}{\left(p_{2}+p_{3}\right)^{2}} & -\frac{p_{2}}{\left(p_{2}+p_{3}\right)^{2}} & 0 \\ 1 & 1 & 1 & 1 \\ 0 & 1 & 0 & 1\end{array}\right]$$
\end{itemize}

\subsubsection{Parameterization 2}
We can also adopt another parameterization. Suppose 
\begin{align*}
	X \sim \operatorname{Bern}(p),  \qquad Y \mid  X\sim \operatorname{Bern}(a+bX) 
\end{align*}%
More specifically,
\begin{align*}
	p(x) &=\exp \left\{x \log \frac{p}{1-p}+\log (1-p)\right\} =\exp \left\{x \cdot \eta_{x}-\log \left(1+e^{\eta_ x}\right)\right\} \quad \text{ where } \eta_x=\log \frac{p}{1-p}  \\ 
	p(y\mid x)&=(a+b x)^{y}(1-a-b x)^{1-y} \\ &
	=\exp \left\{y \log \frac{a+b x}{1-(a+b x)}+\log [1-(a+b x)]\right\}
\end{align*}
The parameter vector of interest is $\theta=(a,b,\eta_x)$. Based on Theorem~\ref{thm:id-par}, we have the following equations. Note that since $X$ is binary, there are at most two distinct values of $X$. Therefore, we have the following two equations:
\begin{align*}
	& \phi_1(\theta)=\log \frac{a+b x_1}{1-\left(a+b x_1\right)}-\log \frac{a+b x_0}{1-\left(a+b x_0\right)} \\
	& \zeta_1(\sigma)=\log \left[1-\left(a+b x_1\right)\right]-\log \left[1-\left(a+b x_0\right)\right]+\left(x_1-x_0\right) \eta_x,\text{ where }x_1=1,x_0=0.
\end{align*}
The resulted Jacobian matrix is: 
$$J=\begin{bmatrix}
	\frac{1}{(a+b)[1-(a+b)]}-\frac{1}{a(1-a)} & \frac{1}{(a+b)[1-(a+b)]} & 0\\
	\frac{-1}{1-(a+b)}+\frac{1}{1-a} & \frac{-1}{1-(a+b)} & x_1-x_0
\end{bmatrix}$$
This concludes that in order to establish target law identification, we need to know at least one parameter in $\{a, b, \eta_x\}$.

It is straightforward to show that \textbf{$p(X\mid Y)$ lies in the exponential family}.


\subsection{$X$ is binary and $Y\mid X$ is normal under canonical link}\label{app:parID-ex-binary_normal}

Suppose 
\begin{align*}
	X \sim \operatorname{Bern}(p), \qquad Y \mid X \sim \N\left(a+b X, \sigma_{y}^{2}\right).  
\end{align*}
More specifically, 
\begin{align*}
	&p(x) = p^{x}(1-p)^{1-x}=\exp \left\{x \cdot \log \frac{p}{1-p}+\log (1-p)\right\}=\exp \left\{x \cdot \eta-\log \left(1+e^{\eta}\right)\right\},\text{ where } \eta=\log \frac{p}{1-p} \\ 
	&p(y\mid x) =\exp \left\{\frac{y(a+b x)-\frac{1}{2}(a+b x)^{2}}{\phi}+\left[-\frac{y^{2}}{2 \phi}-\frac{1}{2} \log \left(2 \pi \phi\right)\right]\right\}, \text{ where } \phi=\sigma_y^2.  
\end{align*}
The unknown parameter vector of interest is $\theta=(a, b, \phi, \eta)$. According to Theorem~\ref{thm:id-par}, $p(X,Y)$ is identifiable if at either $a$ or $\eta$ is known, in addition to knowing one extra parameter in $\theta$. Knowing $\eta$ is equivalent to knowing $p$. 

In order to prove the above claim, we can construct the following equations: (note that when $X$ is binary, we only have at most two distinct values)
$$\begin{aligned}
	&\phi_{1}(\theta)=\frac{\left(a+b x_{1}\right)-\left(a+b x_{0}\right)}{\phi}=\frac{b\left(x_{i}-x_{0}\right)}{\phi}\\
	&\zeta_{1}(\theta)=-\frac{\left(a+b x_{1}\right)^{2}-\left(a+b x_{0}\right)^{2}}{2 \phi}+\eta\left(x_{1}-x_{0}\right),\text{ where }x_1=1,x_0=0. 
\end{aligned}$$
The Jacobian matrix is: 
$$J=\begin{bmatrix}
	0 & \frac{x_{1}-x_{0}}{\phi} & -\frac{b\left( x_{1}-x_{0}\right)}{\phi^{2}} & 0\\
	-\frac{b\left(x_{1}-x_{0}\right)}{\phi} & -\frac{a\left(x_{1}-x_{0}\right)+b\left(x_{1}^{2}-x_{0}^{2}\right)}{\phi} & \frac{\left(a+b x_{1}\right)^{2}-\left(a+b x_{0}\right)^{2}}{2 \phi^{2}} & x_{1}-x_{0}
\end{bmatrix}.$$
After some rank-preserving operations, we get: 
$$\begin{bmatrix}
	0 & x_1-x_0 & x_1-x_0 & 0\\
	1 & a\left(x_{1}-x_{0}\right)+b\left(x_{1}^{2}-x_{0}^{2}\right) & a\left(x_{1}-x_{0}\right)+\frac{b}{2}\left(x_{1}^{2}-x_{0}^{2}\right) & 1
\end{bmatrix}.$$
This concludes the claim that a sufficient set of assumptions for target law identification is knowing either $a$ or $\eta$, in addition to knowing one more parameter in $\theta$. 

Note that in this example, \textbf{$p(X \mid Y)$ is in exponential family} since: 
$$
\scriptsize
	\begin{aligned}
		p(x \mid y)&=\frac{p(y \mid x) p(x)}{p(y)}=\frac{N_{y}\left(a+b x, \sigma_{y}^{2}\right) p^{x}(1-p)^{1-x}}{p(y)}\\
		&=\exp \left\{-\frac{1}{2}\left(\frac{y-(a+b x)}{\sigma_{y}}\right)^{2}+\log \frac{1}{\sqrt{2 \pi} \sigma_{y}}+x \log p+(1-x) \log (1-p)-\log \left[p(y)\right]\right\} \\
		&=\exp \left\{-\frac{1}{2} \frac{\left(x, x^{2}\right)\left(2 a b-2 b y, b^{2}\right)^{T}+(a-y)^{2}}{\sigma_{y}^{2}}+\log \frac{1}{\sqrt{2 \pi} \sigma_{y}}+x \log p+(1-x) \log (1-p)-\log \left[p(y) \right]\right\} \\
		&=\exp \left\{\left(x, x^{2}\right)\left(-\frac{a b-b y}{\sigma_{y}^{2}}+log(\frac{p}{1-p}), -\frac{b^{2}}{2 \sigma_{y}^{2}}\right)^{T}-\frac{(a-y)^{2}}{2 \sigma_{y}^{2}}+\log \frac{1}{\sqrt{2 \pi} \sigma_{y}}+log(1-p)-\log \left[p(y)\right]\right\}.
\end{aligned}
$$


\subsection{$X$ is Poisson and $Y \mid X$ is normal under canonical link}\label{app:parID-ex-poisson_normal}

Suppose 
\begin{align*}
	X \sim \operatorname{ Poisson }(\lambda), \qquad Y\mid X \sim \N\left(a+b x, \sigma_{y}^{2}\right). 
\end{align*}
More specifically, 
\begin{align*}
	&p(y\mid x)=\exp \left\{\frac{y(a+b x)-\frac{1}{2}(a+b x)^{2}}{\phi}+\left[-\frac{y^{2}}{2 \phi}-\frac{1}{2} \log \left(2 \pi \phi\right)\right]\right\}, \text{ where } \phi=\sigma_y^2 \\
	&p(x=k)=\frac{\lambda^{k} e^{-\lambda}}{k !} =\exp \{k \log \lambda-\lambda-\log k !\} =\exp \left\{k \eta_{x}-e^{\eta_{x}}-\log k !\right\},\text{ where } \eta_{x}=\log \lambda
\end{align*}
The unknown parameter vector of interest is $\theta=\left(a, b, \sigma_y^2, \lambda\right)$. According to Theorem~\ref{thm:id-par}, $p(X,Y)$ is identifiable if either $a$ or $\lambda$ is known. 

In order to prove the above claim, we can construct the following equations: 
\begin{align*}
	&\phi_{i}(\theta)=\frac{\left(a+b x_{i}\right)-\left(a+b x_{0}\right)}{\phi}=\frac{b\left(x_{i}-x_{0}\right)}{\phi} \\ 
	&\zeta_{i}(\theta)=-\frac{\left(a+b x_{i}\right)^{2}-\left(a+b x_{0}\right)^{2}}{2 \phi}+\eta_{x}\left(x_{i}-x_{0}\right)+\left(-\log x_{i} !+\log x_{0} !\right), \quad \text{where } i\in (1,\ldots,k)
\end{align*}
The Jacobian matrix is then as follows:
$$J=\begin{bmatrix}
	0 & \frac{x_{1}-x_{0}}{\phi} & -\frac{\left(b x_{1}-x_{0}\right)}{\phi^{2}} & 0\\
	0 & \frac{x_{2}-x_{0}}{\phi} & -\frac{\left(b x_{2}-x_{0}\right)}{\phi^{2}} & 0\\
	\vdots & & & \vdots\\
	0 & \frac{x_{k}-x_{0}}{\phi} & -\frac{\left(b x_{k}-x_{0}\right)}{\phi^{2}} & 0\\
	-\frac{b\left(x_{1}-x_{0}\right)}{\phi} & -\frac{a\left(x_{1}-x_{0}\right)+b\left(x_{1}^{2}-x_{0}^{2}\right)}{\phi} & \frac{\left(a+b x_{1}\right)^{2}-\left(a+b x_{0}\right)^{2}}{2 \phi^{2}} & x_1-x_0\\
	-\frac{b\left(x_{2}-x_{0}\right)}{\phi} & -\frac{a\left(x_{2}-x_{0}\right)+b\left(x_{2}^{2}-x_{0}^{2}\right)}{\phi} & \frac{\left(a+b x_{2}\right)^{2}-\left(a+b x_{0}\right)^{2}}{2 \phi^{2}} & x_2-x_0\\
	\vdots & & & \vdots\\
	-\frac{b\left(x_{k}-x_{0}\right)}{\phi} & -\frac{a\left(x_{k}-x_{0}\right)+b\left(x_{k}^{2}-x_{0}^{2}\right)}{\phi} & \frac{\left(a+b x_{k}\right)^{2}-\left(a+b x_{0}\right)^{2}}{2 \phi^{2}} & x_k-x_0\\
\end{bmatrix}. $$
After some rank-preserving operations, we get: 
$$\begin{bmatrix}
	0 & x_1-x_0 & x_1-x_0 & 0\\
	0 & x_2-x_0 & x_2-x_0 & 0\\
	\vdots & & & \vdots\\
	0 & x_k-x_0 & x_k-x_0 & 0\\
	x_{1}-x_{0} & a\left(x_{1}-x_{0}\right)+b\left(x_{1}^{2}-x_{0}^{2}\right) & a\left(x_{1}-x_{0}\right)+\frac{b}{2}\left(x_{1}^{2}-x_{0}^{2}\right) & x_{1}-x_{0}\\
	x_{2}-x_{0} & a\left(x_{2}-x_{0}\right)+b\left(x_{2}^{2}-x_{0}^{2}\right) & a\left(x_{2}-x_{0}\right)+\frac{b}{2}\left(x_{2}^{2}-x_{0}^{2}\right) & x_{2}-x_{0}\\
	\vdots & & & \vdots\\
	x_{k}-x_{0} & a\left(x_{k}-x_{0}\right)+b\left(x_{k}^{2}-x_{0}^{2}\right) & a\left(x_{k}-x_{0}\right)+\frac{b}{2}\left(x_{k}^{2}-x_{0}^{2}\right) & x_{k}-x_{0}\\
\end{bmatrix}. $$
We need to know either $a$ or $\eta_x$ to establish identifiability.\\

Note that in this example, \textbf{$p(X\mid Y)$ is in the exponential family} since: 
{\small 
	\begin{align*}
		p(x \mid y)&=\frac{p(y \mid x) p(x)}{p(y)}=\frac{N_{y}\left(a+b x, \sigma_{y}^{2}\right) \frac{\lambda^{x} e^{-\lambda}}{x !}}{p(y)}\\
		&=\exp \left\{-\frac{1}{2}\left(\frac{y-(a+b x)}{\sigma_{y}}\right)^{2}+\log \frac{1}{\sqrt{2 \pi} \sigma_{y}}+x \log \lambda-\lambda-\log x !-\log p(y)\right\}\\
		&=\frac{1}{x !} \exp \left\{-\frac{1}{2} \frac{\left(x, x^{2}\right)\left(2 a b-2 b y-2 \sigma_{y}^{2} \log \lambda, b^{2}\right)^{T}+(a-y)^{2}}{\sigma_{y}^{2}}+\log \frac{1}{\sqrt{2 \pi} \sigma_{y}}-\lambda-\log p(y)\right\}\\
		&=\frac{1}{x !} \exp \left\{\left(x, x^{2}\right)\left(-\frac{a b-b y-\sigma_{y}^{2} \log \lambda}{\sigma_{y}^{2}},-\frac{b^{2}}{2 \sigma_{y}^{2}}\right)^{T}-\frac{(a-y)^{2}}{2 \sigma_{y}^{2}}+\log \frac{1}{\sqrt{2 \pi} \sigma_{y}}-\lambda-\log p(y)\right\}. 
	\end{align*}
}


\subsection{$X$ is exponential and $Y \mid X$ is normal under canonical link}\label{app:parID-ex-exponential_normal}

Suppose 
\begin{align*}
	X \sim \operatorname{exponential} (\lambda), \qquad Y\mid X \sim \N\left(a+b x, \sigma_{y}^{2}\right). 
\end{align*}
More specifically, 
\begin{align*}
	p(x) &=\lambda e^{-\lambda x}=\exp \{-\lambda x+\log \lambda\} \\
	p(y\mid x) &=\exp \left\{\frac{y(a+b x)-\frac{1}{2}(a+b x)^{2}}{\phi}+\left[-\frac{y^{2}}{2 \phi}-\frac{1}{2} \log \left(2 \pi \phi\right)\right]\right\}\quad \text{where } \phi=\sigma_y^2 
\end{align*}

The unknown vector of parameters is $\theta=\left(a, b, \phi, \lambda\right)$. According to Theorem~\ref{thm:id-par}, $p(X,Y)$ is identifiable if either $a$ or $\lambda$ is known. 

In order to prove the above claim, we can construct the following equations: 
\begin{align*}
	&\phi_{i}(\theta)=\frac{b\left(x_{i}-x_{0}\right)}{\phi} \\ 
	&\zeta_{i}(\theta)=-\frac{\left(a+b x_{i}\right)^{2}-\left(a+b x_{0}\right)^{2}}{2 \phi}-\lambda\left(x_{i}-x_{0}\right), \quad \text{where } i\in (1,\ldots,k)
\end{align*}
The Jacobian matrix is
$$J=\begin{bmatrix}
	0 & \frac{x_{1}-x_{0}}{\phi} & -\frac{b\left(x_{1}-x_{0}\right)}{\phi^{2}} & 0\\
	\vdots & & & \vdots\\
	0 & \frac{x_{k}-x_{0}}{\phi} & -\frac{b\left(x_{k}-x_{0}\right)}{\phi^{2}} & 0\\
	-\frac{b\left(x_{1}-x_{0}\right)}{\phi} & -\frac{a\left(x_{1}-x_{0}\right)+b\left(x_{1}^{2}-x_{0}^{2}\right)}{\phi} & \frac{\left(a+b x_{1}\right)^{2}-\left(a+b x_{0}\right)^{2}}{2 \phi^{2}} & -(x_1-x_0)\\
	\vdots & & & \vdots\\
	-\frac{b\left(x_{k}-x_{0}\right)}{\phi} & -\frac{a\left(x_{k}-x_{0}\right)+b\left(x_{k}^{2}-x_{0}^{2}\right)}{\phi} & \frac{\left(a+b x_{k}\right)^{2}-\left(a+b x_{0}\right)^{2}}{2 \phi^{2}} & -(x_k-x_0)\\
\end{bmatrix}.$$
After some rank-preserving operations, we get: 
$$\begin{bmatrix}
	0 & x_1-x_0 & x_1-x_0 & 0\\
	\vdots & & & \vdots\\
	0 & x_k-x_0 & x_k-x_0 & 0\\
	x_{1}-x_{0} & -\left[a\left(x_{1}-x_{0}\right)+b\left(x_{1}^{2}-x_{0}^{2}\right)\right] & -\left[a\left(x_{1}-x_{0}\right)+\frac{b}{2}\left(x_{1}^{2}-x_{0}^{2}\right)\right] & x_1-x_0\\
	\vdots & & & \vdots\\
	x_{k}-x_{0} & -\left[a\left(x_{k}-x_{0}\right)+b\left(x_{k}^{2}-x_{0}^{2}\right)\right] & -\left[a\left(x_{k}-x_{0}\right)+\frac{b}{2}\left(x_{k}^{2}-x_{0}^{2}\right)\right] & x_k-x_0\\
\end{bmatrix}.$$
This concludes the initial claim. 

Note that in this example, \textbf{$p(X\mid Y)$ is in the exponential family} since: 
{\small 
	\begin{align*}
		p(x \mid y)=\frac{p(y \mid x) p(x)}{p(y)} & =\frac{N\left((a+b x), \sigma_{y}^{2}\right) \lambda e^{-\lambda x}}{p(y)} \\ & =\exp \left\{-\frac{1}{2}\left(\frac{y-(a+b x)}{\sigma_{y}}\right)^{2}+\log \frac{1}{\sqrt{2 \pi} \sigma_{y}}+\log \lambda-\lambda x-\log p(y)\right\}\\
		&=\exp \left\{-\frac{1}{2} \frac{\left(x, x^{2}\right)\left(2 a b-2 b y-2 \sigma_{y}^{2} \lambda, b^{2}\right)^{T}+(a-y)^{2}}{\sigma_{y}^{2}}+\log \frac{1}{\sqrt{2 \pi} \sigma_{y}}+\log \lambda-\log p(y)\right\}\\
		&=\exp \left\{\left(x, x^{2}\right)\left(-\frac{a b-b y-\sigma_{y}^{2} \lambda}{\sigma_{y}^{2}},-\frac{b^{2}}{2 \sigma_{y}^{2}}\right)^{T}-\frac{(a-y)^{2}}{2 \sigma_{y}^{2}}+\log \frac{1}{\sqrt{2 \pi} \sigma_{y}}+\log \lambda-\log p(y)\right\}. 
	\end{align*} 
}


\subsection{$X$ is exponential and $Y \mid X$ is exponential under canonical link}\label{app:parID-ex-exponential}

Suppose 
\begin{align*}
	X &\sim \operatorname{exponential} (\lambda_x) \\
	Y\mid X &\sim \operatorname{exponential} (\lambda)=\exp \{y(-\lambda)+\log \lambda\}=\exp \{y(a+bx)+\log [-(a+bx)]\}.
\end{align*}
The unknown parameter vector is $\theta=\left(a, b, \lambda_x\right)$. According to Theorem~\ref{thm:id-par} and without any further assumptions on $\theta$, $p(X,Y)$ is identifiable. 

In order to prove the above claim, we can construct the following equations: 
\begin{align*}
	&\phi_{i}(\theta)=b\left(x_{i}-x_{0}\right) \\ 
	&\zeta_{i}(\theta)=\log \left[-(a+b x_{i})\right]-\log \left[-(a+b x_{0})\right]-\lambda_{x}\left(x_{i}-x_{0}\right), \quad  i\in (1, \ldots,k)
\end{align*}
The Jacobian matrix is
$$J=\begin{bmatrix}
	0 & x_1-x_0 & 0\\
	\vdots & & \vdots\\
	0 & x_k-x_0 & 0\\
	\frac{1}{a+b x_{1}}-\frac{1}{a+b x_{0}} & \frac{x_{1}}{a+b x_{1}}-\frac{x_{0}}{a+b x_{0}} & -\left(x_{1}-x_{0}\right)\\
	\vdots & & \vdots\\
	\frac{1}{a+b x_{k}}-\frac{1}{a+b x_{0}} & \frac{x_{k}}{a+b x_{k}}-\frac{x_{0}}{a+b x_{0}} & -\left(x_{k}-x_{0}\right)
\end{bmatrix}.$$
After some rank-preserving operations, we get: 
$$\begin{bmatrix}
	0 & x_1-x_0 & 0\\
	0 & x_2-x_0 & 0\\
	\vdots & & \vdots\\
	0 & x_k-x_0 & 0\\
	\frac{1}{\left(a+b x_{1}\right)\left(a+b x_{0}\right)} & \frac{1}{\left(a+b x_{1}\right)\left(a+b x_{0}\right)} & 1\\
	\frac{1}{\left(a+b x_{2}\right)\left(a+b x_{0}\right)} & \frac{1}{\left(a+b x_{2}\right)\left(a+b x_{0}\right)} & 1\\
	\vdots & & \vdots\\
	\frac{1}{\left(a+b x_{k}\right)\left(a+b x_{0}\right)} & \frac{1}{\left(a+b x_{k}\right)\left(a+b x_{0}\right)} & 1
\end{bmatrix}.$$
This matrix is full rank and thus it concludes the initial claim. 

Note that in this example, \textbf{$p(X\mid Y)$ is not in exponential family (unless $a$ and $b$ are known)}, since: 
\begin{align*}
	p(x \mid y)=\frac{p(y \mid x) p(x)}{p(y)}=\frac{\exp \left\{y(a+b x)+\log [-(a+b x)]+x(-\lambda x)+\log \lambda_{x}\right\}}{p(y)}. 
\end{align*}
The main difficulty is with the term $\log [-(a+b x)]$. 

\newpage
\section{Estimation Proofs}\label{app:est-proofs} 

\subsection{Theorem~\ref{thm:est-order} \quad {\small (Conditional likelihood with order statistics)}}\label{app:sub-est-order}

\begin{proof}
	Denote $l(\theta) = -\frac{2}{N(N-1)} \sum_{1\leq i<k\leq N} R_{x_i}R_{y_i}R_{x_k}R_{y_k} \log \{1+Q_{ik}(\theta)\}$. Following the Taylor expansion, we have
	\[
	0 = \frac{\partial l(\widetilde\theta)}{\partial\theta} = \frac{\partial l(\theta_0)}{\partial\theta} + (\widetilde\theta-\theta_0)\frac{\partial^2 l(\theta_0)}{\partial\theta^2} + o_p(N^{-1/2}).
	\]
	Therefore, 
	\[
	\sqrt{N}(\widetilde\theta-\theta_0) = - \left\{ \frac{\partial^2 l(\theta_0)}{\partial\theta^2} \right\}^{-1} \sqrt{N}\frac{\partial l(\theta_0)}{\partial\theta} + o_p(1).
	\]
	Since both $\frac{\partial^2 l(\theta_0)}{\partial\theta^2}$ and $\frac{\partial l(\theta_0)}{\partial\theta}$ are U-statistics, from the theory of U-statistics, we have
	\[
	\frac{\partial^2 l(\theta_0)}{\partial\theta^2} \xrightarrow{p} A, \mbox{ and } \sqrt{N}\frac{\partial l(\theta_0)}{\partial\theta} \xrightarrow{d} \N(0, B),
	\]
	which completes the proof.
\end{proof}


\subsection{Theorem~\ref{thm:est-gmm} \quad {\small (Generalized estimating equations)}}\label{app:sub-est-gmm}

\begin{proof}
	The proof of (a) is straightforward following the standard argument of generalized estimating equations, so omitted here. In order to find the optimal choice for $f(Y)$, we can compute
	$$\begin{aligned} C & =\E\left\{-\Psi^{\prime}\left(X, Y, R_{x}, R_{y} ; \theta_{0}\right)\right\} \\ & =\E\left[\frac{R_{x} R_{y}}{p\left(R_{y}=1 \mid R_{x}=1, X\right)}  \frac{\partial \E(X \mid Y)}{\partial \theta}\bigg\rvert_{\theta=\theta_0}f(Y)^{T}\right] \\  & =\E\left[R_{x}  \frac{\partial \E(X \mid Y)}{\partial \theta}\bigg\rvert_{\theta=\theta_0}f(Y)^{T}\right]\\
		& =\E\left\{w(Y) a(Y) f(Y)^{T}\right\},\end{aligned}$$
	and
	\begin{align*} D 
		& =\E\left[\frac{R_{x} R_{y}}{p^{2}\left(R_{y}=1 \mid R_{x}=1, X\right)}(X-\E(X \mid Y))^{2}f(Y)f(Y)^{T}\right] \\ 
		& =\E\left[R_{x} \frac{(X-\E(X \mid Y))^{2}}{\pi(X)}f(Y)f(Y)^{T}\right]\\
		&=\E\left[w(Y) \frac{(X-\E(X \mid Y))^{2}}{\pi(X)}f(Y)f(Y)^{T}\right]\\
		&=\E\left[w(Y) E\left[\frac{(X-\E(X \mid Y))^{2}}{\pi(X)}\mid Y\right]f(Y)f(Y)^{T}\right]\\
		&=\E\left[w(Y)b(Y)f(Y)f(Y)^{T}\right],
	\end{align*}

	where $b(Y)=\E\left[\frac{(X-\E(X \mid Y))^{2}}{\pi(X)}\mid Y\right]$ and  $w(Y)=p(R_x=1 \mid Y)$.
	Based on Cauchy-Schwarz inequality, we have
	$$\E\left(u v^{T}\right)\left\{\E\left(v v^{T}\right)\right\}^{-1} \E\left(v u^{T}\right) \lesssim \E\left(u u^{T}\right)$$ with equality hold at $u=v$.
	Here $M \lesssim N$ simply means $M-N$ is negative semi-definite.
	
	\noindent Define $v=\sqrt{w(Y)} \sqrt{b(Y)} f(Y)$ and $ u=\sqrt{\frac{w(Y)}{b(Y)}} a(Y)$, then we have
	\begin{align*}
		&\E\{w(Y) f(Y) a(Y)^T\}\left[\E\{w(Y) b(Y) f(Y) f(Y)^T\}\right]^{-1} \E\{w(Y) a(Y) f(Y)^T\}\lesssim \E\left\{\frac{w(Y)}{b(Y)} a(Y) a(Y)^T\right\}, i.e.,\\
		&\E\left\{\frac{w(Y)}{b(Y)} a(Y) a(Y)^T\right\}^{-1}\E\{w(Y) b(Y) f(Y) f(Y)^T\}\E\{w(Y) f(Y) a(Y)^T\}^{-1}\gtrsim \E\left\{\frac{w(Y)}{b(Y)} a(Y) a(Y)^T\right\}^{-1}.
	\end{align*}
	Note that the right-hand side is irrespective of $f(Y)$. Thus, when $f(Y)=f_{opt}(Y)=\frac{a(Y)}{b(Y)}$, 
	the equality holds, and we have the optimal variance $\left\{\frac{w(Y)}{b(Y)} a(Y) a(Y)^T\right\}^{-1}$.
\end{proof}

\newpage
\section{Additional discussions on estimation} 
\label{app:est-additional}

\subsection{Nonparametric estimation under additional assumptions}
\label{app:sub-est-permutation}

In addition to independence restrictions in display~(\ref{eq:criss_cross_assump}), we assume $p(R_y = 1 \mid R_x, X)$ is not a function of $X$ when $R_x = 0$. This additional assumptions moves us from the criss-cross MNAR model to the permutation model considered by \cite{robins97non-a}. In the permutation model, one can proceed with estimation of arbitrary functions of $X$ and $Y$ as follows. 

Let our parameter of interest be $\beta_h = \E[h(X, Y)]$, which can be identified via the following function of the observed data: 
$$\begin{aligned}
	\beta_h = \E\left[ \frac{R_x \ R_y \ h(X, Y)}{p(R_x = 1\mid Y) \ p(R_y = 1 \mid R_x = 1, X^*)} \right].  
	\label{eq:mid2}
\end{aligned}$$

\noindent The core idea of deriving the efficient influence function (EIF) for $\beta_h$ is to use an intermediate variable that first takes care of the missingness of $X$, and then $Y$ in a sequential manner. Intuitively, this is due to the fact that we can rewrite $\beta_h$ via an intermediate variable $\widetilde{\beta}_h(X, R_x, Y)$ as follows:
\begin{align*}
	\widetilde{\beta}_h(X, R_x, Y) &= \frac{R_x}{p\left(R_x=1 \mid Y \right)} \ h\left(X, Y \right), \qquad 
	\beta_h =  \E\left[\frac{R_y  }{p\left(R_y=1 \mid R_x, X^*\right)} \ \widetilde{\beta}_h(X, R_x, Y) \right]. 
\end{align*}
The claim made by \cite{robins97non-a} is that EIF for $\beta_h$ is equal to the EIF for $ \E\left[\displaystyle \frac{R_y  }{p\left(R_y=1 \mid R_x, X^*\right)} \ \phi(\widetilde{\beta}_h) \right]$, where $\phi(\widetilde{\beta}_h) = \text{EIF}_{\widetilde{\beta}_h} + \E[\widetilde{\beta}_h]$ and  $\text{EIF}_{\widetilde{\beta}_h}$ denotes the efficient influence function for $\E\big[\widetilde{\beta}_h(X, R_x, Y)\big]$. 
Therefore, we first need to derive the EIF for $\E\big[\widetilde{\beta}_h(X, R_x, Y)\big]$. 
$$
\scriptsize 
	\begin{aligned}
		\left.\frac{\partial \E[\widetilde{\beta}_h\left(p_{\varepsilon}\right)]}{\partial \varepsilon}\right|_{\varepsilon=0}&= \left.\frac{\partial}{\partial \varepsilon} \int \frac{R_x h\left(X, Y\right)}{p\left(R_x=1 \mid Y\right)} d p_{\varepsilon}\left(X, Y, R_x\right)\right|_{\varepsilon=0} \\
		&= -\int \frac{R_x h\left(X, Y\right)}{p\left(R_x=1 \mid Y\right)} S\left(R_x \mid Y\right) d p\left(X, Y, R_x\right) 
		+\int \frac{R_x h\left(X, Y\right)}{p\left(R_x=1 \mid Y\right)} S\left(X, Y, R_x\right) d p\left(X, Y, R_x\right) \\
		&= -\int \frac{R_x \E\left[h\left(X, Y\right) \mid R_x=1, Y\right]}{p\left(R_x=1 \mid Y\right)} S\left(R_x \mid Y\right) d p\left(R_x, Y\right)
		+\int\left\{\frac{R_x h\left(X, Y\right)}{p\left(R_x=1 \mid Y\right)}-\E\left[h\left(X, Y\right)\right]\right\} S\left(X, Y, R_x\right) d p\left(X, Y, R_x\right)\\
		& =-\int\left\{\frac{R_x \E\left[h\left(X, Y\right) \mid R_x=1, Y\right]}{p\left(R_x=1 \mid Y\right)}-\E\left[h\left(X, Y\right) \mid R_x=1, Y\right]\right\} S\left(R_x, Y\right) d p\left(R_x, Y\right) \\
		& +\int\left\{\frac{R_x h\left(X, Y\right)}{p\left(R_x=1 \mid Y\right)}-\E\left[h\left(X, Y\right)\right]\right\} S\left(X, Y, R_x\right) d p\left(X, Y, R_x\right)\\
		& =-\int\left\{\frac{R_x \E\left[h\left(X, Y\right) \mid R_x=1, X\right]}{p\left(R_x=1 \mid Y\right)}-\E\left[h\left(X, Y\right) \mid R_x=1, Y\right]\right\} S\left(Y, R_x, X\right) d p\left(R_x, X, Y\right) \\
		&+\int \left\{\frac{R_x h\left(X, Y\right)}{p\left(R_x=1 \mid Y\right)}-\E\left[h\left(X, Y\right)\right]\right\} S\left(X, Y, R_x\right) d p\left(X, Y, R_x\right). 
\end{aligned}
$$

\noindent Therefore, the efficient influence function for $\E[\widetilde{\beta}_h]$, denoted by $\text{EIF}_{\widetilde{\beta}_h}$, is 
as follows
\begin{align*}
	\text{EIF}_{\widetilde{\beta}_h} &= \frac{R_x }{p\left(R_x=1 \mid Y\right)} \Big\{h\left(X, Y\right) - \E\left[h\left(X, Y\right) \mid R_x = 1, Y \right] \Big\} + \Big\{ \E[h(X, Y) \mid R_x = 1, Y] - \E[h(X, Y)] \Big\}. 
\end{align*}
Thus we get: 
\begin{align*}
	\phi(\widetilde{\beta}_h) = \frac{R_x }{p\left(R_x=1 \mid Y\right)} \Big\{h\left(X, Y\right) - \E\left[h\left(X, Y\right) \mid R_x = 1, Y \right] \Big\} + \E\big[ h(X, Y) \mid R_x = 1, Y \big]. 
\end{align*}

Following a similar procedure, we can easily obtain the EIF for $ \E\left[\displaystyle \frac{R_y  }{p\left(R_y=1 \mid R_x, X^*\right)} \ \phi(\widetilde{\beta}_h) \right]$, which yields the EIF for $\beta_h$ as follows: 
\begin{align*}
	\text{EIF}_{\beta_h} = 
	\frac{R_y}{p\left(R_y=1 \mid R_x, X^*\right)}\Big\{ \phi(\widetilde{\beta}_h) \ - \ \E\big[ \phi(\widetilde{\beta}_h) \mid R_y, R_x, X^* \big]  \Big\} 
	+ \Big\{  \  \E\big[ \phi(\widetilde{\beta}_h) \mid R_y = 1, R_x, X^*\big] - \beta_h \Big\}. 
\end{align*}

\vspace{0.5cm}
\subsection{Maximum likelihood estimation} 

In the criss-cross MNAR model, the \textit{observed full data likelihood}, denoted by $\mathcal{L}_\text{obs}(Z; \theta)$, can be written down as follows: 
\begin{align*}
	\mathcal{L}_\text{obs}(X, Y, R; \theta, \psi) 
	&= \prod_{R_x=1, R_y=1} p(X, Y, R_x=1, R_y=1) \times \prod_{R_x=1, R_y=0}  \int p(X, Y, R_x=1, R_y=0) dy \\
	&\ \times \prod_{R_x=0, R_y=1} \int p(X, Y, R_x=0, R_y=1) dx \times \prod_{R_x=0, R_y=0}  \int p(X, Y, R_x=0, R_y=0) dxdy \\
\end{align*}

Under the conditions of Theorem~\ref{thm:id-par} and Condition~\ref{cond:completeness}, one can simply estimate the entire parameter vector of the full law, assuming the parametric forms of the propensity scores in the missingness mechanism are known. 

\newpage
\section{Additional experimental results}\label{app:sims}

\subsection{Simulation results}
\noindent\underline{\textbf{Varying $\rho$.}} We examine the effect of changing the correlation coefficient on the efficiency of the estimators by varying $\rho$ across the range of values from -0.9 to 0.9, with increments of $0.2$. The sample size used is $N=1000$. Table~\ref{table:2} displays the standard deviation (SD) of the three suggested estimators for different values of $\rho$. To avoid distorting the SD patterns after applying the Delta method, we summarize the SD of the direct estimates of each method instead of converting it to OR. The results indicate that both GEE methods provide more efficient estimators when $X$ and $Y$ are highly correlated, but exhibit more estimation uncertainty when the correlation is low. In contrast, the conditional likelihood estimator has less variability when the correlation is low.

\input{table2.tex}

\begin{figure*}[t]
	\centering
	\includegraphics[height=5.3cm, width=16cm]{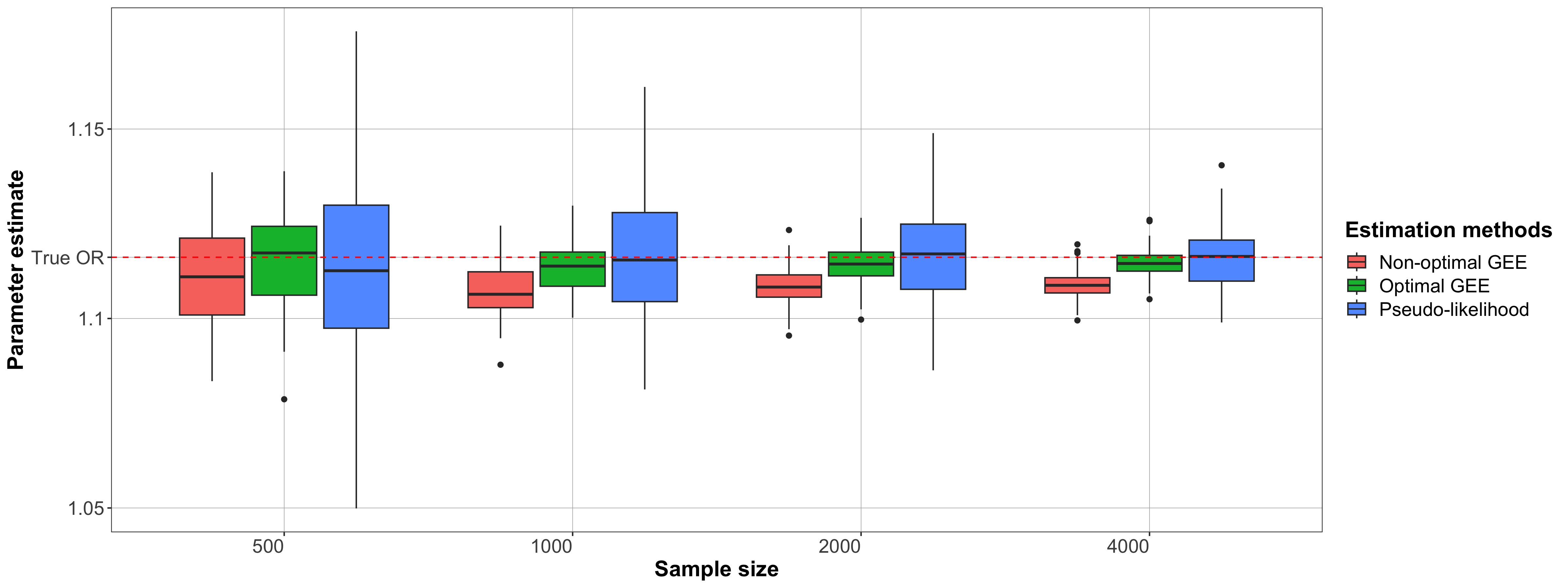}
	\caption{OR estimation under model misspecification.}
	\label{fig:mis}
\end{figure*}

\noindent\underline{\textbf{Model misspecification.}} To understand the behavior of the proposed estimators under model misspecification, we generate data under missing mechanism for $Y$ as $p(R_y=1 \mid X,R_x)= \operatorname{expit}(2-R_x+0.7X+0.2X^2)$. While estimation with GEE is carried out, the relations between $R_y$ and $\{X,R_x\}$ is assumed to be linear. Under model misspecification, Figure~\ref{fig:mis} illustrates that both GEE methods fail to provide an unbiased estimate of the OR despite an increasing sample size. The conditional likelihood still yields unbiased estimates especially with large sample size. Same observation is made in the estimation of $\alpha$ and $\beta$ as shown in Table~\ref{table:3}. Bias and high MSE persist for both methods even with large sample size whereas SD shrinks as sample size increases.

\input{table3.tex}

The simulation results indicate all three methods yield unbiased estimators when the model is correctly specified. GEE methods are more efficient than the conditional likelihood. As expected,  the optimal GEE is consistently more efficient than the non-optimal GEE regardless of the sample size. On the other hand, for OR estimation, the conditional likelihood method is more robust under model misspecification meaning that it yields unbiased estimators even when $p(R_y\mid X,R_x)$ is misspecified. In the presence of a strong correlation between $X$ and $Y$, the GEE estimators exhibit higher efficiency. Conversely, under conditions of weak correlation, the conditional likelihood estimator displays higher efficiency.

\subsection{Real data results}

We also applied our proposed methods to analyze data from the KLIPS dataset, which includes information on monthly income for 2511 regular wage earners in 2005 and 2006. The combined monthly income for these two years has approximately 40\% missing data. Our objective was to investigate whether past income has a lasting effect on future income. We defined $X$ as the logarithm of monthly income in 2005 and $Y$ as the logarithm of monthly income in 2006. Based on empirical data distributions, we assumed that $X$, $Y$, and $X|Y$ are normally distributed. Specifically, we modeled $X|Y$ as $\N(\alpha + \beta Y, \sigma^2)$, where $\sigma^2$ was empirically estimated.

Using our nonparametric identification results, we were able to determine $\alpha$ and $\beta$ without making any additional assumptions. For estimating these parameters, we employed generalized estimating equations (GEEs). Additionally, we used all three methods to estimate $\log(OR)$, where OR represents the odds ratio between the income of the two years. The parameter estimates obtained are summarized in Table~\ref{table:continuous}.

\input{table_realdata_continuous}

The findings presented above indicate a significant and persistent effect of income. Specifically, high income in the past is strongly predictive of high income in the future, and conversely, low income in the past is predictive of low income in the future. These results provide confirmation that the optimal GEE approach outperforms the non-optimal GEE, particularly in terms of higher efficiency when dealing with continuous variable distributions.

\end{document}

%% file: table1.tex
  \providecommand{\huxb}[2]{\arrayrulecolor[RGB]{#1}\global\arrayrulewidth=#2pt}
  \providecommand{\huxvb}[2]{\color[RGB]{#1}\vrule width #2pt}
  \providecommand{\huxtpad}[1]{\rule{0pt}{#1}}
  \providecommand{\huxbpad}[1]{\rule[-#1]{0pt}{#1}}

\begin{table}[t]
\begin{centerbox}
\begin{threeparttable}
\captionsetup{justification=centering,singlelinecheck=off}
\caption{Parameter estimates with varying sample size.}
\label{table:1}
 \setlength{\tabcolsep}{0pt}
\begin{tabular}{l l l l l l l}

\hhline{>{\huxb{0, 0, 0}{1}}=>{\huxb{0, 0, 0}{1}}=>{\huxb{0, 0, 0}{1}}=>{\huxb{0, 0, 0}{1}}=>{\huxb{0, 0, 0}{1}}=>{\huxb{0, 0, 0}{1}}=>{\huxb{0, 0, 0}{1}}=}
\arrayrulecolor{black}

\multicolumn{1}{!{\huxvb{0, 0, 0}{0}}l!{\huxvb{0, 0, 0}{0}}}{\huxtpad{1pt + 1em}\raggedright \hspace{1pt} \textbf{{\fontsize{9pt}{10.8pt}\selectfont }} \hspace{1pt}\huxbpad{1pt}} &
\multicolumn{1}{c!{\huxvb{0, 0, 0}{0}}}{\huxtpad{1pt + 1em}\centering \hspace{1pt} \textbf{{\fontsize{9pt}{10.8pt}\selectfont }} \hspace{1pt}\huxbpad{1pt}} &
\multicolumn{2}{c!{\huxvb{0, 0, 0}{0}}}{\huxtpad{1pt + 1em}\centering \hspace{1pt} \textbf{{\fontsize{9pt}{10.8pt}\selectfont Non-optimal GEE}} \hspace{1pt}\huxbpad{1pt}} &
\multicolumn{1}{c!{\huxvb{0, 0, 0}{0}}}{\huxtpad{1pt + 1em}\centering \hspace{1pt} \textbf{{\fontsize{9pt}{10.8pt}\selectfont }} \hspace{1pt}\huxbpad{1pt}} &
\multicolumn{2}{c!{\huxvb{0, 0, 0}{0}}}{\huxtpad{1pt + 1em}\centering \hspace{1pt} \textbf{{\fontsize{9pt}{10.8pt}\selectfont Optimal GEE}} \hspace{1pt}\huxbpad{1pt}} \tabularnewline[-0.5pt]

\hhline{>{\huxb{255, 255, 255}{0.4}}->{\huxb{255, 255, 255}{0.4}}->{\huxb{0, 0, 0}{0.4}}->{\huxb{0, 0, 0}{0.4}}->{\huxb{255, 255, 255}{0.4}}->{\huxb{0, 0, 0}{0.4}}->{\huxb{0, 0, 0}{0.4}}-}
\arrayrulecolor{black}

\multicolumn{1}{!{\huxvb{0, 0, 0}{0}}l!{\huxvb{0, 0, 0}{0}}}{\huxtpad{1pt + 1em}\raggedright \hspace{1pt} {\fontsize{9pt}{10.8pt}\selectfont N} \hspace{1pt}\huxbpad{1pt}} &
\multicolumn{1}{c!{\huxvb{0, 0, 0}{0}}}{\huxtpad{1pt + 1em}\centering \hspace{1pt} {\fontsize{9pt}{10.8pt}\selectfont Statistics} \hspace{1pt}\huxbpad{1pt}} &
\multicolumn{1}{c!{\huxvb{0, 0, 0}{0}}}{\huxtpad{1pt + 1em}\centering \hspace{1pt} {\fontsize{9pt}{10.8pt}\selectfont \(\alpha\)} \hspace{1pt}\huxbpad{1pt}} &
\multicolumn{1}{c!{\huxvb{0, 0, 0}{0}}}{\huxtpad{1pt + 1em}\centering \hspace{1pt} {\fontsize{9pt}{10.8pt}\selectfont \(\beta\)} \hspace{1pt}\huxbpad{1pt}} &
\multicolumn{1}{c!{\huxvb{0, 0, 0}{0}}}{\huxtpad{1pt + 1em}\centering \hspace{1pt} {\fontsize{9pt}{10.8pt}\selectfont } \hspace{1pt}\huxbpad{1pt}} &
\multicolumn{1}{c!{\huxvb{0, 0, 0}{0}}}{\huxtpad{1pt + 1em}\centering \hspace{1pt} {\fontsize{9pt}{10.8pt}\selectfont \(\alpha\)} \hspace{1pt}\huxbpad{1pt}} &
\multicolumn{1}{c!{\huxvb{0, 0, 0}{0}}}{\huxtpad{1pt + 1em}\centering \hspace{1pt} {\fontsize{9pt}{10.8pt}\selectfont \(\beta\)} \hspace{1pt}\huxbpad{1pt}} \tabularnewline[-0.5pt]

\hhline{>{\huxb{0, 0, 0}{0.4}}->{\huxb{0, 0, 0}{0.4}}->{\huxb{0, 0, 0}{0.4}}->{\huxb{0, 0, 0}{0.4}}->{\huxb{0, 0, 0}{0.4}}->{\huxb{0, 0, 0}{0.4}}->{\huxb{0, 0, 0}{0.4}}-}
\arrayrulecolor{black}

\multicolumn{1}{!{\huxvb{0, 0, 0}{0}}l!{\huxvb{0, 0, 0}{0}}}{\huxtpad{1pt + 1em}\raggedright \hspace{1pt} {\fontsize{9pt}{10.8pt}\selectfont 500} \hspace{1pt}\huxbpad{1pt}} &
\multicolumn{1}{c!{\huxvb{0, 0, 0}{0}}}{\huxtpad{1pt + 1em}\centering \hspace{1pt} {\fontsize{9pt}{10.8pt}\selectfont bias} \hspace{1pt}\huxbpad{1pt}} &
\multicolumn{1}{c!{\huxvb{0, 0, 0}{0}}}{\huxtpad{1pt + 1em}\centering \hspace{1pt} {\fontsize{9pt}{10.8pt}\selectfont 0.1411} \hspace{1pt}\huxbpad{1pt}} &
\multicolumn{1}{c!{\huxvb{0, 0, 0}{0}}}{\huxtpad{1pt + 1em}\centering \hspace{1pt} {\fontsize{9pt}{10.8pt}\selectfont -0.0335} \hspace{1pt}\huxbpad{1pt}} &
\multicolumn{1}{c!{\huxvb{0, 0, 0}{0}}}{\huxtpad{1pt + 1em}\centering \hspace{1pt} {\fontsize{9pt}{10.8pt}\selectfont } \hspace{1pt}\huxbpad{1pt}} &
\multicolumn{1}{c!{\huxvb{0, 0, 0}{0}}}{\huxtpad{1pt + 1em}\centering \hspace{1pt} {\fontsize{9pt}{10.8pt}\selectfont 0.0613} \hspace{1pt}\huxbpad{1pt}} &
\multicolumn{1}{c!{\huxvb{0, 0, 0}{0}}}{\huxtpad{1pt + 1em}\centering \hspace{1pt} {\fontsize{9pt}{10.8pt}\selectfont -0.0028} \hspace{1pt}\huxbpad{1pt}} \tabularnewline[-0.5pt]

\hhline{}
\arrayrulecolor{black}

\multicolumn{1}{!{\huxvb{0, 0, 0}{0}}l!{\huxvb{0, 0, 0}{0}}}{\huxtpad{1pt + 1em}\raggedright \hspace{1pt} {\fontsize{9pt}{10.8pt}\selectfont } \hspace{1pt}\huxbpad{1pt}} &
\multicolumn{1}{c!{\huxvb{0, 0, 0}{0}}}{\huxtpad{1pt + 1em}\centering \hspace{1pt} {\fontsize{9pt}{10.8pt}\selectfont MSE} \hspace{1pt}\huxbpad{1pt}} &
\multicolumn{1}{c!{\huxvb{0, 0, 0}{0}}}{\huxtpad{1pt + 1em}\centering \hspace{1pt} {\fontsize{9pt}{10.8pt}\selectfont 0.0199} \hspace{1pt}\huxbpad{1pt}} &
\multicolumn{1}{c!{\huxvb{0, 0, 0}{0}}}{\huxtpad{1pt + 1em}\centering \hspace{1pt} {\fontsize{9pt}{10.8pt}\selectfont 0.0011} \hspace{1pt}\huxbpad{1pt}} &
\multicolumn{1}{c!{\huxvb{0, 0, 0}{0}}}{\huxtpad{1pt + 1em}\centering \hspace{1pt} {\fontsize{9pt}{10.8pt}\selectfont } \hspace{1pt}\huxbpad{1pt}} &
\multicolumn{1}{c!{\huxvb{0, 0, 0}{0}}}{\huxtpad{1pt + 1em}\centering \hspace{1pt} {\fontsize{9pt}{10.8pt}\selectfont 0.0038} \hspace{1pt}\huxbpad{1pt}} &
\multicolumn{1}{c!{\huxvb{0, 0, 0}{0}}}{\huxtpad{1pt + 1em}\centering \hspace{1pt} {\fontsize{9pt}{10.8pt}\selectfont 0.0000} \hspace{1pt}\huxbpad{1pt}} \tabularnewline[-0.5pt]

\hhline{}
\arrayrulecolor{black}

\multicolumn{1}{!{\huxvb{0, 0, 0}{0}}l!{\huxvb{0, 0, 0}{0}}}{\huxtpad{1pt + 1em}\raggedright \hspace{1pt} {\fontsize{9pt}{10.8pt}\selectfont } \hspace{1pt}\huxbpad{1pt}} &
\multicolumn{1}{c!{\huxvb{0, 0, 0}{0}}}{\huxtpad{1pt + 1em}\centering \hspace{1pt} {\fontsize{9pt}{10.8pt}\selectfont SD} \hspace{1pt}\huxbpad{1pt}} &
\multicolumn{1}{c!{\huxvb{0, 0, 0}{0}}}{\huxtpad{1pt + 1em}\centering \hspace{1pt} {\fontsize{9pt}{10.8pt}\selectfont 0.7980} \hspace{1pt}\huxbpad{1pt}} &
\multicolumn{1}{c!{\huxvb{0, 0, 0}{0}}}{\huxtpad{1pt + 1em}\centering \hspace{1pt} {\fontsize{9pt}{10.8pt}\selectfont 0.3346} \hspace{1pt}\huxbpad{1pt}} &
\multicolumn{1}{c!{\huxvb{0, 0, 0}{0}}}{\huxtpad{1pt + 1em}\centering \hspace{1pt} {\fontsize{9pt}{10.8pt}\selectfont } \hspace{1pt}\huxbpad{1pt}} &
\multicolumn{1}{c!{\huxvb{0, 0, 0}{0}}}{\huxtpad{1pt + 1em}\centering \hspace{1pt} {\fontsize{9pt}{10.8pt}\selectfont 0.7830} \hspace{1pt}\huxbpad{1pt}} &
\multicolumn{1}{c!{\huxvb{0, 0, 0}{0}}}{\huxtpad{1pt + 1em}\centering \hspace{1pt} {\fontsize{9pt}{10.8pt}\selectfont 0.3037} \hspace{1pt}\huxbpad{1pt}} \tabularnewline[-0.5pt]

\hhline{}
\arrayrulecolor{black}

\multicolumn{1}{!{\huxvb{0, 0, 0}{0}}l!{\huxvb{0, 0, 0}{0}}}{\huxtpad{1pt + 1em}\raggedright \hspace{1pt} {\fontsize{9pt}{10.8pt}\selectfont 1000} \hspace{1pt}\huxbpad{1pt}} &
\multicolumn{1}{c!{\huxvb{0, 0, 0}{0}}}{\huxtpad{1pt + 1em}\centering \hspace{1pt} {\fontsize{9pt}{10.8pt}\selectfont bias} \hspace{1pt}\huxbpad{1pt}} &
\multicolumn{1}{c!{\huxvb{0, 0, 0}{0}}}{\huxtpad{1pt + 1em}\centering \hspace{1pt} {\fontsize{9pt}{10.8pt}\selectfont 0.1079} \hspace{1pt}\huxbpad{1pt}} &
\multicolumn{1}{c!{\huxvb{0, 0, 0}{0}}}{\huxtpad{1pt + 1em}\centering \hspace{1pt} {\fontsize{9pt}{10.8pt}\selectfont -0.0281} \hspace{1pt}\huxbpad{1pt}} &
\multicolumn{1}{c!{\huxvb{0, 0, 0}{0}}}{\huxtpad{1pt + 1em}\centering \hspace{1pt} {\fontsize{9pt}{10.8pt}\selectfont } \hspace{1pt}\huxbpad{1pt}} &
\multicolumn{1}{c!{\huxvb{0, 0, 0}{0}}}{\huxtpad{1pt + 1em}\centering \hspace{1pt} {\fontsize{9pt}{10.8pt}\selectfont 0.1010} \hspace{1pt}\huxbpad{1pt}} &
\multicolumn{1}{c!{\huxvb{0, 0, 0}{0}}}{\huxtpad{1pt + 1em}\centering \hspace{1pt} {\fontsize{9pt}{10.8pt}\selectfont -0.0248} \hspace{1pt}\huxbpad{1pt}} \tabularnewline[-0.5pt]

\hhline{}
\arrayrulecolor{black}

\multicolumn{1}{!{\huxvb{0, 0, 0}{0}}l!{\huxvb{0, 0, 0}{0}}}{\huxtpad{1pt + 1em}\raggedright \hspace{1pt} {\fontsize{9pt}{10.8pt}\selectfont } \hspace{1pt}\huxbpad{1pt}} &
\multicolumn{1}{c!{\huxvb{0, 0, 0}{0}}}{\huxtpad{1pt + 1em}\centering \hspace{1pt} {\fontsize{9pt}{10.8pt}\selectfont MSE} \hspace{1pt}\huxbpad{1pt}} &
\multicolumn{1}{c!{\huxvb{0, 0, 0}{0}}}{\huxtpad{1pt + 1em}\centering \hspace{1pt} {\fontsize{9pt}{10.8pt}\selectfont 0.0116} \hspace{1pt}\huxbpad{1pt}} &
\multicolumn{1}{c!{\huxvb{0, 0, 0}{0}}}{\huxtpad{1pt + 1em}\centering \hspace{1pt} {\fontsize{9pt}{10.8pt}\selectfont 0.0008} \hspace{1pt}\huxbpad{1pt}} &
\multicolumn{1}{c!{\huxvb{0, 0, 0}{0}}}{\huxtpad{1pt + 1em}\centering \hspace{1pt} {\fontsize{9pt}{10.8pt}\selectfont } \hspace{1pt}\huxbpad{1pt}} &
\multicolumn{1}{c!{\huxvb{0, 0, 0}{0}}}{\huxtpad{1pt + 1em}\centering \hspace{1pt} {\fontsize{9pt}{10.8pt}\selectfont 0.0102} \hspace{1pt}\huxbpad{1pt}} &
\multicolumn{1}{c!{\huxvb{0, 0, 0}{0}}}{\huxtpad{1pt + 1em}\centering \hspace{1pt} {\fontsize{9pt}{10.8pt}\selectfont 0.0006} \hspace{1pt}\huxbpad{1pt}} \tabularnewline[-0.5pt]

\hhline{}
\arrayrulecolor{black}

\multicolumn{1}{!{\huxvb{0, 0, 0}{0}}l!{\huxvb{0, 0, 0}{0}}}{\huxtpad{1pt + 1em}\raggedright \hspace{1pt} {\fontsize{9pt}{10.8pt}\selectfont } \hspace{1pt}\huxbpad{1pt}} &
\multicolumn{1}{c!{\huxvb{0, 0, 0}{0}}}{\huxtpad{1pt + 1em}\centering \hspace{1pt} {\fontsize{9pt}{10.8pt}\selectfont SD} \hspace{1pt}\huxbpad{1pt}} &
\multicolumn{1}{c!{\huxvb{0, 0, 0}{0}}}{\huxtpad{1pt + 1em}\centering \hspace{1pt} {\fontsize{9pt}{10.8pt}\selectfont 0.6586} \hspace{1pt}\huxbpad{1pt}} &
\multicolumn{1}{c!{\huxvb{0, 0, 0}{0}}}{\huxtpad{1pt + 1em}\centering \hspace{1pt} {\fontsize{9pt}{10.8pt}\selectfont 0.2601} \hspace{1pt}\huxbpad{1pt}} &
\multicolumn{1}{c!{\huxvb{0, 0, 0}{0}}}{\huxtpad{1pt + 1em}\centering \hspace{1pt} {\fontsize{9pt}{10.8pt}\selectfont } \hspace{1pt}\huxbpad{1pt}} &
\multicolumn{1}{c!{\huxvb{0, 0, 0}{0}}}{\huxtpad{1pt + 1em}\centering \hspace{1pt} {\fontsize{9pt}{10.8pt}\selectfont 0.6142} \hspace{1pt}\huxbpad{1pt}} &
\multicolumn{1}{c!{\huxvb{0, 0, 0}{0}}}{\huxtpad{1pt + 1em}\centering \hspace{1pt} {\fontsize{9pt}{10.8pt}\selectfont 0.2467} \hspace{1pt}\huxbpad{1pt}} \tabularnewline[-0.5pt]

\hhline{}
\arrayrulecolor{black}

\multicolumn{1}{!{\huxvb{0, 0, 0}{0}}l!{\huxvb{0, 0, 0}{0}}}{\huxtpad{1pt + 1em}\raggedright \hspace{1pt} {\fontsize{9pt}{10.8pt}\selectfont 2000} \hspace{1pt}\huxbpad{1pt}} &
\multicolumn{1}{c!{\huxvb{0, 0, 0}{0}}}{\huxtpad{1pt + 1em}\centering \hspace{1pt} {\fontsize{9pt}{10.8pt}\selectfont bias} \hspace{1pt}\huxbpad{1pt}} &
\multicolumn{1}{c!{\huxvb{0, 0, 0}{0}}}{\huxtpad{1pt + 1em}\centering \hspace{1pt} {\fontsize{9pt}{10.8pt}\selectfont -0.0820} \hspace{1pt}\huxbpad{1pt}} &
\multicolumn{1}{c!{\huxvb{0, 0, 0}{0}}}{\huxtpad{1pt + 1em}\centering \hspace{1pt} {\fontsize{9pt}{10.8pt}\selectfont 0.0348} \hspace{1pt}\huxbpad{1pt}} &
\multicolumn{1}{c!{\huxvb{0, 0, 0}{0}}}{\huxtpad{1pt + 1em}\centering \hspace{1pt} {\fontsize{9pt}{10.8pt}\selectfont } \hspace{1pt}\huxbpad{1pt}} &
\multicolumn{1}{c!{\huxvb{0, 0, 0}{0}}}{\huxtpad{1pt + 1em}\centering \hspace{1pt} {\fontsize{9pt}{10.8pt}\selectfont -0.0332} \hspace{1pt}\huxbpad{1pt}} &
\multicolumn{1}{c!{\huxvb{0, 0, 0}{0}}}{\huxtpad{1pt + 1em}\centering \hspace{1pt} {\fontsize{9pt}{10.8pt}\selectfont 0.0140} \hspace{1pt}\huxbpad{1pt}} \tabularnewline[-0.5pt]

\hhline{}
\arrayrulecolor{black}

\multicolumn{1}{!{\huxvb{0, 0, 0}{0}}l!{\huxvb{0, 0, 0}{0}}}{\huxtpad{1pt + 1em}\raggedright \hspace{1pt} {\fontsize{9pt}{10.8pt}\selectfont } \hspace{1pt}\huxbpad{1pt}} &
\multicolumn{1}{c!{\huxvb{0, 0, 0}{0}}}{\huxtpad{1pt + 1em}\centering \hspace{1pt} {\fontsize{9pt}{10.8pt}\selectfont MSE} \hspace{1pt}\huxbpad{1pt}} &
\multicolumn{1}{c!{\huxvb{0, 0, 0}{0}}}{\huxtpad{1pt + 1em}\centering \hspace{1pt} {\fontsize{9pt}{10.8pt}\selectfont 0.0067} \hspace{1pt}\huxbpad{1pt}} &
\multicolumn{1}{c!{\huxvb{0, 0, 0}{0}}}{\huxtpad{1pt + 1em}\centering \hspace{1pt} {\fontsize{9pt}{10.8pt}\selectfont 0.0012} \hspace{1pt}\huxbpad{1pt}} &
\multicolumn{1}{c!{\huxvb{0, 0, 0}{0}}}{\huxtpad{1pt + 1em}\centering \hspace{1pt} {\fontsize{9pt}{10.8pt}\selectfont } \hspace{1pt}\huxbpad{1pt}} &
\multicolumn{1}{c!{\huxvb{0, 0, 0}{0}}}{\huxtpad{1pt + 1em}\centering \hspace{1pt} {\fontsize{9pt}{10.8pt}\selectfont 0.0011} \hspace{1pt}\huxbpad{1pt}} &
\multicolumn{1}{c!{\huxvb{0, 0, 0}{0}}}{\huxtpad{1pt + 1em}\centering \hspace{1pt} {\fontsize{9pt}{10.8pt}\selectfont 0.0002} \hspace{1pt}\huxbpad{1pt}} \tabularnewline[-0.5pt]

\hhline{}
\arrayrulecolor{black}

\multicolumn{1}{!{\huxvb{0, 0, 0}{0}}l!{\huxvb{0, 0, 0}{0}}}{\huxtpad{1pt + 1em}\raggedright \hspace{1pt} {\fontsize{9pt}{10.8pt}\selectfont } \hspace{1pt}\huxbpad{1pt}} &
\multicolumn{1}{c!{\huxvb{0, 0, 0}{0}}}{\huxtpad{1pt + 1em}\centering \hspace{1pt} {\fontsize{9pt}{10.8pt}\selectfont SD} \hspace{1pt}\huxbpad{1pt}} &
\multicolumn{1}{c!{\huxvb{0, 0, 0}{0}}}{\huxtpad{1pt + 1em}\centering \hspace{1pt} {\fontsize{9pt}{10.8pt}\selectfont 0.7864} \hspace{1pt}\huxbpad{1pt}} &
\multicolumn{1}{c!{\huxvb{0, 0, 0}{0}}}{\huxtpad{1pt + 1em}\centering \hspace{1pt} {\fontsize{9pt}{10.8pt}\selectfont 0.3081} \hspace{1pt}\huxbpad{1pt}} &
\multicolumn{1}{c!{\huxvb{0, 0, 0}{0}}}{\huxtpad{1pt + 1em}\centering \hspace{1pt} {\fontsize{9pt}{10.8pt}\selectfont } \hspace{1pt}\huxbpad{1pt}} &
\multicolumn{1}{c!{\huxvb{0, 0, 0}{0}}}{\huxtpad{1pt + 1em}\centering \hspace{1pt} {\fontsize{9pt}{10.8pt}\selectfont 0.7043} \hspace{1pt}\huxbpad{1pt}} &
\multicolumn{1}{c!{\huxvb{0, 0, 0}{0}}}{\huxtpad{1pt + 1em}\centering \hspace{1pt} {\fontsize{9pt}{10.8pt}\selectfont 0.2722} \hspace{1pt}\huxbpad{1pt}} \tabularnewline[-0.5pt]

\hhline{}
\arrayrulecolor{black}

\multicolumn{1}{!{\huxvb{0, 0, 0}{0}}l!{\huxvb{0, 0, 0}{0}}}{\huxtpad{1pt + 1em}\raggedright \hspace{1pt} {\fontsize{9pt}{10.8pt}\selectfont 4000} \hspace{1pt}\huxbpad{1pt}} &
\multicolumn{1}{c!{\huxvb{0, 0, 0}{0}}}{\huxtpad{1pt + 1em}\centering \hspace{1pt} {\fontsize{9pt}{10.8pt}\selectfont bias} \hspace{1pt}\huxbpad{1pt}} &
\multicolumn{1}{c!{\huxvb{0, 0, 0}{0}}}{\huxtpad{1pt + 1em}\centering \hspace{1pt} {\fontsize{9pt}{10.8pt}\selectfont -0.0213} \hspace{1pt}\huxbpad{1pt}} &
\multicolumn{1}{c!{\huxvb{0, 0, 0}{0}}}{\huxtpad{1pt + 1em}\centering \hspace{1pt} {\fontsize{9pt}{10.8pt}\selectfont 0.0088} \hspace{1pt}\huxbpad{1pt}} &
\multicolumn{1}{c!{\huxvb{0, 0, 0}{0}}}{\huxtpad{1pt + 1em}\centering \hspace{1pt} {\fontsize{9pt}{10.8pt}\selectfont } \hspace{1pt}\huxbpad{1pt}} &
\multicolumn{1}{c!{\huxvb{0, 0, 0}{0}}}{\huxtpad{1pt + 1em}\centering \hspace{1pt} {\fontsize{9pt}{10.8pt}\selectfont -0.0242} \hspace{1pt}\huxbpad{1pt}} &
\multicolumn{1}{c!{\huxvb{0, 0, 0}{0}}}{\huxtpad{1pt + 1em}\centering \hspace{1pt} {\fontsize{9pt}{10.8pt}\selectfont 0.0097} \hspace{1pt}\huxbpad{1pt}} \tabularnewline[-0.5pt]

\hhline{}
\arrayrulecolor{black}

\multicolumn{1}{!{\huxvb{0, 0, 0}{0}}l!{\huxvb{0, 0, 0}{0}}}{\huxtpad{1pt + 1em}\raggedright \hspace{1pt} {\fontsize{9pt}{10.8pt}\selectfont } \hspace{1pt}\huxbpad{1pt}} &
\multicolumn{1}{c!{\huxvb{0, 0, 0}{0}}}{\huxtpad{1pt + 1em}\centering \hspace{1pt} {\fontsize{9pt}{10.8pt}\selectfont MSE} \hspace{1pt}\huxbpad{1pt}} &
\multicolumn{1}{c!{\huxvb{0, 0, 0}{0}}}{\huxtpad{1pt + 1em}\centering \hspace{1pt} {\fontsize{9pt}{10.8pt}\selectfont 0.0005} \hspace{1pt}\huxbpad{1pt}} &
\multicolumn{1}{c!{\huxvb{0, 0, 0}{0}}}{\huxtpad{1pt + 1em}\centering \hspace{1pt} {\fontsize{9pt}{10.8pt}\selectfont 0.0001} \hspace{1pt}\huxbpad{1pt}} &
\multicolumn{1}{c!{\huxvb{0, 0, 0}{0}}}{\huxtpad{1pt + 1em}\centering \hspace{1pt} {\fontsize{9pt}{10.8pt}\selectfont } \hspace{1pt}\huxbpad{1pt}} &
\multicolumn{1}{c!{\huxvb{0, 0, 0}{0}}}{\huxtpad{1pt + 1em}\centering \hspace{1pt} {\fontsize{9pt}{10.8pt}\selectfont 0.0006} \hspace{1pt}\huxbpad{1pt}} &
\multicolumn{1}{c!{\huxvb{0, 0, 0}{0}}}{\huxtpad{1pt + 1em}\centering \hspace{1pt} {\fontsize{9pt}{10.8pt}\selectfont 0.0001} \hspace{1pt}\huxbpad{1pt}} \tabularnewline[-0.5pt]

\hhline{}
\arrayrulecolor{black}

\multicolumn{1}{!{\huxvb{0, 0, 0}{0}}l!{\huxvb{0, 0, 0}{0}}}{\huxtpad{1pt + 1em}\raggedright \hspace{1pt} {\fontsize{9pt}{10.8pt}\selectfont } \hspace{1pt}\huxbpad{1pt}} &
\multicolumn{1}{c!{\huxvb{0, 0, 0}{0}}}{\huxtpad{1pt + 1em}\centering \hspace{1pt} {\fontsize{9pt}{10.8pt}\selectfont SD} \hspace{1pt}\huxbpad{1pt}} &
\multicolumn{1}{c!{\huxvb{0, 0, 0}{0}}}{\huxtpad{1pt + 1em}\centering \hspace{1pt} {\fontsize{9pt}{10.8pt}\selectfont 0.5989} \hspace{1pt}\huxbpad{1pt}} &
\multicolumn{1}{c!{\huxvb{0, 0, 0}{0}}}{\huxtpad{1pt + 1em}\centering \hspace{1pt} {\fontsize{9pt}{10.8pt}\selectfont 0.2249} \hspace{1pt}\huxbpad{1pt}} &
\multicolumn{1}{c!{\huxvb{0, 0, 0}{0}}}{\huxtpad{1pt + 1em}\centering \hspace{1pt} {\fontsize{9pt}{10.8pt}\selectfont } \hspace{1pt}\huxbpad{1pt}} &
\multicolumn{1}{c!{\huxvb{0, 0, 0}{0}}}{\huxtpad{1pt + 1em}\centering \hspace{1pt} {\fontsize{9pt}{10.8pt}\selectfont 0.4927} \hspace{1pt}\huxbpad{1pt}} &
\multicolumn{1}{c!{\huxvb{0, 0, 0}{0}}}{\huxtpad{1pt + 1em}\centering \hspace{1pt} {\fontsize{9pt}{10.8pt}\selectfont 0.1795} \hspace{1pt}\huxbpad{1pt}} \tabularnewline[-0.5pt]

\hhline{>{\huxb{0, 0, 0}{1}}->{\huxb{0, 0, 0}{1}}->{\huxb{0, 0, 0}{1}}->{\huxb{0, 0, 0}{1}}->{\huxb{0, 0, 0}{1}}->{\huxb{0, 0, 0}{1}}->{\huxb{0, 0, 0}{1}}-}
\arrayrulecolor{black}
\end{tabular}
\end{threeparttable}\par\end{centerbox}

\end{table}

%% file: table_realdata_binary.tex
  \providecommand{\huxb}[2]{\arrayrulecolor[RGB]{#1}\global\arrayrulewidth=#2pt}
  \providecommand{\huxvb}[2]{\color[RGB]{#1}\vrule width #2pt}
  \providecommand{\huxtpad}[1]{\rule{0pt}{#1}}
  \providecommand{\huxbpad}[1]{\rule[-#1]{0pt}{#1}}

\begin{table}[ht]
\begin{centerbox}
\begin{threeparttable}
\captionsetup{justification=centering,singlelinecheck=off}
\caption{Estimates of obesity rates}
\label{table:binary}
 \setlength{\tabcolsep}{0pt}
\begin{tabular}{l l l}

\hhline{>{\huxb{0, 0, 0}{1}}=>{\huxb{0, 0, 0}{1}}=>{\huxb{0, 0, 0}{1}}=}
\arrayrulecolor{black}

\multicolumn{1}{!{\huxvb{0, 0, 0}{0}}c!{\huxvb{0, 0, 0}{0}}}{\huxtpad{1pt + 1em}\centering \hspace{1pt} \textbf{{\fontsize{9pt}{10.8pt}\selectfont }} \hspace{1pt}\huxbpad{1pt}} &
\multicolumn{1}{c!{\huxvb{0, 0, 0}{0}}}{\huxtpad{1pt + 1em}\centering \hspace{1pt} \textbf{{\fontsize{9pt}{10.8pt}\selectfont Girls}} \hspace{1pt}\huxbpad{1pt}} &
\multicolumn{1}{c!{\huxvb{0, 0, 0}{0}}}{\huxtpad{1pt + 1em}\centering \hspace{1pt} \textbf{{\fontsize{9pt}{10.8pt}\selectfont Boys}} \hspace{1pt}\huxbpad{1pt}} \tabularnewline[-0.5pt]

\hhline{>{\huxb{0, 0, 0}{0.4}}->{\huxb{0, 0, 0}{0.4}}->{\huxb{0, 0, 0}{0.4}}-}
\arrayrulecolor{black}

\multicolumn{1}{!{\huxvb{0, 0, 0}{0}}c!{\huxvb{0, 0, 0}{0}}}{\huxtpad{1pt + 1em}\centering \hspace{1pt} \textit{{\fontsize{9pt}{10.8pt}\selectfont }} \hspace{1pt}\huxbpad{1pt}} &
\multicolumn{2}{c!{\huxvb{0, 0, 0}{0}}}{\huxtpad{1pt + 1em}\centering \hspace{1pt} \textit{{\fontsize{9pt}{20.8pt}\selectfont Non-optimal GEE}} \hspace{1pt}\huxbpad{1pt}} \tabularnewline[-0.5pt]

\hhline{}
\arrayrulecolor{black}

\multicolumn{1}{!{\huxvb{0, 0, 0}{0}}c!{\huxvb{0, 0, 0}{0}}}{\huxtpad{1pt + 1em}\centering \hspace{1pt} {\fontsize{9pt}{10.8pt}\selectfont \(\theta_{11}\)} \hspace{1pt}\huxbpad{1pt}} &
\multicolumn{1}{c!{\huxvb{0, 0, 0}{0}}}{\huxtpad{1pt + 1em}\centering \hspace{1pt} {\fontsize{9pt}{10.8pt}\selectfont 0.723 (-)} \hspace{1pt}\huxbpad{1pt}} &
\multicolumn{1}{c!{\huxvb{0, 0, 0}{0}}}{\huxtpad{1pt + 1em}\centering \hspace{1pt} {\fontsize{9pt}{10.8pt}\selectfont 0.71 (-)} \hspace{1pt}\huxbpad{1pt}} \tabularnewline[-0.5pt]

\hhline{}
\arrayrulecolor{black}

\multicolumn{1}{!{\huxvb{0, 0, 0}{0}}c!{\huxvb{0, 0, 0}{0}}}{\huxtpad{1pt + 1em}\centering \hspace{1pt} {\fontsize{9pt}{10.8pt}\selectfont \(\theta_{12}\)} \hspace{1pt}\huxbpad{1pt}} &
\multicolumn{1}{c!{\huxvb{0, 0, 0}{0}}}{\huxtpad{1pt + 1em}\centering \hspace{1pt} {\fontsize{9pt}{10.8pt}\selectfont 0.081 (0.067)} \hspace{1pt}\huxbpad{1pt}} &
\multicolumn{1}{c!{\huxvb{0, 0, 0}{0}}}{\huxtpad{1pt + 1em}\centering \hspace{1pt} {\fontsize{9pt}{10.8pt}\selectfont 0.097 (0.074)} \hspace{1pt}\huxbpad{1pt}} \tabularnewline[-0.5pt]

\hhline{}
\arrayrulecolor{black}

\multicolumn{1}{!{\huxvb{0, 0, 0}{0}}c!{\huxvb{0, 0, 0}{0}}}{\huxtpad{1pt + 1em}\centering \hspace{1pt} {\fontsize{9pt}{10.8pt}\selectfont \(\theta_{21}\)} \hspace{1pt}\huxbpad{1pt}} &
\multicolumn{1}{c!{\huxvb{0, 0, 0}{0}}}{\huxtpad{1pt + 1em}\centering \hspace{1pt} {\fontsize{9pt}{10.8pt}\selectfont 0.078 (0.067)} \hspace{1pt}\huxbpad{1pt}} &
\multicolumn{1}{c!{\huxvb{0, 0, 0}{0}}}{\huxtpad{1pt + 1em}\centering \hspace{1pt} {\fontsize{9pt}{10.8pt}\selectfont 0.075 (0.066)} \hspace{1pt}\huxbpad{1pt}} \tabularnewline[-0.5pt]

\hhline{}
\arrayrulecolor{black}

\multicolumn{1}{!{\huxvb{0, 0, 0}{0}}c!{\huxvb{0, 0, 0}{0}}}{\huxtpad{1pt + 1em}\centering \hspace{1pt} {\fontsize{9pt}{10.8pt}\selectfont \(\theta_{22}\)} \hspace{1pt}\huxbpad{1pt}} &
\multicolumn{1}{c!{\huxvb{0, 0, 0}{0}}}{\huxtpad{1pt + 1em}\centering \hspace{1pt} {\fontsize{9pt}{10.8pt}\selectfont 0.118 (0.055)} \hspace{1pt}\huxbpad{1pt}} &
\multicolumn{1}{c!{\huxvb{0, 0, 0}{0}}}{\huxtpad{1pt + 1em}\centering \hspace{1pt} {\fontsize{9pt}{10.8pt}\selectfont 0.118 (0.059)} \hspace{1pt}\huxbpad{1pt}} \tabularnewline[-0.5pt]

\hhline{}
\arrayrulecolor{black}

\multicolumn{1}{!{\huxvb{0, 0, 0}{0}}c!{\huxvb{0, 0, 0}{0}}}{\huxtpad{1pt + 1em}\centering \hspace{1pt} {\fontsize{9pt}{10.8pt}\selectfont log(OR)} \hspace{1pt}\huxbpad{1pt}} &
\multicolumn{1}{c!{\huxvb{0, 0, 0}{0}}}{\huxtpad{1pt + 1em}\centering \hspace{1pt} {\fontsize{9pt}{10.8pt}\selectfont 2.6 (0.194)} \hspace{1pt}\huxbpad{1pt}} &
\multicolumn{1}{c!{\huxvb{0, 0, 0}{0}}}{\huxtpad{1pt + 1em}\centering \hspace{1pt} {\fontsize{9pt}{10.8pt}\selectfont 2.442 (0.179)} \hspace{1pt}\huxbpad{1pt}} \tabularnewline[-0.5pt]

\hhline{}
\arrayrulecolor{black}

\multicolumn{1}{!{\huxvb{0, 0, 0}{0}}c!{\huxvb{0, 0, 0}{0}}}{\huxtpad{1pt + 1em}\centering \hspace{1pt} \textit{{\fontsize{9pt}{10.8pt}\selectfont }} \hspace{1pt}\huxbpad{1pt}} &
\multicolumn{2}{c!{\huxvb{0, 0, 0}{0}}}{\huxtpad{1pt + 1em}\centering \hspace{1pt} \textit{{\fontsize{9pt}{10.8pt}\selectfont Pseudo-likelihood}} \hspace{1pt}\huxbpad{1pt}} \tabularnewline[-0.5pt]

\hhline{}
\arrayrulecolor{black}

\multicolumn{1}{!{\huxvb{0, 0, 0}{0}}c!{\huxvb{0, 0, 0}{0}}}{\huxtpad{1pt + 1em}\centering \hspace{1pt} {\fontsize{9pt}{10.8pt}\selectfont log(OR)} \hspace{1pt}\huxbpad{1pt}} &
\multicolumn{1}{c!{\huxvb{0, 0, 0}{0}}}{\huxtpad{1pt + 1em}\centering \hspace{1pt} {\fontsize{9pt}{10.8pt}\selectfont 2.6 (0.014)} \hspace{1pt}\huxbpad{1pt}} &
\multicolumn{1}{c!{\huxvb{0, 0, 0}{0}}}{\huxtpad{1pt + 1em}\centering \hspace{1pt} {\fontsize{9pt}{10.8pt}\selectfont 2.442 (0.012)} \hspace{1pt}\huxbpad{1pt}} \tabularnewline[-0.5pt]

\hhline{>{\huxb{0, 0, 0}{1}}->{\huxb{0, 0, 0}{1}}->{\huxb{0, 0, 0}{1}}-}
\arrayrulecolor{black}
\end{tabular}
\end{threeparttable}\par\end{centerbox}

\end{table}

%% file: table2.tex
  \providecommand{\huxb}[2]{\arrayrulecolor[RGB]{#1}\global\arrayrulewidth=#2pt}
  \providecommand{\huxvb}[2]{\color[RGB]{#1}\vrule width #2pt}
  \providecommand{\huxtpad}[1]{\rule{0pt}{#1}}
  \providecommand{\huxbpad}[1]{\rule[-#1]{0pt}{#1}}

\begin{table}[h]
\begin{centerbox}
\begin{threeparttable}
\captionsetup{justification=centering,singlelinecheck=off}
\caption{Standard deviation of estimators with varying correlation between $X$ and $Y$}
\label{table:2}
 \setlength{\tabcolsep}{0pt}
\begin{tabular}{l l l l}

\hhline{>{\huxb{0, 0, 0}{1}}=>{\huxb{0, 0, 0}{1}}=>{\huxb{0, 0, 0}{1}}=>{\huxb{0, 0, 0}{1}}=}
\arrayrulecolor{black}

\multicolumn{1}{!{\huxvb{0, 0, 0}{0}}l!{\huxvb{0, 0, 0}{0}}}{\huxtpad{1pt + 1em}\raggedright \hspace{1pt} \textbf{{\fontsize{9pt}{10.8pt}\selectfont \(\rho\)}} \hspace{1pt}\huxbpad{1pt}} &
\multicolumn{1}{c!{\huxvb{0, 0, 0}{0}}}{\huxtpad{1pt + 1em}\centering \hspace{1pt} \textbf{{\fontsize{9pt}{10.8pt}\selectfont \(\beta\)
(non-optimal GEE)}} \hspace{1pt}\huxbpad{1pt}} &
\multicolumn{1}{c!{\huxvb{0, 0, 0}{0}}}{\huxtpad{1pt + 1em}\centering \hspace{1pt} \textbf{{\fontsize{9pt}{10.8pt}\selectfont \(\beta\)
(optimal GEE)}} \hspace{1pt}\huxbpad{1pt}} &
\multicolumn{1}{c!{\huxvb{0, 0, 0}{0}}}{\huxtpad{1pt + 1em}\centering \hspace{1pt} \textbf{{\fontsize{9pt}{10.8pt}\selectfont logOR \newline (conditional likelihood)}} \hspace{1pt}\huxbpad{1pt}} \tabularnewline[-0.5pt]

\hhline{>{\huxb{0, 0, 0}{0.4}}->{\huxb{0, 0, 0}{0.4}}->{\huxb{0, 0, 0}{0.4}}->{\huxb{0, 0, 0}{0.4}}-}
\arrayrulecolor{black}

\multicolumn{1}{!{\huxvb{0, 0, 0}{0}}l!{\huxvb{0, 0, 0}{0}}}{\huxtpad{1pt + 1em}\raggedright \hspace{1pt} {\fontsize{9pt}{10.8pt}\selectfont -0.9} \hspace{1pt}\huxbpad{1pt}} &
\multicolumn{1}{c!{\huxvb{0, 0, 0}{0}}}{\huxtpad{1pt + 1em}\centering \hspace{1pt} {\fontsize{9pt}{10.8pt}\selectfont 0.0468} \hspace{1pt}\huxbpad{1pt}} &
\multicolumn{1}{c!{\huxvb{0, 0, 0}{0}}}{\huxtpad{1pt + 1em}\centering \hspace{1pt} {\fontsize{9pt}{10.8pt}\selectfont 0.0354} \hspace{1pt}\huxbpad{1pt}} &
\multicolumn{1}{c!{\huxvb{0, 0, 0}{0}}}{\huxtpad{1pt + 1em}\centering \hspace{1pt} {\fontsize{9pt}{10.8pt}\selectfont 0.1272} \hspace{1pt}\huxbpad{1pt}} \tabularnewline[-0.5pt]

\hhline{}
\arrayrulecolor{black}

\multicolumn{1}{!{\huxvb{0, 0, 0}{0}}l!{\huxvb{0, 0, 0}{0}}}{\huxtpad{1pt + 1em}\raggedright \hspace{1pt} {\fontsize{9pt}{10.8pt}\selectfont -0.7} \hspace{1pt}\huxbpad{1pt}} &
\multicolumn{1}{c!{\huxvb{0, 0, 0}{0}}}{\huxtpad{1pt + 1em}\centering \hspace{1pt} {\fontsize{9pt}{10.8pt}\selectfont 0.0678} \hspace{1pt}\huxbpad{1pt}} &
\multicolumn{1}{c!{\huxvb{0, 0, 0}{0}}}{\huxtpad{1pt + 1em}\centering \hspace{1pt} {\fontsize{9pt}{10.8pt}\selectfont 0.0622} \hspace{1pt}\huxbpad{1pt}} &
\multicolumn{1}{c!{\huxvb{0, 0, 0}{0}}}{\huxtpad{1pt + 1em}\centering \hspace{1pt} {\fontsize{9pt}{10.8pt}\selectfont 0.0470} \hspace{1pt}\huxbpad{1pt}} \tabularnewline[-0.5pt]

\hhline{}
\arrayrulecolor{black}

\multicolumn{1}{!{\huxvb{0, 0, 0}{0}}l!{\huxvb{0, 0, 0}{0}}}{\huxtpad{1pt + 1em}\raggedright \hspace{1pt} {\fontsize{9pt}{10.8pt}\selectfont -0.5} \hspace{1pt}\huxbpad{1pt}} &
\multicolumn{1}{c!{\huxvb{0, 0, 0}{0}}}{\huxtpad{1pt + 1em}\centering \hspace{1pt} {\fontsize{9pt}{10.8pt}\selectfont 0.0847} \hspace{1pt}\huxbpad{1pt}} &
\multicolumn{1}{c!{\huxvb{0, 0, 0}{0}}}{\huxtpad{1pt + 1em}\centering \hspace{1pt} {\fontsize{9pt}{10.8pt}\selectfont 0.1033} \hspace{1pt}\huxbpad{1pt}} &
\multicolumn{1}{c!{\huxvb{0, 0, 0}{0}}}{\huxtpad{1pt + 1em}\centering \hspace{1pt} {\fontsize{9pt}{10.8pt}\selectfont 0.0268} \hspace{1pt}\huxbpad{1pt}} \tabularnewline[-0.5pt]

\hhline{}
\arrayrulecolor{black}

\multicolumn{1}{!{\huxvb{0, 0, 0}{0}}l!{\huxvb{0, 0, 0}{0}}}{\huxtpad{1pt + 1em}\raggedright \hspace{1pt} {\fontsize{9pt}{10.8pt}\selectfont -0.3} \hspace{1pt}\huxbpad{1pt}} &
\multicolumn{1}{c!{\huxvb{0, 0, 0}{0}}}{\huxtpad{1pt + 1em}\centering \hspace{1pt} {\fontsize{9pt}{10.8pt}\selectfont 0.108} \hspace{1pt}\huxbpad{1pt}} &
\multicolumn{1}{c!{\huxvb{0, 0, 0}{0}}}{\huxtpad{1pt + 1em}\centering \hspace{1pt} {\fontsize{9pt}{10.8pt}\selectfont 0.1319} \hspace{1pt}\huxbpad{1pt}} &
\multicolumn{1}{c!{\huxvb{0, 0, 0}{0}}}{\huxtpad{1pt + 1em}\centering \hspace{1pt} {\fontsize{9pt}{10.8pt}\selectfont 0.0206} \hspace{1pt}\huxbpad{1pt}} \tabularnewline[-0.5pt]

\hhline{}
\arrayrulecolor{black}

\multicolumn{1}{!{\huxvb{0, 0, 0}{0}}l!{\huxvb{0, 0, 0}{0}}}{\huxtpad{1pt + 1em}\raggedright \hspace{1pt} {\fontsize{9pt}{10.8pt}\selectfont -0.1} \hspace{1pt}\huxbpad{1pt}} &
\multicolumn{1}{c!{\huxvb{0, 0, 0}{0}}}{\huxtpad{1pt + 1em}\centering \hspace{1pt} {\fontsize{9pt}{10.8pt}\selectfont 0.127} \hspace{1pt}\huxbpad{1pt}} &
\multicolumn{1}{c!{\huxvb{0, 0, 0}{0}}}{\huxtpad{1pt + 1em}\centering \hspace{1pt} {\fontsize{9pt}{10.8pt}\selectfont 0.1023} \hspace{1pt}\huxbpad{1pt}} &
\multicolumn{1}{c!{\huxvb{0, 0, 0}{0}}}{\huxtpad{1pt + 1em}\centering \hspace{1pt} {\fontsize{9pt}{10.8pt}\selectfont 0.0201} \hspace{1pt}\huxbpad{1pt}} \tabularnewline[-0.5pt]

\hhline{}
\arrayrulecolor{black}

\multicolumn{1}{!{\huxvb{0, 0, 0}{0}}l!{\huxvb{0, 0, 0}{0}}}{\huxtpad{1pt + 1em}\raggedright \hspace{1pt} {\fontsize{9pt}{10.8pt}\selectfont 0.1} \hspace{1pt}\huxbpad{1pt}} &
\multicolumn{1}{c!{\huxvb{0, 0, 0}{0}}}{\huxtpad{1pt + 1em}\centering \hspace{1pt} {\fontsize{9pt}{10.8pt}\selectfont 0.118} \hspace{1pt}\huxbpad{1pt}} &
\multicolumn{1}{c!{\huxvb{0, 0, 0}{0}}}{\huxtpad{1pt + 1em}\centering \hspace{1pt} {\fontsize{9pt}{10.8pt}\selectfont 0.0979} \hspace{1pt}\huxbpad{1pt}} &
\multicolumn{1}{c!{\huxvb{0, 0, 0}{0}}}{\huxtpad{1pt + 1em}\centering \hspace{1pt} {\fontsize{9pt}{10.8pt}\selectfont 0.0179} \hspace{1pt}\huxbpad{1pt}} \tabularnewline[-0.5pt]

\hhline{}
\arrayrulecolor{black}

\multicolumn{1}{!{\huxvb{0, 0, 0}{0}}l!{\huxvb{0, 0, 0}{0}}}{\huxtpad{1pt + 1em}\raggedright \hspace{1pt} {\fontsize{9pt}{10.8pt}\selectfont 0.3} \hspace{1pt}\huxbpad{1pt}} &
\multicolumn{1}{c!{\huxvb{0, 0, 0}{0}}}{\huxtpad{1pt + 1em}\centering \hspace{1pt} {\fontsize{9pt}{10.8pt}\selectfont 0.154} \hspace{1pt}\huxbpad{1pt}} &
\multicolumn{1}{c!{\huxvb{0, 0, 0}{0}}}{\huxtpad{1pt + 1em}\centering \hspace{1pt} {\fontsize{9pt}{10.8pt}\selectfont 0.0783} \hspace{1pt}\huxbpad{1pt}} &
\multicolumn{1}{c!{\huxvb{0, 0, 0}{0}}}{\huxtpad{1pt + 1em}\centering \hspace{1pt} {\fontsize{9pt}{10.8pt}\selectfont 0.0189} \hspace{1pt}\huxbpad{1pt}} \tabularnewline[-0.5pt]

\hhline{}
\arrayrulecolor{black}

\multicolumn{1}{!{\huxvb{0, 0, 0}{0}}l!{\huxvb{0, 0, 0}{0}}}{\huxtpad{1pt + 1em}\raggedright \hspace{1pt} {\fontsize{9pt}{10.8pt}\selectfont 0.5} \hspace{1pt}\huxbpad{1pt}} &
\multicolumn{1}{c!{\huxvb{0, 0, 0}{0}}}{\huxtpad{1pt + 1em}\centering \hspace{1pt} {\fontsize{9pt}{10.8pt}\selectfont 0.0877} \hspace{1pt}\huxbpad{1pt}} &
\multicolumn{1}{c!{\huxvb{0, 0, 0}{0}}}{\huxtpad{1pt + 1em}\centering \hspace{1pt} {\fontsize{9pt}{10.8pt}\selectfont 0.0535} \hspace{1pt}\huxbpad{1pt}} &
\multicolumn{1}{c!{\huxvb{0, 0, 0}{0}}}{\huxtpad{1pt + 1em}\centering \hspace{1pt} {\fontsize{9pt}{10.8pt}\selectfont 0.0267} \hspace{1pt}\huxbpad{1pt}} \tabularnewline[-0.5pt]

\hhline{}
\arrayrulecolor{black}

\multicolumn{1}{!{\huxvb{0, 0, 0}{0}}l!{\huxvb{0, 0, 0}{0}}}{\huxtpad{1pt + 1em}\raggedright \hspace{1pt} {\fontsize{9pt}{10.8pt}\selectfont 0.7} \hspace{1pt}\huxbpad{1pt}} &
\multicolumn{1}{c!{\huxvb{0, 0, 0}{0}}}{\huxtpad{1pt + 1em}\centering \hspace{1pt} {\fontsize{9pt}{10.8pt}\selectfont 0.0628} \hspace{1pt}\huxbpad{1pt}} &
\multicolumn{1}{c!{\huxvb{0, 0, 0}{0}}}{\huxtpad{1pt + 1em}\centering \hspace{1pt} {\fontsize{9pt}{10.8pt}\selectfont 0.0432} \hspace{1pt}\huxbpad{1pt}} &
\multicolumn{1}{c!{\huxvb{0, 0, 0}{0}}}{\huxtpad{1pt + 1em}\centering \hspace{1pt} {\fontsize{9pt}{10.8pt}\selectfont 0.0413} \hspace{1pt}\huxbpad{1pt}} \tabularnewline[-0.5pt]

\hhline{}
\arrayrulecolor{black}

\multicolumn{1}{!{\huxvb{0, 0, 0}{0}}l!{\huxvb{0, 0, 0}{0}}}{\huxtpad{1pt + 1em}\raggedright \hspace{1pt} {\fontsize{9pt}{10.8pt}\selectfont 0.9} \hspace{1pt}\huxbpad{1pt}} &
\multicolumn{1}{c!{\huxvb{0, 0, 0}{0}}}{\huxtpad{1pt + 1em}\centering \hspace{1pt} {\fontsize{9pt}{10.8pt}\selectfont 0.0296} \hspace{1pt}\huxbpad{1pt}} &
\multicolumn{1}{c!{\huxvb{0, 0, 0}{0}}}{\huxtpad{1pt + 1em}\centering \hspace{1pt} {\fontsize{9pt}{10.8pt}\selectfont 0.0211} \hspace{1pt}\huxbpad{1pt}} &
\multicolumn{1}{c!{\huxvb{0, 0, 0}{0}}}{\huxtpad{1pt + 1em}\centering \hspace{1pt} {\fontsize{9pt}{10.8pt}\selectfont 0.0917} \hspace{1pt}\huxbpad{1pt}} \tabularnewline[-0.5pt]

\hhline{>{\huxb{0, 0, 0}{1}}->{\huxb{0, 0, 0}{1}}->{\huxb{0, 0, 0}{1}}->{\huxb{0, 0, 0}{1}}-}
\arrayrulecolor{black}
\end{tabular}
\end{threeparttable}\par\end{centerbox}

\end{table}

%% file: table3.tex
  \providecommand{\huxb}[2]{\arrayrulecolor[RGB]{#1}\global\arrayrulewidth=#2pt}
  \providecommand{\huxvb}[2]{\color[RGB]{#1}\vrule width #2pt}
  \providecommand{\huxtpad}[1]{\rule{0pt}{#1}}
  \providecommand{\huxbpad}[1]{\rule[-#1]{0pt}{#1}}

\begin{table}[ht]
\begin{centerbox}
\begin{threeparttable}
\captionsetup{justification=centering,singlelinecheck=off}
\caption{Estimation under model misspecification}
\label{table:3}
 \setlength{\tabcolsep}{0pt}
\begin{tabular}{l l l l l l l}

\hhline{>{\huxb{0, 0, 0}{1}}=>{\huxb{0, 0, 0}{1}}=>{\huxb{0, 0, 0}{1}}=>{\huxb{0, 0, 0}{1}}=>{\huxb{0, 0, 0}{1}}=>{\huxb{0, 0, 0}{1}}=>{\huxb{0, 0, 0}{1}}=}
\arrayrulecolor{black}

\multicolumn{1}{!{\huxvb{0, 0, 0}{0}}l!{\huxvb{0, 0, 0}{0}}}{\huxtpad{1pt + 1em}\raggedright \hspace{1pt} \textbf{{\fontsize{9pt}{10.8pt}\selectfont }} \hspace{1pt}\huxbpad{1pt}} &
\multicolumn{1}{c!{\huxvb{0, 0, 0}{0}}}{\huxtpad{1pt + 1em}\centering \hspace{1pt} \textbf{{\fontsize{9pt}{10.8pt}\selectfont }} \hspace{1pt}\huxbpad{1pt}} &
\multicolumn{2}{c!{\huxvb{0, 0, 0}{0}}}{\huxtpad{1pt + 1em}\centering \hspace{1pt} \textbf{{\fontsize{9pt}{10.8pt}\selectfont Non-optimal GEE}} \hspace{1pt}\huxbpad{1pt}} &
\multicolumn{1}{c!{\huxvb{0, 0, 0}{0}}}{\huxtpad{1pt + 1em}\centering \hspace{1pt} \textbf{{\fontsize{9pt}{10.8pt}\selectfont }} \hspace{1pt}\huxbpad{1pt}} &
\multicolumn{2}{c!{\huxvb{0, 0, 0}{0}}}{\huxtpad{1pt + 1em}\centering \hspace{1pt} \textbf{{\fontsize{9pt}{10.8pt}\selectfont Optimal GEE}} \hspace{1pt}\huxbpad{1pt}} \tabularnewline[-0.5pt]

\hhline{>{\huxb{255, 255, 255}{0.4}}->{\huxb{255, 255, 255}{0.4}}->{\huxb{0, 0, 0}{0.4}}->{\huxb{0, 0, 0}{0.4}}->{\huxb{255, 255, 255}{0.4}}->{\huxb{0, 0, 0}{0.4}}->{\huxb{0, 0, 0}{0.4}}-}
\arrayrulecolor{black}

\multicolumn{1}{!{\huxvb{0, 0, 0}{0}}l!{\huxvb{0, 0, 0}{0}}}{\huxtpad{1pt + 1em}\raggedright \hspace{1pt} {\fontsize{9pt}{10.8pt}\selectfont N} \hspace{1pt}\huxbpad{1pt}} &
\multicolumn{1}{c!{\huxvb{0, 0, 0}{0}}}{\huxtpad{1pt + 1em}\centering \hspace{1pt} {\fontsize{9pt}{10.8pt}\selectfont Statistics} \hspace{1pt}\huxbpad{1pt}} &
\multicolumn{1}{c!{\huxvb{0, 0, 0}{0}}}{\huxtpad{1pt + 1em}\centering \hspace{1pt} {\fontsize{9pt}{10.8pt}\selectfont \(\alpha\)} \hspace{1pt}\huxbpad{1pt}} &
\multicolumn{1}{c!{\huxvb{0, 0, 0}{0}}}{\huxtpad{1pt + 1em}\centering \hspace{1pt} {\fontsize{9pt}{10.8pt}\selectfont \(\beta\)} \hspace{1pt}\huxbpad{1pt}} &
\multicolumn{1}{c!{\huxvb{0, 0, 0}{0}}}{\huxtpad{1pt + 1em}\centering \hspace{1pt} {\fontsize{9pt}{10.8pt}\selectfont } \hspace{1pt}\huxbpad{1pt}} &
\multicolumn{1}{c!{\huxvb{0, 0, 0}{0}}}{\huxtpad{1pt + 1em}\centering \hspace{1pt} {\fontsize{9pt}{10.8pt}\selectfont \(\alpha\)} \hspace{1pt}\huxbpad{1pt}} &
\multicolumn{1}{c!{\huxvb{0, 0, 0}{0}}}{\huxtpad{1pt + 1em}\centering \hspace{1pt} {\fontsize{9pt}{10.8pt}\selectfont \(\beta\)} \hspace{1pt}\huxbpad{1pt}} \tabularnewline[-0.5pt]

\hhline{>{\huxb{0, 0, 0}{0.4}}->{\huxb{0, 0, 0}{0.4}}->{\huxb{0, 0, 0}{0.4}}->{\huxb{0, 0, 0}{0.4}}->{\huxb{0, 0, 0}{0.4}}->{\huxb{0, 0, 0}{0.4}}->{\huxb{0, 0, 0}{0.4}}-}
\arrayrulecolor{black}

\multicolumn{1}{!{\huxvb{0, 0, 0}{0}}l!{\huxvb{0, 0, 0}{0}}}{\huxtpad{1pt + 1em}\raggedright \hspace{1pt} {\fontsize{9pt}{10.8pt}\selectfont 500} \hspace{1pt}\huxbpad{1pt}} &
\multicolumn{1}{c!{\huxvb{0, 0, 0}{0}}}{\huxtpad{1pt + 1em}\centering \hspace{1pt} {\fontsize{9pt}{10.8pt}\selectfont bias} \hspace{1pt}\huxbpad{1pt}} &
\multicolumn{1}{c!{\huxvb{0, 0, 0}{0}}}{\huxtpad{1pt + 1em}\centering \hspace{1pt} {\fontsize{9pt}{10.8pt}\selectfont -0.3435} \hspace{1pt}\huxbpad{1pt}} &
\multicolumn{1}{c!{\huxvb{0, 0, 0}{0}}}{\huxtpad{1pt + 1em}\centering \hspace{1pt} {\fontsize{9pt}{10.8pt}\selectfont 0.1260} \hspace{1pt}\huxbpad{1pt}} &
\multicolumn{1}{c!{\huxvb{0, 0, 0}{0}}}{\huxtpad{1pt + 1em}\centering \hspace{1pt} {\fontsize{9pt}{10.8pt}\selectfont } \hspace{1pt}\huxbpad{1pt}} &
\multicolumn{1}{c!{\huxvb{0, 0, 0}{0}}}{\huxtpad{1pt + 1em}\centering \hspace{1pt} {\fontsize{9pt}{10.8pt}\selectfont -0.3352} \hspace{1pt}\huxbpad{1pt}} &
\multicolumn{1}{c!{\huxvb{0, 0, 0}{0}}}{\huxtpad{1pt + 1em}\centering \hspace{1pt} {\fontsize{9pt}{10.8pt}\selectfont 0.1224} \hspace{1pt}\huxbpad{1pt}} \tabularnewline[-0.5pt]

\hhline{}
\arrayrulecolor{black}

\multicolumn{1}{!{\huxvb{0, 0, 0}{0}}l!{\huxvb{0, 0, 0}{0}}}{\huxtpad{1pt + 1em}\raggedright \hspace{1pt} {\fontsize{9pt}{10.8pt}\selectfont } \hspace{1pt}\huxbpad{1pt}} &
\multicolumn{1}{c!{\huxvb{0, 0, 0}{0}}}{\huxtpad{1pt + 1em}\centering \hspace{1pt} {\fontsize{9pt}{10.8pt}\selectfont MSE} \hspace{1pt}\huxbpad{1pt}} &
\multicolumn{1}{c!{\huxvb{0, 0, 0}{0}}}{\huxtpad{1pt + 1em}\centering \hspace{1pt} {\fontsize{9pt}{10.8pt}\selectfont 0.1180} \hspace{1pt}\huxbpad{1pt}} &
\multicolumn{1}{c!{\huxvb{0, 0, 0}{0}}}{\huxtpad{1pt + 1em}\centering \hspace{1pt} {\fontsize{9pt}{10.8pt}\selectfont 0.0159} \hspace{1pt}\huxbpad{1pt}} &
\multicolumn{1}{c!{\huxvb{0, 0, 0}{0}}}{\huxtpad{1pt + 1em}\centering \hspace{1pt} {\fontsize{9pt}{10.8pt}\selectfont } \hspace{1pt}\huxbpad{1pt}} &
\multicolumn{1}{c!{\huxvb{0, 0, 0}{0}}}{\huxtpad{1pt + 1em}\centering \hspace{1pt} {\fontsize{9pt}{10.8pt}\selectfont 0.1124} \hspace{1pt}\huxbpad{1pt}} &
\multicolumn{1}{c!{\huxvb{0, 0, 0}{0}}}{\huxtpad{1pt + 1em}\centering \hspace{1pt} {\fontsize{9pt}{10.8pt}\selectfont 0.0150} \hspace{1pt}\huxbpad{1pt}} \tabularnewline[-0.5pt]

\hhline{}
\arrayrulecolor{black}

\multicolumn{1}{!{\huxvb{0, 0, 0}{0}}l!{\huxvb{0, 0, 0}{0}}}{\huxtpad{1pt + 1em}\raggedright \hspace{1pt} {\fontsize{9pt}{10.8pt}\selectfont } \hspace{1pt}\huxbpad{1pt}} &
\multicolumn{1}{c!{\huxvb{0, 0, 0}{0}}}{\huxtpad{1pt + 1em}\centering \hspace{1pt} {\fontsize{9pt}{10.8pt}\selectfont SD} \hspace{1pt}\huxbpad{1pt}} &
\multicolumn{1}{c!{\huxvb{0, 0, 0}{0}}}{\huxtpad{1pt + 1em}\centering \hspace{1pt} {\fontsize{9pt}{10.8pt}\selectfont 0.4557} \hspace{1pt}\huxbpad{1pt}} &
\multicolumn{1}{c!{\huxvb{0, 0, 0}{0}}}{\huxtpad{1pt + 1em}\centering \hspace{1pt} {\fontsize{9pt}{10.8pt}\selectfont 0.1966} \hspace{1pt}\huxbpad{1pt}} &
\multicolumn{1}{c!{\huxvb{0, 0, 0}{0}}}{\huxtpad{1pt + 1em}\centering \hspace{1pt} {\fontsize{9pt}{10.8pt}\selectfont } \hspace{1pt}\huxbpad{1pt}} &
\multicolumn{1}{c!{\huxvb{0, 0, 0}{0}}}{\huxtpad{1pt + 1em}\centering \hspace{1pt} {\fontsize{9pt}{10.8pt}\selectfont 0.4483} \hspace{1pt}\huxbpad{1pt}} &
\multicolumn{1}{c!{\huxvb{0, 0, 0}{0}}}{\huxtpad{1pt + 1em}\centering \hspace{1pt} {\fontsize{9pt}{10.8pt}\selectfont 0.1930} \hspace{1pt}\huxbpad{1pt}} \tabularnewline[-0.5pt]

\hhline{}
\arrayrulecolor{black}

\multicolumn{1}{!{\huxvb{0, 0, 0}{0}}l!{\huxvb{0, 0, 0}{0}}}{\huxtpad{1pt + 1em}\raggedright \hspace{1pt} {\fontsize{9pt}{10.8pt}\selectfont 1000} \hspace{1pt}\huxbpad{1pt}} &
\multicolumn{1}{c!{\huxvb{0, 0, 0}{0}}}{\huxtpad{1pt + 1em}\centering \hspace{1pt} {\fontsize{9pt}{10.8pt}\selectfont bias} \hspace{1pt}\huxbpad{1pt}} &
\multicolumn{1}{c!{\huxvb{0, 0, 0}{0}}}{\huxtpad{1pt + 1em}\centering \hspace{1pt} {\fontsize{9pt}{10.8pt}\selectfont -0.4667} \hspace{1pt}\huxbpad{1pt}} &
\multicolumn{1}{c!{\huxvb{0, 0, 0}{0}}}{\huxtpad{1pt + 1em}\centering \hspace{1pt} {\fontsize{9pt}{10.8pt}\selectfont 0.1607} \hspace{1pt}\huxbpad{1pt}} &
\multicolumn{1}{c!{\huxvb{0, 0, 0}{0}}}{\huxtpad{1pt + 1em}\centering \hspace{1pt} {\fontsize{9pt}{10.8pt}\selectfont } \hspace{1pt}\huxbpad{1pt}} &
\multicolumn{1}{c!{\huxvb{0, 0, 0}{0}}}{\huxtpad{1pt + 1em}\centering \hspace{1pt} {\fontsize{9pt}{10.8pt}\selectfont -0.4606} \hspace{1pt}\huxbpad{1pt}} &
\multicolumn{1}{c!{\huxvb{0, 0, 0}{0}}}{\huxtpad{1pt + 1em}\centering \hspace{1pt} {\fontsize{9pt}{10.8pt}\selectfont 0.1578} \hspace{1pt}\huxbpad{1pt}} \tabularnewline[-0.5pt]

\hhline{}
\arrayrulecolor{black}

\multicolumn{1}{!{\huxvb{0, 0, 0}{0}}l!{\huxvb{0, 0, 0}{0}}}{\huxtpad{1pt + 1em}\raggedright \hspace{1pt} {\fontsize{9pt}{10.8pt}\selectfont } \hspace{1pt}\huxbpad{1pt}} &
\multicolumn{1}{c!{\huxvb{0, 0, 0}{0}}}{\huxtpad{1pt + 1em}\centering \hspace{1pt} {\fontsize{9pt}{10.8pt}\selectfont MSE} \hspace{1pt}\huxbpad{1pt}} &
\multicolumn{1}{c!{\huxvb{0, 0, 0}{0}}}{\huxtpad{1pt + 1em}\centering \hspace{1pt} {\fontsize{9pt}{10.8pt}\selectfont 0.2178} \hspace{1pt}\huxbpad{1pt}} &
\multicolumn{1}{c!{\huxvb{0, 0, 0}{0}}}{\huxtpad{1pt + 1em}\centering \hspace{1pt} {\fontsize{9pt}{10.8pt}\selectfont 0.0258} \hspace{1pt}\huxbpad{1pt}} &
\multicolumn{1}{c!{\huxvb{0, 0, 0}{0}}}{\huxtpad{1pt + 1em}\centering \hspace{1pt} {\fontsize{9pt}{10.8pt}\selectfont } \hspace{1pt}\huxbpad{1pt}} &
\multicolumn{1}{c!{\huxvb{0, 0, 0}{0}}}{\huxtpad{1pt + 1em}\centering \hspace{1pt} {\fontsize{9pt}{10.8pt}\selectfont 0.2122} \hspace{1pt}\huxbpad{1pt}} &
\multicolumn{1}{c!{\huxvb{0, 0, 0}{0}}}{\huxtpad{1pt + 1em}\centering \hspace{1pt} {\fontsize{9pt}{10.8pt}\selectfont 0.0249} \hspace{1pt}\huxbpad{1pt}} \tabularnewline[-0.5pt]

\hhline{}
\arrayrulecolor{black}

\multicolumn{1}{!{\huxvb{0, 0, 0}{0}}l!{\huxvb{0, 0, 0}{0}}}{\huxtpad{1pt + 1em}\raggedright \hspace{1pt} {\fontsize{9pt}{10.8pt}\selectfont } \hspace{1pt}\huxbpad{1pt}} &
\multicolumn{1}{c!{\huxvb{0, 0, 0}{0}}}{\huxtpad{1pt + 1em}\centering \hspace{1pt} {\fontsize{9pt}{10.8pt}\selectfont SD} \hspace{1pt}\huxbpad{1pt}} &
\multicolumn{1}{c!{\huxvb{0, 0, 0}{0}}}{\huxtpad{1pt + 1em}\centering \hspace{1pt} {\fontsize{9pt}{10.8pt}\selectfont 0.3254} \hspace{1pt}\huxbpad{1pt}} &
\multicolumn{1}{c!{\huxvb{0, 0, 0}{0}}}{\huxtpad{1pt + 1em}\centering \hspace{1pt} {\fontsize{9pt}{10.8pt}\selectfont 0.1397} \hspace{1pt}\huxbpad{1pt}} &
\multicolumn{1}{c!{\huxvb{0, 0, 0}{0}}}{\huxtpad{1pt + 1em}\centering \hspace{1pt} {\fontsize{9pt}{10.8pt}\selectfont } \hspace{1pt}\huxbpad{1pt}} &
\multicolumn{1}{c!{\huxvb{0, 0, 0}{0}}}{\huxtpad{1pt + 1em}\centering \hspace{1pt} {\fontsize{9pt}{10.8pt}\selectfont 0.3160} \hspace{1pt}\huxbpad{1pt}} &
\multicolumn{1}{c!{\huxvb{0, 0, 0}{0}}}{\huxtpad{1pt + 1em}\centering \hspace{1pt} {\fontsize{9pt}{10.8pt}\selectfont 0.1346} \hspace{1pt}\huxbpad{1pt}} \tabularnewline[-0.5pt]

\hhline{}
\arrayrulecolor{black}

\multicolumn{1}{!{\huxvb{0, 0, 0}{0}}l!{\huxvb{0, 0, 0}{0}}}{\huxtpad{1pt + 1em}\raggedright \hspace{1pt} {\fontsize{9pt}{10.8pt}\selectfont 2000} \hspace{1pt}\huxbpad{1pt}} &
\multicolumn{1}{c!{\huxvb{0, 0, 0}{0}}}{\huxtpad{1pt + 1em}\centering \hspace{1pt} {\fontsize{9pt}{10.8pt}\selectfont bias} \hspace{1pt}\huxbpad{1pt}} &
\multicolumn{1}{c!{\huxvb{0, 0, 0}{0}}}{\huxtpad{1pt + 1em}\centering \hspace{1pt} {\fontsize{9pt}{10.8pt}\selectfont -0.4859} \hspace{1pt}\huxbpad{1pt}} &
\multicolumn{1}{c!{\huxvb{0, 0, 0}{0}}}{\huxtpad{1pt + 1em}\centering \hspace{1pt} {\fontsize{9pt}{10.8pt}\selectfont 0.1737} \hspace{1pt}\huxbpad{1pt}} &
\multicolumn{1}{c!{\huxvb{0, 0, 0}{0}}}{\huxtpad{1pt + 1em}\centering \hspace{1pt} {\fontsize{9pt}{10.8pt}\selectfont } \hspace{1pt}\huxbpad{1pt}} &
\multicolumn{1}{c!{\huxvb{0, 0, 0}{0}}}{\huxtpad{1pt + 1em}\centering \hspace{1pt} {\fontsize{9pt}{10.8pt}\selectfont -0.4747} \hspace{1pt}\huxbpad{1pt}} &
\multicolumn{1}{c!{\huxvb{0, 0, 0}{0}}}{\huxtpad{1pt + 1em}\centering \hspace{1pt} {\fontsize{9pt}{10.8pt}\selectfont 0.1689} \hspace{1pt}\huxbpad{1pt}} \tabularnewline[-0.5pt]

\hhline{}
\arrayrulecolor{black}

\multicolumn{1}{!{\huxvb{0, 0, 0}{0}}l!{\huxvb{0, 0, 0}{0}}}{\huxtpad{1pt + 1em}\raggedright \hspace{1pt} {\fontsize{9pt}{10.8pt}\selectfont } \hspace{1pt}\huxbpad{1pt}} &
\multicolumn{1}{c!{\huxvb{0, 0, 0}{0}}}{\huxtpad{1pt + 1em}\centering \hspace{1pt} {\fontsize{9pt}{10.8pt}\selectfont MSE} \hspace{1pt}\huxbpad{1pt}} &
\multicolumn{1}{c!{\huxvb{0, 0, 0}{0}}}{\huxtpad{1pt + 1em}\centering \hspace{1pt} {\fontsize{9pt}{10.8pt}\selectfont 0.2361} \hspace{1pt}\huxbpad{1pt}} &
\multicolumn{1}{c!{\huxvb{0, 0, 0}{0}}}{\huxtpad{1pt + 1em}\centering \hspace{1pt} {\fontsize{9pt}{10.8pt}\selectfont 0.0302} \hspace{1pt}\huxbpad{1pt}} &
\multicolumn{1}{c!{\huxvb{0, 0, 0}{0}}}{\huxtpad{1pt + 1em}\centering \hspace{1pt} {\fontsize{9pt}{10.8pt}\selectfont } \hspace{1pt}\huxbpad{1pt}} &
\multicolumn{1}{c!{\huxvb{0, 0, 0}{0}}}{\huxtpad{1pt + 1em}\centering \hspace{1pt} {\fontsize{9pt}{10.8pt}\selectfont 0.2253} \hspace{1pt}\huxbpad{1pt}} &
\multicolumn{1}{c!{\huxvb{0, 0, 0}{0}}}{\huxtpad{1pt + 1em}\centering \hspace{1pt} {\fontsize{9pt}{10.8pt}\selectfont 0.0285} \hspace{1pt}\huxbpad{1pt}} \tabularnewline[-0.5pt]

\hhline{}
\arrayrulecolor{black}

\multicolumn{1}{!{\huxvb{0, 0, 0}{0}}l!{\huxvb{0, 0, 0}{0}}}{\huxtpad{1pt + 1em}\raggedright \hspace{1pt} {\fontsize{9pt}{10.8pt}\selectfont } \hspace{1pt}\huxbpad{1pt}} &
\multicolumn{1}{c!{\huxvb{0, 0, 0}{0}}}{\huxtpad{1pt + 1em}\centering \hspace{1pt} {\fontsize{9pt}{10.8pt}\selectfont SD} \hspace{1pt}\huxbpad{1pt}} &
\multicolumn{1}{c!{\huxvb{0, 0, 0}{0}}}{\huxtpad{1pt + 1em}\centering \hspace{1pt} {\fontsize{9pt}{10.8pt}\selectfont 0.2343} \hspace{1pt}\huxbpad{1pt}} &
\multicolumn{1}{c!{\huxvb{0, 0, 0}{0}}}{\huxtpad{1pt + 1em}\centering \hspace{1pt} {\fontsize{9pt}{10.8pt}\selectfont 0.1041} \hspace{1pt}\huxbpad{1pt}} &
\multicolumn{1}{c!{\huxvb{0, 0, 0}{0}}}{\huxtpad{1pt + 1em}\centering \hspace{1pt} {\fontsize{9pt}{10.8pt}\selectfont } \hspace{1pt}\huxbpad{1pt}} &
\multicolumn{1}{c!{\huxvb{0, 0, 0}{0}}}{\huxtpad{1pt + 1em}\centering \hspace{1pt} {\fontsize{9pt}{10.8pt}\selectfont 0.2358} \hspace{1pt}\huxbpad{1pt}} &
\multicolumn{1}{c!{\huxvb{0, 0, 0}{0}}}{\huxtpad{1pt + 1em}\centering \hspace{1pt} {\fontsize{9pt}{10.8pt}\selectfont 0.1042} \hspace{1pt}\huxbpad{1pt}} \tabularnewline[-0.5pt]

\hhline{}
\arrayrulecolor{black}

\multicolumn{1}{!{\huxvb{0, 0, 0}{0}}l!{\huxvb{0, 0, 0}{0}}}{\huxtpad{1pt + 1em}\raggedright \hspace{1pt} {\fontsize{9pt}{10.8pt}\selectfont 4000} \hspace{1pt}\huxbpad{1pt}} &
\multicolumn{1}{c!{\huxvb{0, 0, 0}{0}}}{\huxtpad{1pt + 1em}\centering \hspace{1pt} {\fontsize{9pt}{10.8pt}\selectfont bias} \hspace{1pt}\huxbpad{1pt}} &
\multicolumn{1}{c!{\huxvb{0, 0, 0}{0}}}{\huxtpad{1pt + 1em}\centering \hspace{1pt} {\fontsize{9pt}{10.8pt}\selectfont -0.4497} \hspace{1pt}\huxbpad{1pt}} &
\multicolumn{1}{c!{\huxvb{0, 0, 0}{0}}}{\huxtpad{1pt + 1em}\centering \hspace{1pt} {\fontsize{9pt}{10.8pt}\selectfont 0.1616} \hspace{1pt}\huxbpad{1pt}} &
\multicolumn{1}{c!{\huxvb{0, 0, 0}{0}}}{\huxtpad{1pt + 1em}\centering \hspace{1pt} {\fontsize{9pt}{10.8pt}\selectfont } \hspace{1pt}\huxbpad{1pt}} &
\multicolumn{1}{c!{\huxvb{0, 0, 0}{0}}}{\huxtpad{1pt + 1em}\centering \hspace{1pt} {\fontsize{9pt}{10.8pt}\selectfont -0.4387} \hspace{1pt}\huxbpad{1pt}} &
\multicolumn{1}{c!{\huxvb{0, 0, 0}{0}}}{\huxtpad{1pt + 1em}\centering \hspace{1pt} {\fontsize{9pt}{10.8pt}\selectfont 0.1568} \hspace{1pt}\huxbpad{1pt}} \tabularnewline[-0.5pt]

\hhline{}
\arrayrulecolor{black}

\multicolumn{1}{!{\huxvb{0, 0, 0}{0}}l!{\huxvb{0, 0, 0}{0}}}{\huxtpad{1pt + 1em}\raggedright \hspace{1pt} {\fontsize{9pt}{10.8pt}\selectfont } \hspace{1pt}\huxbpad{1pt}} &
\multicolumn{1}{c!{\huxvb{0, 0, 0}{0}}}{\huxtpad{1pt + 1em}\centering \hspace{1pt} {\fontsize{9pt}{10.8pt}\selectfont MSE} \hspace{1pt}\huxbpad{1pt}} &
\multicolumn{1}{c!{\huxvb{0, 0, 0}{0}}}{\huxtpad{1pt + 1em}\centering \hspace{1pt} {\fontsize{9pt}{10.8pt}\selectfont 0.2022} \hspace{1pt}\huxbpad{1pt}} &
\multicolumn{1}{c!{\huxvb{0, 0, 0}{0}}}{\huxtpad{1pt + 1em}\centering \hspace{1pt} {\fontsize{9pt}{10.8pt}\selectfont 0.0261} \hspace{1pt}\huxbpad{1pt}} &
\multicolumn{1}{c!{\huxvb{0, 0, 0}{0}}}{\huxtpad{1pt + 1em}\centering \hspace{1pt} {\fontsize{9pt}{10.8pt}\selectfont } \hspace{1pt}\huxbpad{1pt}} &
\multicolumn{1}{c!{\huxvb{0, 0, 0}{0}}}{\huxtpad{1pt + 1em}\centering \hspace{1pt} {\fontsize{9pt}{10.8pt}\selectfont 0.1924} \hspace{1pt}\huxbpad{1pt}} &
\multicolumn{1}{c!{\huxvb{0, 0, 0}{0}}}{\huxtpad{1pt + 1em}\centering \hspace{1pt} {\fontsize{9pt}{10.8pt}\selectfont 0.0246} \hspace{1pt}\huxbpad{1pt}} \tabularnewline[-0.5pt]

\hhline{}
\arrayrulecolor{black}

\multicolumn{1}{!{\huxvb{0, 0, 0}{0}}l!{\huxvb{0, 0, 0}{0}}}{\huxtpad{1pt + 1em}\raggedright \hspace{1pt} {\fontsize{9pt}{10.8pt}\selectfont } \hspace{1pt}\huxbpad{1pt}} &
\multicolumn{1}{c!{\huxvb{0, 0, 0}{0}}}{\huxtpad{1pt + 1em}\centering \hspace{1pt} {\fontsize{9pt}{10.8pt}\selectfont SD} \hspace{1pt}\huxbpad{1pt}} &
\multicolumn{1}{c!{\huxvb{0, 0, 0}{0}}}{\huxtpad{1pt + 1em}\centering \hspace{1pt} {\fontsize{9pt}{10.8pt}\selectfont 0.1524} \hspace{1pt}\huxbpad{1pt}} &
\multicolumn{1}{c!{\huxvb{0, 0, 0}{0}}}{\huxtpad{1pt + 1em}\centering \hspace{1pt} {\fontsize{9pt}{10.8pt}\selectfont 0.0689} \hspace{1pt}\huxbpad{1pt}} &
\multicolumn{1}{c!{\huxvb{0, 0, 0}{0}}}{\huxtpad{1pt + 1em}\centering \hspace{1pt} {\fontsize{9pt}{10.8pt}\selectfont } \hspace{1pt}\huxbpad{1pt}} &
\multicolumn{1}{c!{\huxvb{0, 0, 0}{0}}}{\huxtpad{1pt + 1em}\centering \hspace{1pt} {\fontsize{9pt}{10.8pt}\selectfont 0.1487} \hspace{1pt}\huxbpad{1pt}} &
\multicolumn{1}{c!{\huxvb{0, 0, 0}{0}}}{\huxtpad{1pt + 1em}\centering \hspace{1pt} {\fontsize{9pt}{10.8pt}\selectfont 0.0673} \hspace{1pt}\huxbpad{1pt}} \tabularnewline[-0.5pt]

\hhline{>{\huxb{0, 0, 0}{1}}->{\huxb{0, 0, 0}{1}}->{\huxb{0, 0, 0}{1}}->{\huxb{0, 0, 0}{1}}->{\huxb{0, 0, 0}{1}}->{\huxb{0, 0, 0}{1}}->{\huxb{0, 0, 0}{1}}-}
\arrayrulecolor{black}
\end{tabular}
\end{threeparttable}\par\end{centerbox}

\end{table}

%% file: table_realdata_continuous.tex
  \providecommand{\huxb}[2]{\arrayrulecolor[RGB]{#1}\global\arrayrulewidth=#2pt}
  \providecommand{\huxvb}[2]{\color[RGB]{#1}\vrule width #2pt}
  \providecommand{\huxtpad}[1]{\rule{0pt}{#1}}
  \providecommand{\huxbpad}[1]{\rule[-#1]{0pt}{#1}}

\begin{table}[ht]
\begin{centerbox}
\begin{threeparttable}
\captionsetup{justification=centering,singlelinecheck=off}
\caption{Parameter estimates for KLIPS data}
\label{table:continuous}
 \setlength{\tabcolsep}{0pt}
\begin{tabular}{l l l l}

\hhline{>{\huxb{0, 0, 0}{1}}=>{\huxb{0, 0, 0}{1}}=>{\huxb{0, 0, 0}{1}}=>{\huxb{0, 0, 0}{1}}=}
\arrayrulecolor{black}

\multicolumn{1}{!{\huxvb{0, 0, 0}{0}}l!{\huxvb{0, 0, 0}{0}}}{\huxtpad{1pt + 1em}\raggedright \hspace{1pt} \textbf{{\fontsize{9pt}{10.8pt}\selectfont }} \hspace{1pt}\huxbpad{1pt}} &
\multicolumn{1}{c!{\huxvb{0, 0, 0}{0}}}{\huxtpad{1pt + 1em}\centering \hspace{1pt} \textbf{{\fontsize{9pt}{10.8pt}\selectfont \(\alpha\)}} \hspace{1pt}\huxbpad{1pt}} &
\multicolumn{1}{c!{\huxvb{0, 0, 0}{0}}}{\huxtpad{1pt + 1em}\centering \hspace{1pt} \textbf{{\fontsize{9pt}{10.8pt}\selectfont \(\beta\)}} \hspace{1pt}\huxbpad{1pt}} &
\multicolumn{1}{c!{\huxvb{0, 0, 0}{0}}}{\huxtpad{1pt + 1em}\centering \hspace{1pt} \textbf{{\fontsize{9pt}{10.8pt}\selectfont log(OR)}} \hspace{1pt}\huxbpad{1pt}} \tabularnewline[-0.5pt]

\hhline{>{\huxb{0, 0, 0}{0.4}}->{\huxb{0, 0, 0}{0.4}}->{\huxb{0, 0, 0}{0.4}}->{\huxb{0, 0, 0}{0.4}}-}
\arrayrulecolor{black}

\multicolumn{1}{!{\huxvb{0, 0, 0}{0}}l!{\huxvb{0, 0, 0}{0}}}{\huxtpad{1pt + 1em}\raggedright \hspace{1pt} \textbf{{\fontsize{9pt}{10.8pt}\selectfont Non-optimal GEE}} \hspace{1pt}\huxbpad{1pt}} &
\multicolumn{1}{c!{\huxvb{0, 0, 0}{0}}}{\huxtpad{1pt + 1em}\centering \hspace{1pt} {\fontsize{9pt}{10.8pt}\selectfont 0.25 (0.289)} \hspace{1pt}\huxbpad{1pt}} &
\multicolumn{1}{c!{\huxvb{0, 0, 0}{0}}}{\huxtpad{1pt + 1em}\centering \hspace{1pt} {\fontsize{9pt}{10.8pt}\selectfont 0.923 (0.055)} \hspace{1pt}\huxbpad{1pt}} &
\multicolumn{1}{c!{\huxvb{0, 0, 0}{0}}}{\huxtpad{1pt + 1em}\centering \hspace{1pt} {\fontsize{9pt}{10.8pt}\selectfont 12.621 (0.706)} \hspace{1pt}\huxbpad{1pt}} \tabularnewline[-0.5pt]

\hhline{}
\arrayrulecolor{black}

\multicolumn{1}{!{\huxvb{0, 0, 0}{0}}l!{\huxvb{0, 0, 0}{0}}}{\huxtpad{1pt + 1em}\raggedright \hspace{1pt} \textbf{{\fontsize{9pt}{10.8pt}\selectfont Optimal GEE}} \hspace{1pt}\huxbpad{1pt}} &
\multicolumn{1}{c!{\huxvb{0, 0, 0}{0}}}{\huxtpad{1pt + 1em}\centering \hspace{1pt} {\fontsize{9pt}{10.8pt}\selectfont 0.348 (0.153)} \hspace{1pt}\huxbpad{1pt}} &
\multicolumn{1}{c!{\huxvb{0, 0, 0}{0}}}{\huxtpad{1pt + 1em}\centering \hspace{1pt} {\fontsize{9pt}{10.8pt}\selectfont 0.905 (0.029)} \hspace{1pt}\huxbpad{1pt}} &
\multicolumn{1}{c!{\huxvb{0, 0, 0}{0}}}{\huxtpad{1pt + 1em}\centering \hspace{1pt} {\fontsize{9pt}{10.8pt}\selectfont 12.364 (0.376)} \hspace{1pt}\huxbpad{1pt}} \tabularnewline[-0.5pt]

\hhline{}
\arrayrulecolor{black}

\multicolumn{1}{!{\huxvb{0, 0, 0}{0}}l!{\huxvb{0, 0, 0}{0}}}{\huxtpad{1pt + 1em}\raggedright \hspace{1pt} \textbf{{\fontsize{9pt}{10.8pt}\selectfont Pseudo-likelihood}} \hspace{1pt}\huxbpad{1pt}} &
\multicolumn{1}{c!{\huxvb{0, 0, 0}{0}}}{\huxtpad{1pt + 1em}\centering \hspace{1pt} {\fontsize{9pt}{10.8pt}\selectfont } \hspace{1pt}\huxbpad{1pt}} &
\multicolumn{1}{c!{\huxvb{0, 0, 0}{0}}}{\huxtpad{1pt + 1em}\centering \hspace{1pt} {\fontsize{9pt}{10.8pt}\selectfont } \hspace{1pt}\huxbpad{1pt}} &
\multicolumn{1}{c!{\huxvb{0, 0, 0}{0}}}{\huxtpad{1pt + 1em}\centering \hspace{1pt} {\fontsize{9pt}{10.8pt}\selectfont 10.467 (0.025)} \hspace{1pt}\huxbpad{1pt}} \tabularnewline[-0.5pt]

\hhline{>{\huxb{0, 0, 0}{1}}->{\huxb{0, 0, 0}{1}}->{\huxb{0, 0, 0}{1}}->{\huxb{0, 0, 0}{1}}-}
\arrayrulecolor{black}
\end{tabular}
\end{threeparttable}\par\end{centerbox}

\end{table}